\documentclass[twocolumn,preprintnumbers,superscriptaddress,nofootinbib,aps,prd,floatfix]{revtex4-2}
\pdfoutput=1
\usepackage{enumerate}
\usepackage{amsmath,amssymb}
\usepackage{graphicx}
\usepackage{slashed}
\usepackage{xspace,slashed}
\usepackage{hyperref}
\hypersetup{colorlinks=true, citecolor=blue, urlcolor=blue, linkcolor=blue}
\usepackage[normalem]{ulem}
\usepackage{subfigure,orcidlink}
\usepackage{multirow,array}
\usepackage{float}
\usepackage{hyperref}
\hypersetup{colorlinks=true, citecolor=blue, urlcolor=blue, linkcolor=blue}
\usepackage{tabularray}
\UseTblrLibrary{booktabs}

\begin{document}	
	\title{Asymmetric self-interacting dark matter with a canonical seesaw model}
	%%%%%%%%%%%%%%%%%%%%%%%%%%%%%%%%%%%%%%%%%%%%%%%%%%%%%%%%%%%%%%%%%%%%%%%%%%%%   Authors   %%%%%%%%%%%%%%%%%%%%%%%%%%%%%%%%%%%%%%%%%%%%%%%%%%%%%%%%%%%%%%%%%%%%%%%%%%%%%%%
	\author{Debasish Borah\orcidlink{https://orcid.org/0000-0001-8375-282X}}\email{dborah@iitg.ac.in}
	\affiliation{Department of Physics, Indian Institute of Technology Guwahati, Assam 781039, India}
	
	\author{Satyabrata Mahapatra\orcidlink{https://orcid.org/0000-0002-4000-5071}}
	\email{satyabrata@g.skku.edu}
	\affiliation{Department of Physics and Institute of Basic Science, Sungkyunkwan University, Suwon 16419, Korea}
	
	\author{Partha Kumar Paul\orcidlink{https://orcid.org/0000-0002-9107-5635}}
	\email{ph22resch11012@iith.ac.in}
	\affiliation{Department of Physics, Indian Institute of Technology Hyderabad,\\ Kandi, Telangana-502285, India.}
	
	\author{Narendra Sahu\orcidlink{https://orcid.org/0000-0002-9675-0484}}
	\email{nsahu@phy.iith.ac.in}
	\affiliation{Department of Physics, Indian Institute of Technology Hyderabad,\\ Kandi, Telangana-502285, India.}
	
	\author{Prashant Shukla\orcidlink{https://orcid.org/0000-0001-8118-5331}}
	\email{pshukla@barc.gov.in}
	\affiliation{Nuclear Physics Division, Bhabha Atomic Research Centre,
		Mumbai, 400085, India.}
	\affiliation{Homi Bhabha National Institute, Anushakti Nagar, Mumbai,
		400094, India.}
	
	%%%%%%%%%%%%%%%%%%%%%%%%%%%%%%%%%%%%%%%%%%%%%%%%%%%%%%%%%%%%%%%%%%%
	%%% abstract %%%%%	
	\begin{abstract}
		We study the possibility of generating dark matter (DM) and baryon asymmetry of the Universe (BAU) simultaneously in an asymmetric DM framework, which also alleviates the small-scale structure issues of cold DM. While the thermal relic of such self-interacting DM remains under-abundant due to efficient annihilation into light mediators, a nonzero asymmetry in the dark sector can lead to the survival of the required DM in the Universe. The existence of a light mediator leads to the required self-interactions of DM at small scales while keeping DM properties similar to cold DM at large scales. It also ensures that the symmetric DM component annihilates away, leaving the asymmetric part in the spirit of cogenesis. The particle physics implementation is done in canonical seesaw models of light neutrino mass, connecting it to the origin of DM and BAU. In particular, we consider type-I and type-III seesaw origin of neutrino mass for simplicity and minimality of the field content. We show that the desired self-interactions and relic of DM together with BAU while satisfying relevant constraints lead to strict limits on DM mass $\mathcal{O}({\rm GeV}) \lesssim M_{\rm DM}  \lesssim460 $ GeV. In spite of being a high-scale seesaw, the models remain verifiable in different experiments, including direct and indirect DM searches as well as colliders.
	\end{abstract}	
	
	\maketitle
	%\flushbottom
	%%%%%%%%%%%%%%%%%%%%%%%%%%%%%%%%%%%%%%%%%%%%%%%%%%%%%%%%%%%%%%%%%%%
	%%%%% Introduction %%%%%%%%
	
	\section{Introduction}
	\label{intro}
	
	Dark matter (DM), an enigmatic form of matter devoid of luminous and baryonic compositions, comprises a substantial fraction of the Universe, posing a longstanding puzzle in both cosmology and particle physics. Its presence, inferred from gravitational effects \cite{Zwicky:1933gu,Rubin:1970zza,Clowe:2006eq, Planck:2018vyg}, holds profound implications for our comprehension of the Universe's evolution. The prevailing model in cosmology, known as the $\Lambda$CDM model, has achieved notable success in elucidating several crucial aspects of the observable Universe. Here, $\Lambda$ represents the cosmological constant or dark energy, while CDM denotes cold DM, a pressureless, collisionless fluid essential for initiating the formation of gravitational potential wells crucial for structure formation. However, persistent discrepancies arise at smaller scales between observations and the predictions of collisionless CDM, referred to as the too-big-to-fail, missing satellite, and core-cusp problems in the literature~\cite{Tulin:2017ara, Bullock:2017xww}. Self-interacting DM (SIDM) emerges as a compelling solution to these discrepancies~\cite{Spergel:1999mh,Buckley:2009in, Feng:2009hw, Feng:2009mn, Loeb:2010gj, Zavala:2012us, Vogelsberger:2012ku}, with the requisite self-interaction parametrized in terms of the cross section to the mass ratio: $\sigma/M_{\rm DM} \sim (0.1-100)~ {\rm cm}^2/{\rm g}$. DM self-interactions, facilitated by light force carriers, induce not only substantial self-interaction but also engender velocity-dependent interactions, exhibiting enhanced effectiveness in smaller halos characterized by lower velocity dispersion. However, their efficacy diminishes in larger halos with higher velocity dispersion, consistent with the collisionless picture of CDM \cite{Buckley:2009in, Feng:2009hw, Feng:2009mn, Loeb:2010gj, Bringmann:2016din, Kaplinghat:2015aga, Aarssen:2012fx, Tulin:2013teo}. Nonetheless, the considerable coupling of DM with light mediators, necessary for self-interaction, leads to significant DM annihilation rates, often resulting in a deficit of relic abundance, particularly in the low DM mass regime~\cite{Borah:2022ask}.
	
	On the other hand, observational data from the cosmic microwave background (CMB) and astrophysical observations consistent with the big bang nucleosynthesis (BBN) predictions reveal a remarkable coincidence -- the present-day abundances of DM and baryonic matter are very similar, with the ratio of their energy densities being approximately 5 ({\it i.e.} $\rho_{
		\rm DM}\simeq 5\rho_{B}$)~\cite{Planck:2018vyg}. More precisely, CMB experiments determine the
	relic abundance of DM and baryon to be $\Omega_{\rm DM}h^2=0.12\pm0.0012$ and $\Omega_{B}h^2=0.02237\pm0.00015$, respectively~\cite{Planck:2018vyg}. While there exist different well-motivated frameworks like the weakly interacting massive particle  paradigm of DM \cite{Kolb:1990vq, Jungman:1995df, Bertone:2004pz, Feng:2010gw, Arcadi:2017kky, Roszkowski:2017nbc} and baryogenesis \cite{Weinberg:1979bt, Kolb:1979qa}, leptogenesis \cite{Fukugita:1986hr} to explain the BAU, the remarkable similarity $\rho_{
		\rm DM}\simeq 5\rho_{B}$ may be indicative of a different dynamical origin or cogenesis mechanism. The asymmetric DM (ADM)\footnote{See \cite{Petraki:2013wwa,Zurek:2013wia} for reviews of ADM scenarios.} paradigm~\cite{Nussinov:1985xr, Kaplan:2009ag, Davoudiasl:2012uw, DuttaBanik:2020vfr, Barman:2021ost, Cui:2020dly, Falkowski:2011xh,  Patel:2022xyv, Biswas:2018sib,Narendra:2018vfw,Nagata:2016knk, Arina:2011cu, Arina:2012fb, Arina:2012aj, Narendra:2019cyt,Mahapatra:2023dbr,Borah:2022qln, Borah:2023qag} is one of the most popular cogenesis mechanisms where there exists an asymmetry in the number density of DM over anti-DM, similar to baryons. Since the asymmetries in dark and visible sectors have a common origin, ADM can naturally explain similar abundances of DM and visible matter. Many of such ADM scenarios rely on the leptogenesis route to BAU and can also explain the origin of light neutrino mass and mixing, another observed phenomenon \cite{ParticleDataGroup:2020ssz} which remains unexplained in the standard model (SM).
	
	The ADM paradigm not only provides a cogenesis of baryon and DM but can also lead to the correct thermal relic of SIDM with light mediator which otherwise remain thermally under-abundant in the low mass regime \cite{Borah:2022ask}. While it is possible to have a hybrid of thermal and non-thermal contribution to SIDM relic \cite{Dutta:2021wbn, Borah:2021pet, Borah:2021rbx, Borah:2021qmi}, ADM provides a more minimal setup with other motivations related to cogenesis. A conserved quantum number in the dark sector equivalent to an asymmetry helps in generating the observed DM relic, which cannot be depleted beyond a certain limit, in spite of efficient annihilation of DM and anti-DM particles into light mediators \cite{Iminniyaz:2011yp}. 
	It is noteworthy that in ADM scenarios, the observed DM relic density can be obtained without violating the stringent constraints from the CMB and indirect searches on DM annihilation into charged final states or photons~\cite{Slatyer:2015jla}, in contrast to conventional DM models. This is primarily due to the key feature of ADM whereby it can result in a reduced rate of DM-anti DM annihilation compared to symmetric DM models, owing to the suppressed population of DM antiparticles in the ADM relic density \cite{Lin:2011gj, Baldes:2017gzu}.
	
	Drawing inspiration from these considerations, in this article, we explore asymmetric self-interacting DM (ASIDM) which provides cogenesis together with DM self-interactions. While some earlier works \cite{Petraki:2014uza, Chen:2023rrl, Dutta:2022knf,Ghosh:2021qbo} considered the possibility of ASIDM, we consider a minimal setup within canonical seesaw mechanisms namely, type-I \cite{Minkowski:1977sc, GellMann:1980vs, Mohapatra:1979ia, Schechter:1980gr, Schechter:1981cv} and type-III seesaw \cite{Foot:1988aq}. While it is possible to implement the idea of ASIDM within type-II seesaw mechanism \cite{Mohapatra:1980yp, Schechter:1981cv, Wetterich:1981bx, Lazarides:1980nt, Brahmachari:1997cq} as well, we consider the other two canonical seesaw mechanisms only such that the field content remains minimal. Thus, our setup can accommodate several observed phenomena in the Universe, namely, nonzero neutrino mass, DM, and the BAU, together with providing solutions to the baryon-DM coincidence puzzle and small-scale structure issues of cold DM.  We consider the DM relic to arise purely from the asymmetric component, while the net lepton asymmetry is converted to the observed baryon asymmetry through electroweak sphaleron processes. We illustrate the generation of the requisite lepton asymmetry and the observed DM relic abundance by considering both weak washout and strong washout scenarios in both the type-I and type-III frameworks. This is achieved by solving the Boltzmann equation, which accounts for all relevant processes involved in asymmetry generation and transfer between the visible and dark sectors. We show that the requirement of DM relic, DM self-interactions, and DM-baryon cogenesis, together with all relevant phenomenological constraints, require DM mass to remain within sub-TeV ballpark $\mathcal{O}(\rm GeV) \leq M_{\rm DM} \lesssim 460 $ GeV. For stable DM, the models can have observable direct detection prospects like DM-nucleon or DM-electron scatterings as well as collider productions of TeV scale particles. While stable ADM does not have promising indirect detection prospects due to inefficient annihilation rates at late epochs, a long-lived ADM can show up at gamma-ray or neutrino telescopes due to its decay. We find the parameter space of the model consistent with all phenomenological constraints while keeping the detection prospects promising at different experiments.
	
	The structure of the article is as follows. Section \ref{sidm} provides a concise overview of DM self-interaction within the most minimal setup. Subsequently, Sec. \ref{asidm} outlines the implementation of asymmetric SIDM, detailing its integration into both type-I and type-III seesaw frameworks and exploring the intricacies of cogenesis. In Sec. \ref{unstableDM}, we discuss the possibility of unstable DM and elucidate the constraints on DM lifetime. Following this, in Sec. \ref{detection}, we examine the intriguing detection prospects for asymmetric self-interacting DM through both direct and indirect search experiments. Finally, we conclude in Sec. \ref{conclusion} with several technical details in the appendices. %%%%%%%%%%%%%%%%%%%%%%%%%%%%%%%%%%%%%%%%%%%%%%%%%%%%%%%%%%%%%%%%%%%
	\section{Thermal relic of SIDM}\label{sidm} 
	We consider DM to be a fermionic singlet, denoted as $\chi$ ({\it i.e. $\chi \equiv {\rm DM}$}), characterized as odd under an additional discrete symmetry $\mathcal{Z}_2$, which guarantees its stability.  One light scalar mediator, represented by $\Phi$, which transforms trivially under $\mathcal{Z}_2$ is invoked to mediate the DM self-interactions. 
	The Lagrangian governing self-interaction can be written as:
	\begin{equation}
		\mathcal{L}_{\rm self-int.}= -\lambda_{_{\rm DM}} \Phi \bar{\chi}\chi + {\rm H.c.}
	\end{equation}
	
	\begin{figure}[h]
		\centering
		\includegraphics[scale=0.8]{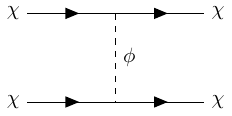}~~
		\includegraphics[scale=0.8]{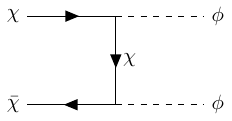}
		\caption{DM self-interaction mediated by light scalar $\phi$ (left) and DM annihilation to a pair of light mediator, $\phi$, (right) where $\phi$ is the physical scalar as given in Eq (\ref{eq:physicalscalar}). }\label{fig:selfint}
	\end{figure}
	
	In this scenario, the non-relativistic self-interaction of DM is effectively described by a Yukawa-type potential $\frac{\lambda^2_{\rm DM}}{4\pi r}e^{-M_{\phi}r}$. The details on the non-relativistic self-interaction
	cross section is given in Appendix~\ref{app:selfint}. The Feynman diagram for DM self-interaction and the dominant number changing process of DM {\it i.e.} DM annihilation into the light mediator $\phi$,  are shown in Fig. \ref{fig:selfint}. 
	
	\begin{figure}[t]
		\centering
		\includegraphics[scale=0.45]{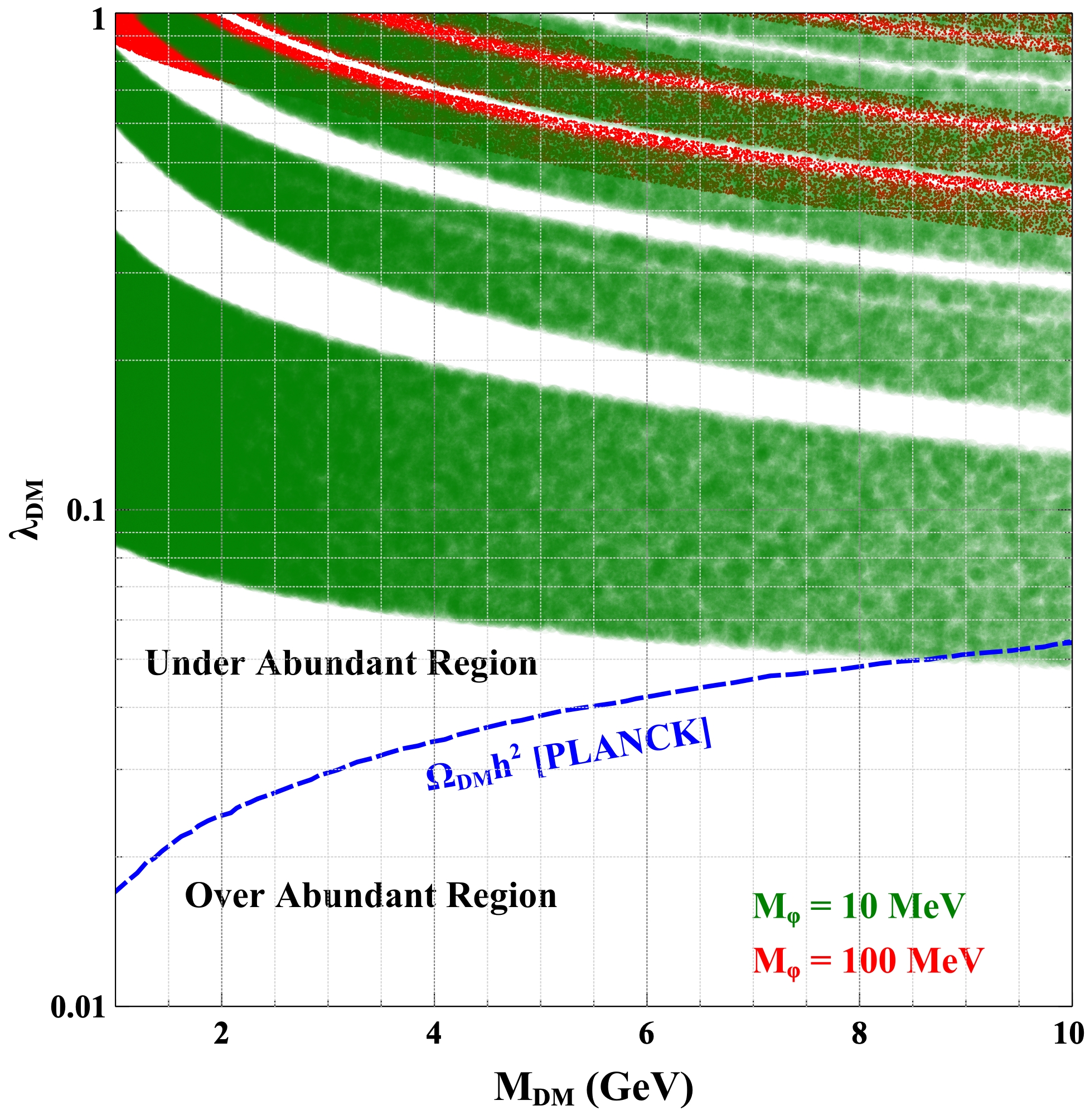}
		\caption{The parameter space is depicted in $\lambda_{_{\rm DM}}$ vs $M_{\rm DM}$ plane for required DM self-interaction: $\sigma/M_{\rm DM} \in [0.1-100] {\rm cm}^2/{\rm g}$.
			The green shaded region corresponds to $M_\phi$= 10 MeV, and the red shaded region corresponds to $M_\phi$= 100 MeV. The blue dashed line corresponds to the correct relic abundance of DM observed by Planck.}\label{fig:gdvsmd}
	\end{figure}
	
	In Fig. \ref{fig:gdvsmd}, we illustrate the parameter space in the $\lambda_{\rm DM}$-$M_{\rm DM}$ plane capable of generating the desired DM self-interactions for two distinct mediator mass values: $M_{\phi} = 10$ MeV (green patches) and 100 MeV (red patches). As discussed in Appendix \ref{app:selfint}, depending on the masses of DM and mediator $\phi$, the coupling $\lambda_{\rm DM}$ and DM velocity, the SIDM parameter space can lie in one of the three regimes namely, the Born regime, classical regime, and resonance regime. In the resonance regime, because of the quantum mechanical resonance features, there are gaps in the viable parameter space, as shown in Fig. \ref{fig:gdvsmd}\footnote{Similar features can also be found in the resonance regime in the plane of $M_{\phi}$-$M_{\rm DM}$ as shown in Fig \ref{fig:DDplot}.}.
	The blue dashed line represents the contour of the correct thermal relic abundance of DM, which is solely decided by its annihilation to $\phi$, the cross section for which, ignoring the mass of $\phi$, is given by
	\begin{equation}
		\langle \sigma v \rangle_{\chi \bar{\chi}\to \phi\phi} \sim \frac{\lambda^4_{\rm DM}}{16 \pi M^2_{\rm DM}}.
	\end{equation}
	Evidently, in the low mass regime, the parameter space meeting the self-interaction criteria results in an under-abundant thermal DM relic. This deficit can be avoided in an asymmetric SIDM setup, which we will discuss in the next section.
	
	The scalar potential of the model containing singlet scalar $\Phi$ and the SM Higgs doublet $H$ is given by
	\begin{eqnarray}
		V(H, \Phi)&=&\frac{1}{2} \mu^2_\Phi \Phi^2 -\mu_H^2 (H^\dagger H)  + \lambda_H(H^\dagger H)^2 + \frac{\lambda_{\Phi}}{4}\Phi^4\nonumber \\&+&\frac{\mu_1}{\sqrt{2}}\Phi(H^\dagger H)+\frac{1}{2}\lambda_{H\Phi}(H^\dagger H)\Phi^2\,
		\label{Eq:scalarpot}
	\end{eqnarray}
	Parametrizing the scalars $H$ and $\Phi$ as
	\begin{equation}
		H=\frac{1}{\sqrt{2}}\begin{pmatrix}
			0\\
			v_0+h
		\end{pmatrix}\,,~~ \Phi=\phi+u\label{eq:physicalscalar}
	\end{equation}
	the vacuum expectation values (VEVs) are estimated to be $$v_0 \simeq\sqrt{ \frac{2 \mu_H^2-\lambda_{H\Phi}u^2-\sqrt{2}\mu_1u}{2\lambda_H}}, ~u\simeq-\frac{\frac{\mu_1}{\sqrt{2}}v^2_0}{2\mu^2_\Phi+\lambda_{H\Phi}v^2_0}.$$
	The mass-squared matrix spanning the $h$ and $\phi$ is given by
	\begin{equation}
		M^2_{h\phi}\simeq\begin{pmatrix}
			2\lambda_H v_0^2 & \frac{\mu_1}{\sqrt{2}} v_0+\lambda_{H\Phi}uv_0\\
			\frac{\mu_1}{\sqrt{2}} v_0+\lambda_{H\Phi}uv_0 &  2\lambda_{\Phi}u^2-\frac{1}{2\sqrt{2}}\frac{\mu_{1}v^2_0}{u}
		\end{pmatrix}.
	\end{equation}
	Diagonalizing the above mass matrix, we get the mass eigenstates $h_1$ with mass  $M_{h_1}=125$ GeV and $h_2$ with mass $M_{h_2}$. The $h-\phi$ mixing angle is then given by  
	\begin{equation}
		\tan 2\gamma \sim \frac{2(\frac{\mu_1}{\sqrt{2}} v_0+\lambda_{H\Phi}uv_0)}{2\lambda_H v_0^2-2\lambda_{\Phi}u^2+\frac{1}{2\sqrt{2}}\frac{\mu_{1}v^2_0}{u}}.\,
	\end{equation}
	As we discuss in Sec. \ref{detection}, the $\phi-h$ mixing opens up a window for direct detection of DM. For all practical purposes, we use a small mixing angle, $\gamma$, for which $M_{h_2} \simeq M_{\phi}$.
	%%%%%%%%%%%%%%%%%%%%%%%%%%%%%%%%%%%%%%%%%%%%%%%%%%%%%%%%%%%%%%%%%%%	
	\section{Asymmetric SIDM in seesaw frameworks}\label{asidm}
	
	To successfully realize the ASIDM scenario within the type-I seesaw framework, which extends the standard model (SM) by introducing right-handed neutrinos (RHNs) $N_{R_{i}}$ to facilitate light neutrino mass generation, it is imperative to introduce an additional dark singlet scalar ($\rho$) into the particle spectrum alongside the fermionic DM particle $\chi$. Similarly, within the type-III seesaw scenario, where the SM is extended with $SU(2)_L$ triplet fermions $\Sigma$ to achieve light neutrino mass generation, ensuring the gauge invariance of the Lagrangian mandates the inclusion of an extra triplet scalar ($\Delta$) to enable the realization of the ASIDM scenario. In both the type-I and type-III frameworks, the simultaneous $CP$-violating out-of-equilibrium decay of the RHNs (or triplet fermions) into both the visible and dark sectors generates a net lepton asymmetry and an asymmetry in the DM $\chi$. Here we discuss the details of ASIDM implementation in these two canonical seesaw models.
	%%%%%%%%%%%%%%%%%%%%%%%%%%%%%%%%%%%%%%%%%%%%%%%%%%%%%%%%%%%%%%
	\subsection{Model I (decay of RHN)}\label{rhn}
	The fermion sector of the SM is extended by incorporating three right-handed neutrinos ($N_{R_1}, N_{R_2}, N_{R_3}$) in addition to the singlet Dirac fermion ($\chi$) of bare mass $M_\chi$. The right-handed neutrinos are assigned a lepton number of 0 ($L=0$), while $\chi$ is designated a lepton number of 1 ($L=1$). In order to allow the coupling of $\chi$ with RHNs, we introduce a real singlet scalar $\rho$, keeping it heavier than DM. An additional $\mathcal{Z}_2$ symmetry under which $\chi, \rho$ are odd while other fields are even is imposed to ensure the stability of DM. The real singlet scalar $\Phi$, even under $\mathcal{Z}_2$ symmetry, gives rise to DM self-interactions discussed before.
	
	The relevant Lagrangian is given by
	\begin{eqnarray}
		\mathcal{L}& \supset &- \frac{1}{2} M_{N_R} \overline{N_R^c}N_R -  y_{N} \overline{L}  \Tilde{H} N_R -
		y_\chi \overline{N_R}~\rho~\chi + {\rm H.c.}  \nonumber\\
		\label{lagrangian1} 
	\end{eqnarray}
	The out-of-equilibrium decay of the lightest RHN into $L \, H$ and $\chi \, \rho$ can create lepton and DM asymmetries which will be discussed in details below.
	%%%%%%%%%%%%%%%%%%%%%%%%%%%%%%%%%%%%%%%%%%%%%%%%%%%%%%%%%%%%%%
	\subsection{Model II (decay of fermion triplet)}\label{triplet}
	In this model, we extend the SM with three hypercharge-less fermion triplets ($\Sigma_1, \Sigma_2, \Sigma_3$), one gauge singlet Dirac fermion $\chi$ with bare mass $M_\chi$, and one hypercharge-less triplet scalar $\Delta$. We assign the fermion triplet zero lepton number and the $\chi$ lepton number one. To get the velocity-dependent self-interaction of the DM, we include one light scalar mediator, $\phi$. An additional discrete symmetry, $\mathcal{Z}_2$, is imposed, under which $\chi$ and $\Delta$ are odd, while all other fields are even. $\chi$ is chosen to be the lightest $\mathcal{Z}_2$-odd particle ensuring DM stability.
	
	The relevant Lagrangian in this setup is given by
	\begin{eqnarray}
		\mathcal{L}& \supset &- \frac{1}{2} {\rm Tr}[M_\Sigma \overline{\Sigma^c}\Sigma] -\sqrt{2}y_{_\Sigma} \overline{L} \Sigma \Tilde{H} - 
		y_\chi {\rm Tr}[\overline{\Sigma}~\Delta~\chi] + {\rm H.c.}  \nonumber\\
		\label{lagrangian2} 
	\end{eqnarray}
	
	For details of neutrino mass generation in both these scenarios, please refer to Appendix~\ref{app:numass}.
	
	%%%%%%%%%%%%%%%%%%%%%%%%%%%%%%%%%%%%%%%%
	
	\subsubsection{\bf{ASIDM and cogenesis}}\label{cogenesis}
	
	The additional fermion $x (\equiv N_R~ {\rm or} ~\Sigma)$ simultaneously decays into both dark sector ($\chi, y (\equiv \rho ~ {\rm or} ~\Delta)$) and the visible sector ($L, H$). If the three Sakharov conditions \cite{Sakharov:1967dj} are satisfied, then this can lead to the generation of an asymmetry in both the dark sector and the visible sector. The symmetric portion of the DM is depleted through annihilation into a light mediator ensured by sizeable DM coupling to mediator $\Phi$ required for self-interactions, while the residual asymmetry contributes to the DM relic as asymmetric DM.
	We consider a mass hierarchy among the fermion mass states as $M_{x_1}<M_{x_2}<M_{x_3}$. Given that $x_1$ is the lightest among the three, any asymmetry resulting from the decay of $x_2$ and $x_3$ at high temperatures will be washed out by the interactions of $x_1$ efficient at lower temperatures. Hence, for pragmatic purposes, we solely focus on interactions involving $x_1$, which will ultimately generate asymmetry in both sectors. The asymmetry in both sectors originates from the decay of $x_1$. However, this produced asymmetry undergoes partial attenuation due to the washout processes, such as inverse decay and scatterings that violate the lepton number by 1 and 2 units. In addition, one should also incorporate the lepton number conserving scatterings into the analysis to take into account the transfer of asymmetry between the two sectors.
	
	Here it is worth noting that, in the type-III scenario, the presence of additional gauge interactions associated with the fermion triplets solely impacts the number density of these triplets. As these gauge interactions conserve the lepton number, they do not directly contribute to the asymmetry. As outlined in earlier works on type-III seesaw leptogenesis \cite{Hambye:2003rt, Hambye:2012fh}, the gauge interactions facilitate the rapid attainment of thermal equilibrium for triplets with masses up to $10^{14}$ GeV, assuming a dynamical initial abundance. Even if the initial abundance starts at zero, due to these gauge interactions, the system behaves akin to the case with thermal initial abundance. Contrarily, in the type-I scenario, such a phenomenon does not occur. RHNs lack gauge interactions, and, thus, the asymmetries are influenced by the initial conditions of the RHN number density.
	
	Depending on the Yukawa coupling and $M_{\Sigma_1}$, the triplets predominantly maintain equilibrium via either gauge or Yukawa interactions. The Yukawa interactions are parametrized by a parameter known as effective neutrino mass
	\begin{eqnarray}
		\Tilde{m}_1=\frac{(y_{_\Sigma}^\dagger y_{_\Sigma})_{11}v^2_0}{2M_{\Sigma_1}}.
	\end{eqnarray}
	On the other hand, the rates of gauge interactions depend solely on the electroweak gauge coupling $g$ and $M_{\Sigma_1}$. For lower (higher) $M_{\Sigma_1}$ values, gauge processes maintain equilibrium for a longer (shorter) duration. The prolonged equilibrium of triplets delays their decoupling and subsequent decay, resulting in a lower number density of triplets available for asymmetry production.
	This poses challenges in realizing the observed baryon asymmetry for very small $\Tilde{m}_1$, as in such a scenario, asymmetry suppression due to reduced abundance of the triplet by virtue of gauge interactions becomes prominent. This scenario is referred to as the \textquotedblleft
	gauge regime\textquotedblright. Conversely, for larger $\Tilde{m}_1$ values, equilibrium is established via the inverse decays, leading to its dominance over gauge interactions. In such cases, triplets remain in equilibrium mainly due to Yukawa interactions, hence termed as the \textquotedblleft
	Yukawa regime\textquotedblright.  Here, it is worth mentioning that in the gauge regime, the inverse decay washout effect becomes negligible, and similarly, in the Yukawa regime, the gauge annihilation terms can be ignored.  Thus, in the Yukawa regime, the efficiency in the type-III setup is similar to that of a decaying RHN in the type-I setup. Additionally, for small values of $\Tilde{m}_1$, typically $\Tilde{m}_1<10^{-3}$, where Yukawa interactions are subdued compared to gauge interactions, the weak washout regime entirely lies within the gauge regime across all masses of $M_\Sigma$.
	
	In the type-I seesaw scenario, the Boltzmann equations (BEs) governing the generation of asymmetries as well as the abundance of the  $N_{R_1}$, can be written as
	\begin{widetext}
		\begin{eqnarray}
			\frac{d Y_{N_{R_1}}}{ d z}&=&- \frac{\Gamma_D}{{\rm H^\prime} z} (Y_{N_{R_1}} - Y_{N_{R_1}}^{\rm eq})- \frac{\Gamma_{\Delta L=1}}{{\rm H^\prime}z} (Y_{N_{R_1}} - Y_{N_{R_1}}^{\rm eq})- \frac{(\Gamma_{\Delta L=0}+\Gamma^\prime_{\Delta L=2})}{{\rm H^\prime}z} \frac{(Y^2_{N_{R_1}} - (Y_{N_{R_1}}^{\rm eq})^2)}{Y_{N_{R_1}}^{\rm eq}}\label{eq:BEY}
		\end{eqnarray}
		\begin{eqnarray}
			\frac{d Y_{\Delta L}}{d  z}&=&\epsilon_L~ \frac{\Gamma_D}{{\rm H^\prime}z} ~(Y_{N_{R_1}} - Y_{N_{R_1}}^{eq}) -\left( \frac{1}{2} \frac{\Gamma_D}{{\rm H^\prime}z} ~\frac{Y_{N_{R_1}}^{\rm eq}}{Y_L^{\rm eq}}~ {\rm Br}_L + \frac{\Gamma^W_{\Delta L=1}+\Gamma^W_{\Delta L=2}}{{\rm H^\prime}z} \right) Y_{\Delta L}\nonumber\\&-&\frac{\Gamma_{N_{R_1}}}{\rm H_1} {\rm Br}_L {\rm Br}_\chi \left(I_{T_+}(z)(Y_{\Delta L}+Y_{\Delta \chi})+ I_{T_-}(z)(Y_{\Delta L}-Y_{\Delta \chi})\right)\label{eq:BEYl}
		\end{eqnarray}
		\begin{eqnarray}
			\frac{d Y_{\Delta \chi}}{d z}&=&\epsilon_\chi~ \frac{\Gamma_D}{{\rm H^\prime}z} ~(Y_{N_{R_1}} - Y_{N_{R_1}}^{\rm eq}) -\left( \frac{1}{2} \frac{\Gamma_D}{{\rm H^\prime}z} ~\frac{Y_{N_{R_1}}^{\rm eq}}{Y_\chi^{\rm eq}}~ {\rm Br}_\chi + \frac{\Gamma^W_{\Delta L=1}+\Gamma^W_{\Delta L=2}}{{\rm H^\prime}z} \right) ~Y_{\Delta \chi}\nonumber\\&-&\frac{\Gamma_{N_{R_1}}}{\rm H_1} {\rm Br}_L {\rm Br}_\chi \left(I_{T_+}(z)(Y_{\Delta \chi}+Y_{\Delta L})+ I_{T_-}(z)(Y_{\Delta \chi}-Y_{\Delta L})\right)
			\label{eq:BEYx}
		\end{eqnarray}
	\end{widetext}
	where $z=M_{N_{R_1}}/T$, ${\rm H}^\prime={\rm H_1} z^{-2}$, is the Hubble parameter with ${\rm H_1}=1.66\sqrt{g_*}M^2_{N_{R_1}}/M_{\rm pl}$, $g_*$ is the effective relativistic degrees of freedom and $M_{\rm pl}$ is the Planck mass. Here, $Y_p=n_p/n_\gamma$ denotes the comoving number density of particle $p$ with $n_p, n_\gamma$ denoting the number density of $p$ and the photon number density of the Universe, respectively. Similarly, $Y_{\Delta p}=(n_p-n_{\overline{p}})/n_\gamma$ denotes the comoving density of the asymmetry in $p$. ${\rm Br}_p$ denotes the branching ratio of the mother particle's decay into particle $p$.
	
\begin{figure*}[tbh]
		\centering \includegraphics[width=3.5cm,height=2.2cm]{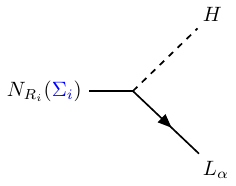}
		\includegraphics[width=3.5cm,height=2.2cm]{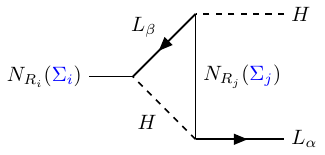}
		\includegraphics[width=3.5cm,height=2.2cm]{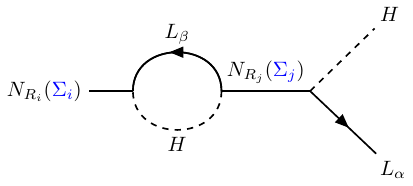}
		\includegraphics[width=3.5cm,height=2.2cm]{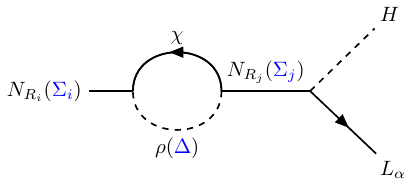}\\
		\includegraphics[width=3.5cm,height=2.2cm]{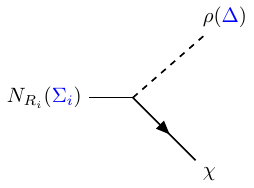}
		\includegraphics[width=3.5cm,height=2.2cm]{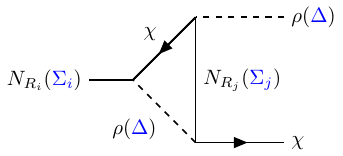}
		\includegraphics[width=3.5cm,height=2.2cm]{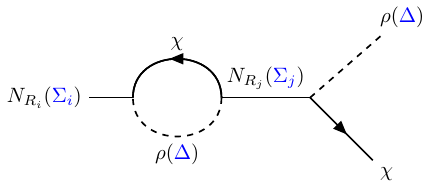}
		\includegraphics[width=3.5cm,height=2.2cm]{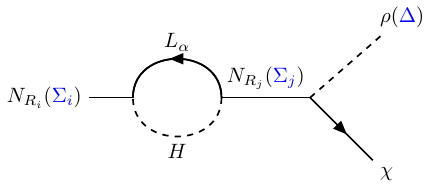}
		\caption{The visible sector (dark sector) nonzero $CP$ asymmetry arises due to the interference of the tree-level diagram with the loop-level diagrams, as shown in the top (bottom) panel.}
		\label{fig:cpasymm}
	\end{figure*}

	Similarly, the Boltzmann equations (BEs) governing the asymmetries as well as the abundance of the BSM fermion, $\Sigma_1$, are given as,
	\begin{widetext}
		\begin{eqnarray}
			\frac{d Y_{\Sigma_{1}}}{ d z}&=&- \frac{\Gamma_D}{{\rm H^\prime} z} (Y_{\Sigma_{1}} - Y_{\Sigma_{1}}^{\rm eq})- \frac{\Gamma_{\Delta L=1}}{{\rm H^\prime}z} (Y_{N_{R_1}} - Y_{N_{R_1}}^{\rm eq})-\frac{(\Gamma_{\Delta L=0}+\Gamma_A)}{{\rm H^\prime}z} \frac{(Y^2_{\Sigma_{1}} - (Y_{\Sigma_{1}}^{\rm eq})^2)}{Y_{\Sigma_{1}}^{\rm eq}}\label{eq:BEY3}
		\end{eqnarray}
		\begin{eqnarray}
			\frac{d Y_{\Delta L}}{d  z}&=&\epsilon_L~ \frac{\Gamma_D}{{\rm H^\prime}z} ~(Y_{\Sigma_{1}} - Y_{\Sigma_{1}}^{\rm eq}) -\left( \frac{1}{2} \frac{\Gamma_D}{{\rm H^\prime}z} ~\frac{Y_{\Sigma_{1}}^{\rm eq}}{Y_L^{\rm eq}}~ {\rm Br}_L + \frac{\Gamma^W_{\Delta L=1}+\Gamma^W_{\Delta L=2}}{{\rm H^\prime}z} \right) Y_{\Delta L}\nonumber\\&-&\frac{\Gamma_{\Sigma_{1}}}{\rm H_1} {\rm Br}_L {\rm Br}_\chi \left(I_{T_+}(z)(Y_{\Delta L}+Y_{\Delta \chi})+ I_{T_-}(z)(Y_{\Delta L}-Y_{\Delta \chi})\right)\label{eq:BEYl3}
		\end{eqnarray}
		\begin{eqnarray}
			\frac{d Y_{\Delta \chi}}{d z}&=&\epsilon_\chi~ \frac{\Gamma_D}{{\rm H^\prime}z} ~(Y_{\Sigma_{1}} - Y_{\Sigma_{1}}^{\rm eq}) -\left( \frac{1}{2} \frac{\Gamma_D}{{\rm H^\prime}z} ~\frac{Y_{\Sigma_{1}}^{\rm eq}}{Y_\chi^{\rm eq}}~ {\rm Br}_\chi + \frac{\Gamma^W_{\Delta L=1}+\Gamma^W_{\Delta L=2}}{{\rm H^\prime}z}\right) ~Y_{\Delta \chi}\nonumber\\&-&\frac{\Gamma_{\Sigma_{1}}}{\rm H_1} {\rm Br}_L {\rm Br}_\chi \left(I_{T_+}(z)(Y_{\Delta \chi}+Y_{\Delta L})+ I_{T_-}(z)(Y_{\Delta \chi}-Y_{\Delta L})\right)
			\label{eq:BEYx3}
		\end{eqnarray}
	\end{widetext}
	where $z=M_{\Sigma_{1}}/T$, and definitions of ${\rm H^\prime}, Y_p, Y_{\Delta p}$ remain the same as before. In both sets of BEs for type-I and type-III seesaw, $\Gamma_D=\Gamma_{x}*K_1(z)/K_2(z)$, $\Gamma_x$ is the total decay width of mother particle $x$ while $K_1(z)$, $K_2(z)$ being the modified Bessel functions of first and second kind, respectively. $\epsilon_L$ and $\epsilon_\chi$, respectively, are the visible sector and dark sector $CP$ asymmetry parameters arising from the interference of the tree level and one loop level diagram shown in Fig \ref{fig:cpasymm} are  defined as
	\begin{eqnarray}  \epsilon_L=\frac{\Gamma_{x\rightarrow LH}-\Gamma_{x\rightarrow \bar{L}H^\dagger}}{\Gamma_x};\epsilon_\chi=\frac{\Gamma_{x\rightarrow \chi y}-\Gamma_{x\rightarrow \bar{\chi}y^\dagger}}{\Gamma_x}.
	\end{eqnarray}
	In the hierarchical scenario, the CP asymmetries are estimated as \cite{Hambye:2003rt,Falkowski:2011xh}
	\begin{eqnarray}
		\epsilon_L&=&-\frac{1}{16\pi}\frac{M_1}{((y^\dagger y)_{11}+y_{\chi_1} y^*_{\chi_1})}\sum_j\frac{1}{M_j}\big\{3(1){\rm Im}[(y^\dagger y)^2_{1j}]\nonumber\\&+&2{\rm Im}[(y^\dagger y)_{1j}y_{\chi_1}y^*_{\chi_j}]\big\},\label{eq:cpL}
	\end{eqnarray}
	\begin{eqnarray}
		\epsilon_\chi&=&-\frac{1}{16\pi}\frac{M_1}{((y^\dagger y)_{11}+y_{\chi_1} y^*_{\chi_1})}\sum_j\frac{1}{M_j}\big\{3(1){\rm Im}[(y_{\chi_1}y^*_{\chi_j})^2]\nonumber\\&+&2{\rm Im}[(y^\dagger y)_{1j}y_{\chi_1}y^*_{\chi_j}]\big\},\label{eq:cp_X}
	\end{eqnarray}
	where the factor 3(1) comes in the case of right-handed neutrino(triplet fermion) decay.
	
	In the BEs, $\Gamma_{\Delta L=1}$, $\Gamma_{\Delta L=2}$, and $\Gamma_{\Delta L=0}$ are the interaction 
	rates for $\Delta L=1$, $\Delta L=2$ and lepton number conserving scatterings, respectively, that affects the abundance of $N_{R_1}(\Sigma_1)$ \footnote{$\Gamma^\prime_{\Delta L=2}$ corresponds to the process $N_{R_1}N_{R_1}\rightarrow\chi\chi$}. The $\Delta L=0$, $\Delta L=1$, and $\Delta L=2$ processes have been shown explicitly in Appendices~\ref{ap:delL0}, \ref{ap:delL1}, and \ref{ap:delL2}, respectively. In Eq (\ref{eq:BEY}), $\Gamma^\prime_{\Delta L=2}$ corresponds to the process $N_{R_1}N_{R_1}\rightarrow\chi\chi$. $\Gamma_A$ denotes the gauge interactions of the triplet fermion shown in Appendix~\ref{ap:gaugepro}. Here, it is worth noting that $\Gamma^W_{\Delta L=1}$, $\Gamma^W_{\Delta L=2}$ are the scattering rates of the $\Delta L=1,2$ washout processes that contribute to the diminishing of the asymmetries. For instance, if we consider the process, $N_{R_1} N_{R_1}\rightarrow \chi \chi$, then the interaction rate is expressed as $\Gamma = \langle\sigma v\rangle_{N_{R_1} N_{R_1}\rightarrow \chi \chi} \times n _{N_{R_1}}$. This rate governs the evolution of the $N_{R_1}$ abundance.  Additionally, since this process violates the lepton number by two units, it contributes to the washout of the asymmetries. For these washout processes, the interaction rates will depend on the number density of $\chi$. Thus, the interaction rate for the washout process can be written as $\Gamma^W = \langle \sigma v\rangle_{N_{R_1} N_{R_1}\rightarrow \chi \chi} \times n^{\rm eq}_\chi$. 
	The last terms in Eqs. \eqref{eq:BEYl}, \eqref{eq:BEYx}, \eqref{eq:BEYl3}, and \eqref{eq:BEYx3} represent the $\Delta L=0$ processes that transfer the asymmetries from one sector to another. These are shown in Appendix~\ref{ap:delL0transfer}.  We have used \texttt{calcHEP}~\cite{Belyaev:2012qa} to calculate the scattering cross sections.  The analytical expressions for $I_{T_+}$ and $I_{T_-}$ are adopted from \cite{Falkowski:2011xh}. 
	
	The produced lepton asymmetry then gets converted to the observed baryon asymmetry [$\eta_{_B}\equiv {(n_B-n_{\overline{B}})}/{n_\gamma}=(6.1\pm0.3)\times10^{-10}$] by the electroweak sphalerons \cite{Kuzmin:1985mm}. In the type-I seesaw case, the baryon asymmetry and the lepton asymmetries are related as:
	\begin{eqnarray}
		\eta_{_B}(z\rightarrow\infty)&=&\frac{C_{L\rightarrow B}}{f}Y_{\Delta L}(z\rightarrow\infty)\nonumber\\&=&\frac{C_{L\rightarrow B}}{f}\epsilon_L \kappa_{L}(z\rightarrow\infty)Y^{\rm eq}_{N_{R_1}}(z\rightarrow0),\nonumber\\
	\end{eqnarray}
	Here, $C_{L\rightarrow B}$ is the lepton to baryon conversion factor, which in the type-I case is calculated to be $-0.518519$ and $f=g^{*}_{s}/g^{0}_{s}=29.1560$ is the dilution factor calculated assuming standard photon production from the onset of leptogenesis
	till recombination.\footnote{$g^*_s=106.75+\frac{7}{8}\times2+\frac{7}{8}\times4+1+1=114$, is the relativistic degrees of freedom (d.o.f) at the onset of leptogenesis for the type-I seesaw scenario and $g^*_0=3.91$ is the relativistic d.o.f today.} For details of $C_{L\rightarrow B}$ calculation, please refer to Appendix~\ref{app:conv}. Consequently, the required lepton asymmetry falls within the range $\{3.2613\times10^{-8},3.5987\times10^{-8}\}$. Here, $\kappa_{L}$ and $\kappa_{\chi}$ are the visible sector and dark sector efficiency factors defined as
	\begin{eqnarray}
		\kappa_L=\frac{Y_{\Delta L}}{\epsilon_L Y^{\rm eq}_{x}|_{T\gg M_x}},\kappa_\chi=\frac{Y_{\Delta \chi}}{\epsilon_\chi Y^{\rm eq}_{x}|_{T\gg M_x}}
	\end{eqnarray}
	
	Conversely, in the type-III seesaw scenario, the relation between the baryon asymmetry and the lepton asymmetries is expressed as
	\begin{eqnarray}
		\eta_{_B}(z\rightarrow\infty)&=&3\frac{C_{L\rightarrow B}}{f}Y_{\Delta L}(z\rightarrow\infty)\nonumber\\&=&3\frac{C_{L\rightarrow B}}{f}\epsilon_L \kappa_{L}(z\rightarrow\infty)Y^{eq}_{\Sigma_{_1}}(z\rightarrow0).\nonumber\\
	\end{eqnarray}
	In our model, $C_{L\rightarrow B}=-0.518519$, which is explicitly calculated in Appendix~\ref{app:conv}. Here $f$ is 30.5627.\footnote{$g^*_s=106.75+\frac{7}{8}\times6+\frac{7}{8}\times4+3+1=119.5$, is the relativistic degrees of freedom (d.o.f) at the onset of leptogenesis in the type-III seesaw scenario, $g^*_0=3.91$ is the present day relativistic d.o.f.. } Thus, in this case, the required lepton asymmetry ranges from $\{1.1396\times10^{-8},1.2574\times10^{-8}\}$.
	
	Here, it is worth noting that the ratio of the DM relic density and baryon relic density can be expressed as 
	\begin{eqnarray}
		\label{eq:ratio}	R\equiv\frac{\Omega_{\rm DM}h^2}{\Omega_{B}h^2}=\frac{f}{(3)C_{L\rightarrow B}}\frac{M_{\rm DM}}{m_p}\frac{\epsilon_\chi}{\epsilon_L}\frac{\kappa_{\chi}}{\kappa_{L}},
	\end{eqnarray}
	where $m_p$ is the proton mass and the factor $3$ in the denominator only appears in the type-III seesaw scenario.
	As mentioned earlier, given the ratio $R\sim 5$, one might naively anticipate that $M_{\rm DM}\sim 5 m_{p}$, assuming a similarity in the asymmetry of the number densities between DM and baryons. However, as indicated by Eq.~\eqref{eq:ratio}, the resulting asymmetries in the visible and dark sectors can vary significantly depending on the $CP$ asymmetry parameters $\epsilon_\chi$ and $\epsilon_L$ and corresponding efficiency parameters. Consequently, there exists the possibility for the DM mass to be viable across a wide range of masses.
	
	\begin{figure}[b]
		\includegraphics[scale=0.5]{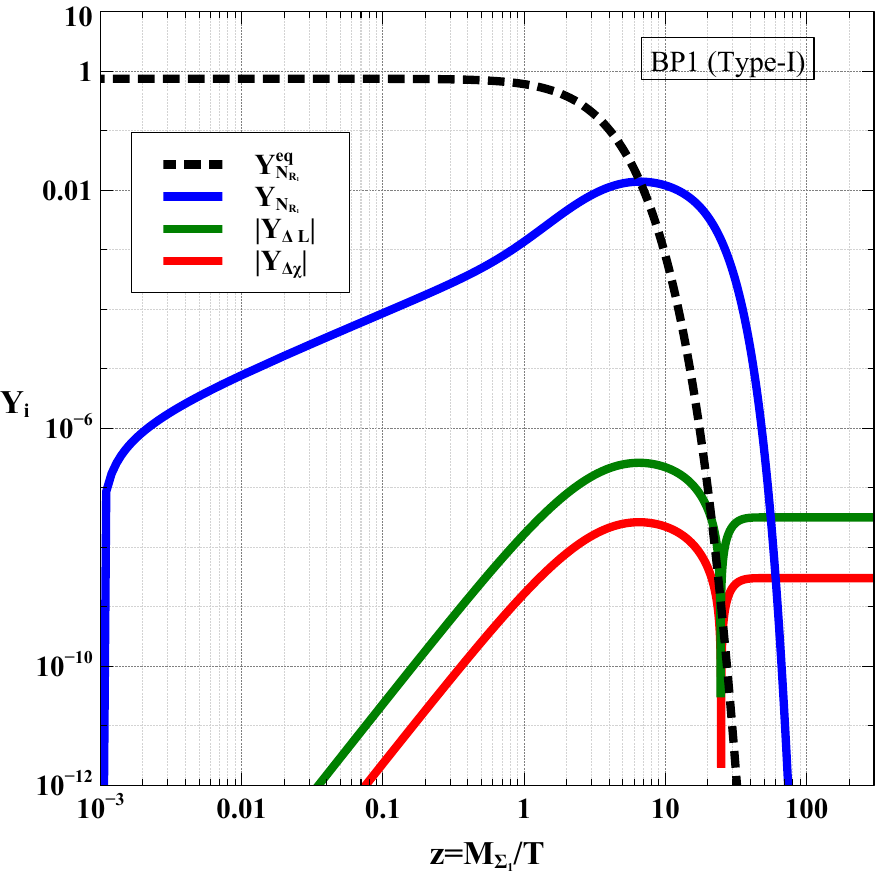}
		\caption{Cosmological evolution of the RHN abundance, visible sector, and dark sector asymmetries in case of weak washout in type-I seesaw. The final asymmetries are $Y_{\Delta L}=3.34696\times10^{-8}$, $Y_{\Delta \chi}=3.01718\times10^{-9}$, $R=5.40394$.}\label{fig:type1weak}
	\end{figure}
	
	\begin{table*}[t]
		\centering
		\caption{Benchmark points chosen for showcasing the evolution of visible sector and dark sector asymmetries.
		}
		\resizebox{18cm}{!}{
			\begin{tblr}{
					colspec={lllllllll},
					row{1}={font=\bfseries},
					column{1}={font=\itshape},
					row{1}={bg=gray!50},row{2}={bg=gray!40},row{3}={bg=gray!30},row{4}={bg=gray!20},row{5}={bg=gray!10},
				}
				\toprule BPs&$M_1\rm(GeV)$& $y_{\chi_1}$ &$y_{\chi_2}$ &$y_{\chi_3}$ & $z_a$ & $\epsilon_L$ & $\epsilon_\chi$  \\
				\toprule
				BP1(Type I)& $5.00\times10^{12}$ & $5.7488\times10^{-4}$ & $(6.4931+i1.6841)\times10^{-3}$ & $(19.7791+i0.4963)\times10^{-2}$ & $1.5256+i7.3652\times10^{-3}$ & $2.20\times10^{-5}$ & $2.08\times10^{-6}$\\
				
				BP2(Type I)&$5.00\times10^{12}$ & $8.1301\times10^{-3}$ & $(6.2571+i2.3119)\times10^{-2}$ & $0.3983+i2.1705$& $0.2212+i6.5979\times10^{-3}$ &$3.91\times10^{-4}$&$3.48\times10^{-7}$\\
				
				BP1(Type III)& $1.53\times10^{12}$&$7.6650\times10^{-4}$&$(5.4113+i3.1460)\times10^{-4}$&$0.2907+i8.8866\times10^{-2}$&$(9.0627+i1.0406)\times10^{-3}$&$1.16\times10^{-5}$&$2.94\times10^{-8}$\\
				
				BP2(Type III)&$6.28\times10^{12}$&$1.8223\times10^{-3}$&$2.2.3007+i2.9800$&$2.8471+i0.1452$&$1.1593\times10^{-2}+i0.2903$&$2.91\times10^{-5}$&$7.30\times10^{-6}$\\
				\bottomrule
		\end{tblr}}
		\label{tab:tab1}
	\end{table*}
	
	We now calculate the visible sector Yukawa coupling using the Casas-Ibarra  parametrization as mentioned in Eqs. \eqref{Eq:casasibarra1} and \eqref{Eq:casasibarra}. The best-fit values of the neutrino oscillation parameters are used to obtain the Yukawa couplings in the normal ordering from \cite{deSalas:2020pgw}. The heavier fermion masses are fixed at $M_2=10M_1$ and $M_3=50M_1$. We have considered a general form of the rotation matrix,  $\mathcal{R}\equiv\mathcal{R}_{12}\mathcal{R}_{23}\mathcal{R}_{13}$ in the Casas-Ibarra parametrization with a complex angle $z_a$. The two Majorana phases in the Pontecorvo–Maki–Nakagawa–Sakata (PMNS) matrix and the lightest SM neutrino mass state are set to zero. We then compute the visible and dark sector $CP$ asymmetry parameters by varying the dark sector Yukawa coupling using Eqs (\ref{eq:cpL}) and (\ref{eq:cp_X}). We have shown the benchmark points in Table \ref{tab:tab1}.
	
	We begin by solving the Boltzmann equations for the type-I case in both weak and strong washout regimes. For the weak washout regime, the parameters are varied, and we consider the benchmark point BP1 (Type-I) with the following values: $\{M_{N_{R_1}}=5\times10^{12}~{\rm GeV}, (y_{N}^\dagger y_N)_{11}=2.9979\times10^{-6},y_{\chi_1}=5.7488\times10^{-4},\epsilon_L=2.20\times10^{-5},\epsilon_\chi=2.08\times10^{-6}, M_\chi=1~ {\rm GeV}\}$. In this case, the branching ratios of $N_{R_1}$ decaying into the dark and visible sectors are $94.776\%$ and $5.224\%$, respectively.
	We illustrate the evolution of the $N_{R_1}$ abundance and the asymmetries in Fig. \ref{fig:type1weak}. Starting with zero initial abundance of the RHN, it is produced through inverse decay and scattering processes. The asymmetries in both sectors begin to build up as soon as a sufficient amount of RHNs is produced. Clearly, due to the chosen Yukawa couplings, the scenario falls into the weak washout regime. We can see that the $N_{R_1}$ abundance reaches equilibrium around $z\sim 7$ and immediately departs from equilibrium. Subsequently, the number density experiences Boltzmann suppression, and the washout effects are no longer effective; hence, the asymmetry saturates. 
	
	\begin{figure}[b]
		\includegraphics[scale=0.5]{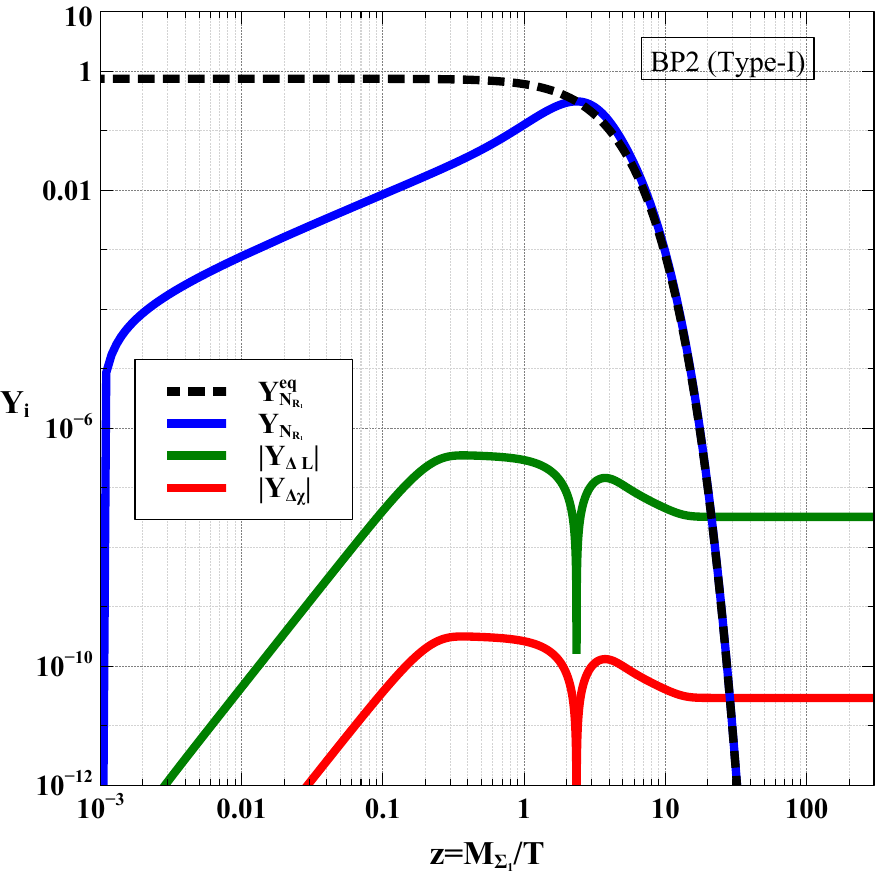}
		\caption{Cosmological evolution of the RHN abundance, visible sector, and dark sector asymmetries in case of strong washout in type-I seesaw. The final asymmetries are $Y_{\Delta L}=3.33907\times10^{-8}$, $Y_{\Delta \chi}=2.98828\times10^{-11}$, $R=5.36483$.}\label{fig:type1strong}
	\end{figure}
	
	Similarly, we solve the Boltzmann equations in the strong washout regime and illustrate the evolution of the asymmetries in the visible and dark sectors along with the abundance of RHN in Fig. \ref{fig:type1strong}. In this case, the parameters are fixed at $\{M_{N_{R_1}}=5\times10^{12}~{\rm GeV}, (y^\dagger_N y_N)_{11}=2.9874\times10^{-4}, y_{\chi_1}=8.1301\times10^{-3}, \epsilon_L=3.91\times10^{-4}, \epsilon_\chi=3.48\times10^{-7}, M_\chi=100~{\rm GeV}\}$. We observe that the RHN is produced through inverse decay and scattering processes and reaches equilibrium around $z\sim 2$. However, due to the larger Yukawa couplings, the RHN remains in equilibrium. In this case, the branching ratios into the visible and dark sectors are $90.0391\%$ and $9.9609\%$, respectively. As the washout processes are strong, the produced asymmetries in both sectors get reduced by some factors and finally saturate after $z\sim 20$. 
	
	\begin{figure}[t]
		\includegraphics[scale=0.5]{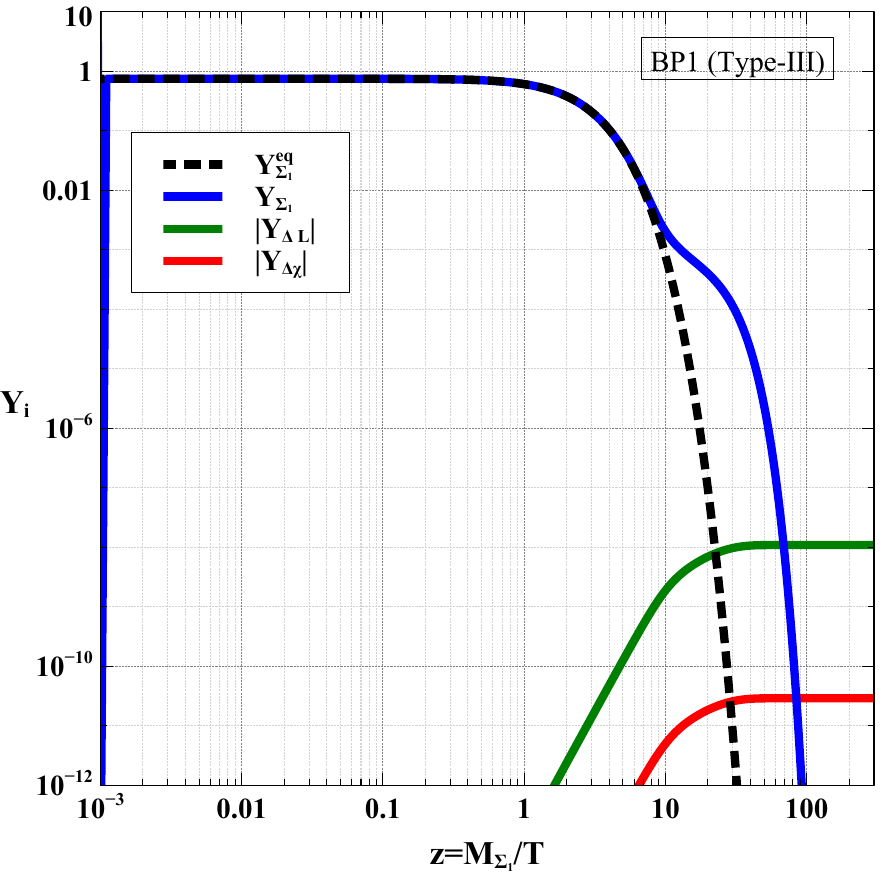}
		\caption{Cosmological evolution of the fermion triplet $\Sigma_1$ abundance, visible sector, and dark sector asymmetries in case of weak washout in type-III seesaw.  The final asymmetries are $Y_{\Delta L}=1.19869\times10^{-8}$, $Y_{\Delta \chi}=3.03807\times10^{-11}$, $R=5.30878$.}\label{fig:type3weak}
	\end{figure}
	
	Transitioning to the type-III scenario, we analyze the evolution of the triplet fermion abundance and the resulting asymmetries in both the dark and visible sectors, as depicted in Fig. \ref{fig:type3weak}. For this analysis, we set the DM mass at $M_\chi= 100$ GeV and establish the parameters as $\{M_{\Sigma_1}=1.532\times10^{12} {\rm GeV}, (y^\dagger_\Sigma y_\Sigma)_{11}=2.4538\times10^{-7}, y_{\chi_1}=7.6650\times10^{-4}, \epsilon_L=1.16\times10^{-5}, \epsilon_\chi=2.94\times10^{-8}\}$, taking into account all relevant processes.
	As previously mentioned, gauge interactions ensure that the triplets $\Sigma$ maintain thermal equilibrium. Despite beginning with an initial abundance of zero, the triplets rapidly achieve thermal equilibrium, characterizing this as the gauge regime. In this regime, the triplets remain in equilibrium solely due to gauge interactions. However, around $z\sim 10$, these interactions become subdominant compared to the Hubble rate, leading the triplets to fall out of equilibrium and subsequently decay. By $z\sim 70$, the triplet population is sufficiently depleted, and the asymmetries in both sectors reach their final saturated values. Since this falls within the weak washout regime, the final asymmetries do not experience any suppression.
	\begin{figure}[t]
		\includegraphics[scale=0.5]{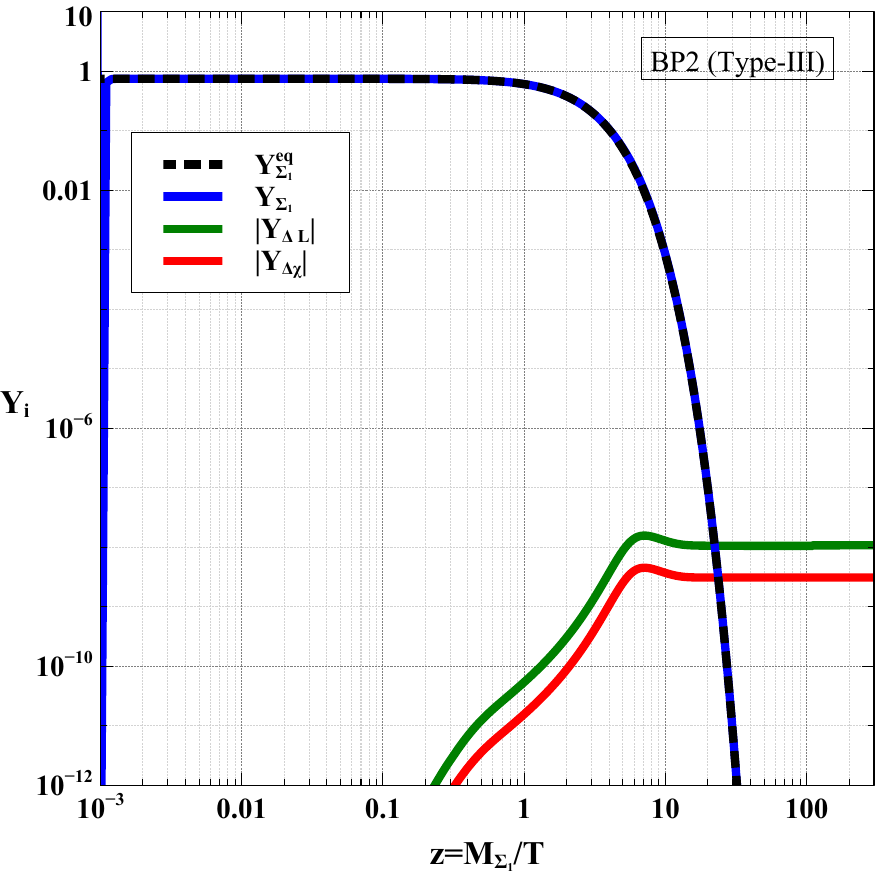}
		\caption{Cosmological evolution of the triplet abundance, visible sector, and dark sector asymmetries in case of strong washout in type-III seesaw.  The final asymmetries are $Y_{\Delta L}=1.18126\times10^{-8}$, $Y_{\Delta \chi}=3.05164\times10^{-9}$, $R=5.41115$.}\label{fig:type3strong}
	\end{figure}
	
	In Fig. \ref{fig:type3strong}, we illustrate the evolution of the triplet fermion abundance and the resulting asymmetries in both sectors for the type-III case within the strong washout regime. The parameters are set as $\{M_{\Sigma_1}=6.28\times10^{12} ~{\rm GeV}, (y^\dagger_\Sigma y_\Sigma)_{11}=1.22006\times10^{-3}, y_{\chi_1}=1.8223\times10^{-3}, \epsilon_L=2.91\times10^{-5}, \epsilon_\chi=7.30\times10^{-6}, M_{\chi}=1 ~{\rm GeV}\}$. This parameter set falls within the Yukawa regime. In this regime, the triplet enters equilibrium rapidly due to gauge interactions, but its subsequent evolution is governed by the Yukawa interactions, which maintain its equilibrium state. In this scenario, the branching ratios for the triplet decaying into the visible and dark sectors are $99.8641\%$ and $0.1359\%$, respectively. Due to the substantial branching into the visible sector, the washout effects from inverse decay processes are significant, compounded by additional washouts arising from the lepton number violating scattering processes. Consequently, we observe a suppression in the final asymmetries, as depicted in Fig. \ref{fig:type3strong}.
	%%%%%%%%%%%%%%%%%%%%%%%%%%%%%%%%%%%%%%%%%%%%%%%%%%%%%%%%%%%%%%%%%%%
	\section{Unstable ASIDM}\label{unstableDM}
	In the above discussion, we have imposed an additional discrete $\mathcal{Z}_2$ symmetry to ensure the stability of the DM particle $\chi$, where only $\chi$ and the scalar $\rho$ (in the type-I scenario) or $\Delta$ (in the type-III scenario) are odd while all other particles transform trivially. However, if we allow soft $\mathcal{Z}_2$ symmetry-breaking terms in the scalar potential, the situation changes after electroweak symmetry breaking. When the SM Higgs field acquires a nonzero VEV, the scalars $\rho$ or $\Delta$ also acquire an induced VEV due to the presence of interaction terms such as $\mu_\rho \rho H^\dagger H$ or $\mu_\Delta H^\dagger (\Vec{\sigma}\Vec{\Delta}H)$, which softly break the $\mathcal{Z}_2$ symmetry. This not only opens up new detection prospects of DM but can also erase the asymmetries associated with the scalars, such that they do not affect visible and DM asymmetries via late decay. This ensures that the DM relic is decided by the asymmetry generated at the seesaw scale, with the symmetric part getting annihilated away. Here, it is worth noting that the annihilation of the symmetric component of DM freezes out well before BBN. 
	
	As mentioned above, the soft breaking of the $\mathcal{Z}_2$ symmetry has crucial consequences regarding the stability of the DM particle $\chi$, opening up its decay modes. When the scalars $\rho$ ($\Delta$) in model I (model II) acquire nonzero vacuum expectation values, it induces a small mixing between $\chi$ and the right-handed neutrino $N_{R_1}$ (or the triplet fermion $\Sigma_1$). Consequently, this mixing propagates to the light neutrino sector due to the seesaw mechanism. As a result, an effective mixing between the DM particle $\chi$ and the light neutrinos is generated.
	This effective mixing between $\chi$ and neutrinos opens up various decay channels for the DM particle, including $\nu ~ Z$, $e^-~ W^+$, $\phi~\nu$, and $\nu ~\bar{f}f$. The Feynman diagrams illustrating these decay modes are shown in Fig. \ref{fig:xtonff}, the decay rates for which are given in Appendix \ref{ap:decayrate}.
	\begin{figure}[h]
		\centering	\includegraphics[scale=0.6]{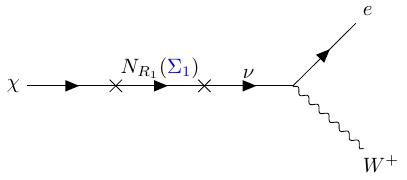}	\includegraphics[scale=0.6]{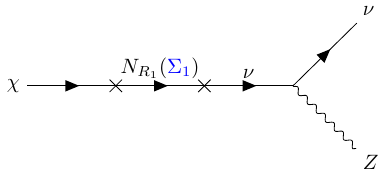}	
		\includegraphics[scale=0.6]{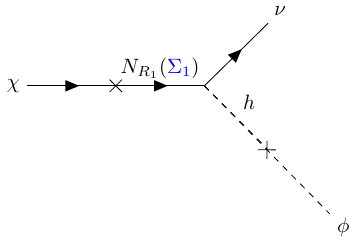}	\includegraphics[scale=0.6]{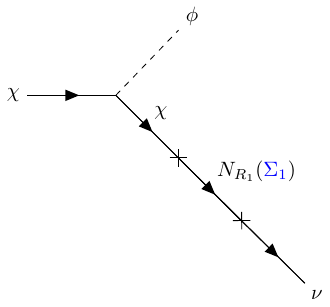}
		\includegraphics[scale=0.6]{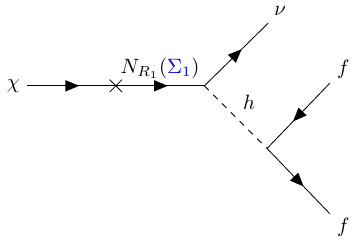}
		\caption{Feynman diagram of different decay modes of $\chi$.}\label{fig:xtonff}
	\end{figure} 
	\begin{figure}[h]
		\centering
		\includegraphics[scale=0.45]{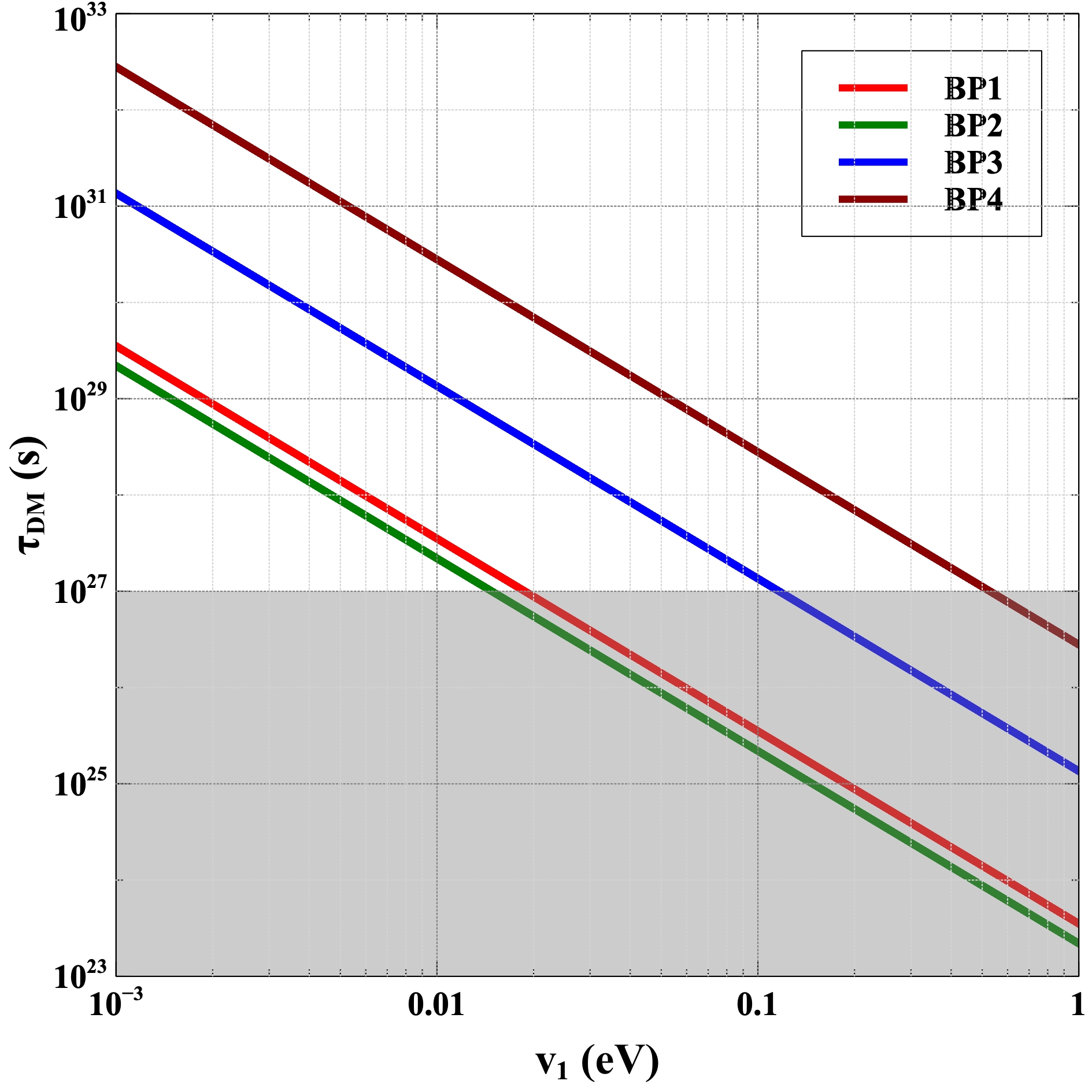}
		\caption{Lifetime of the DM vs VEV of the scalar $\rho$ ($\Delta$) for four benchmark points from Table \ref{tab:tab2}. Here we have fixed $y_\chi=10^{-3}$, $M_{R_1}$($M_{\Sigma_{1}})=10^{12}$ GeV.}	\label{fig:tauvsv1}
	\end{figure}
	\begin{table*}[t]
		\centering
		\caption{Benchmark points that simultaneously satisfy DM self-interaction, direct detection, and BBN constraints.
		}
		\resizebox{15cm}{!}{
			\begin{tblr}{
					colspec={lllllll},
					row{1}={font=\bfseries},
					column{1}={font=\itshape},
					row{1}={bg=gray!50},row{2}={bg=gray!40},row{3}={bg=gray!30},row{4}={bg=gray!20},row{5}={bg=gray!10},
				}
				\toprule BPs&$M_{\rm DM}\rm(GeV)$&$M_{\phi} \rm(MeV)$&$\lambda_{\rm DM}$ & $\sin\gamma$& $\sigma^{\rm SI}_{\rm DM-N}~(\rm cm^2)$ &$\tau_{\phi}\rm (s)$\\
				\toprule
				BP1&
				$0.1$&$67.54$ & $1.26$ & $2.02\times10^{-3}$& $5.71\times10^{-37}$& $1.39\times10^{-5}$ \\
				
				BP2&
				$1$&$223.00$ & $2.89$ & $8.67\times10^{-5}$& $1.33\times10^{-39}$& $1.64\times10^{-6}$\\
				
				BP3&
				$10$&$586.27$ & $1.11$ & $1.88\times10^{-8}$& $6.11\times10^{-49}$& $5.31\times10^{-1}$ \\
				
				BP4&
				$100$&$2531.10$ & $0.72$ & $2.64\times10^{-7}$& $1.69\times10^{-49}$& $5.12\times10^{-4}$\\
				\bottomrule
		\end{tblr}}
		\label{tab:tab2}
	\end{table*}
	For $\chi$ to qualify as a viable DM candidate, its lifetime ($\tau_{\rm DM}$) should be greater than $10^{27}$ seconds \cite{Baring:2015sza}, which is a conservative lower limit obtained using the gamma-ray data from the Fermi-LAT observation of Milky Way satellite dSphs. The dominant decay modes of the DM particle depend on its mass range. When the DM mass lies between 100 MeV and $M_W$, it predominantly decays into neutrinos ($\nu$) and scalar particles ($\phi$). In this range, the decay width is inversely proportional to the DM mass. Consequently, as the DM mass increases, its lifetime also increases. Thus, as the DM lifetime is inversely proportional to the square of the VEV $v_1$, to maintain a constant lifetime, the value of $v_1$ must increase as the DM mass increases. In the mass range between $M_W$ and $M_Z$, the DM predominantly decays into $W^+$ and $e^-$. In this case, the decay width is proportional to the DM mass. As the DM mass increases, the lifetime increases, and to maintain a fixed lifetime, the value of $v_1$ must decrease.
	When the DM mass falls within the range of $M_Z$ to $M_{h_1}$, the dominant decay mode is into $Z$ and $\nu$. Here, the decay width is also proportional to the DM mass, and a similar behavior is observed as in the previous range.
	Once the DM mass exceeds the Higgs mass, the $h$ and $\nu$ decay channel opens up, further increasing the decay width. Consequently, the value of $v_1$ decreases as the DM mass increases in this regime. In Fig \ref{fig:tauvsv1}, we have shown the DM lifetime as a function of VEV of the $\mathcal{Z}_2$ even scalar for four benchmark points as mentioned in Table \ref{tab:tab2}. These benchmark points simultaneously satisfy DM self-interaction,  direct detection, and BBN constraints. The gray-shaded region is excluded by the DM lifetime constraint. As mentioned above, the lifetime of the DM is $\propto v_1^{-2}$; for a fixed value of the DM mass, the lifetime decreases as $v_1$ increases. As considered in the BP3, for a 10 GeV DM, the $v_1$ has to be smaller than 0.1 eV so as to satisfy the DM lifetime constraints. Thus, depending on the value of the $v_1$, a wide range of DM masses can be accommodated while satisfying the DM lifetime constraint.
	%%%%%%%%%%%%%%%%%%%%%%%%%%%%%%%%%%%%%%%%%%%%%%%%%%%%%%%%%%%%%%%%%%%
	\section{Detection Prospects}
	\label{detection}
	In this section, we discuss the prospects of detecting the ASIDM of our frameworks in direct and indirect DM search experiments.
	
	%%%%%%%%%%%%%%%%%%%%%%%%%%%%%%%%%%%%%%%%
	\subsection{Direct detection}
	\begin{figure}[h]
		\centering
		\includegraphics[scale=1]{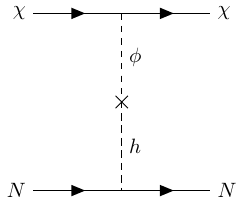}
		\caption{The spin-independent scattering of DM-nucleon ($N$) via Higgs portal.}
		\label{fig:DDdiag}
	\end{figure}
	
	The possibility of spin-independent (SI) DM-nucleon elastic scattering allows for the detection of DM in terrestrial laboratories. In this setup, SI elastic scattering of DM is achievable in terrestrial laboratories via $h$-$\phi$ mixing, where DM particles can scatter off target nuclei. The Feynman diagram for this process is shown in Fig. \ref{fig:DDdiag}.
	
	The spin-independent elastic scattering cross section of DM per nucleon can be calculated as \cite{Ellis:2008hf}
	\begin{equation}
		\sigma^{\rm SI}_{{\rm DM}-{\rm N}}=\frac{\mu_r^2}{\pi A^2}[Z f_p + (A-Z) f_n]^2\label{eq:dd1}
	\end{equation}
	where $\mu_r=\frac{M_\chi m_n}{M_\chi+m_n}$ is the reduced mass, $m_n$ is the nucleon (proton or
	neutron) mass, $A$ is the mass number of target nucleus, $Z$ is the atomic number of target nucleus. The $f_p$ and $f_n$ are the interaction strengths of proton and neutron with DM, respectively, and are given as
	
	\begin{equation}
		f_{p,n}=\sum_{q=u,d,s} f_{T_q}^{p,n} \alpha_q \frac{m_{p,n}}{m_q}+\frac{2}{27} f_{T_G}^{p,n} \sum_{q=c,t,b}\alpha_q \frac{m_{p,n}}{m_q}\label{eq:dd2}
	\end{equation}
	where
	
	\begin{equation}
		\alpha_q=\lambda_{_{{\rm DM}}}*\frac{m_q}{v_0}*\sin(\gamma)\cos(\gamma)(\frac{1}{M^2_{\phi}}-\frac{1}{M^2_{h_1}})\label{eq:dd3}
	\end{equation}
	with $\gamma$ being the mixing angle between $h$ and $\phi$. The values of $f_{T_q}^{p,n}$, $f_{T_G}^{p,n}$ can be found in \cite{Hoferichter:2017olk}. 
	\begin{figure}[t]
		\centering		
		\includegraphics[scale=0.45]{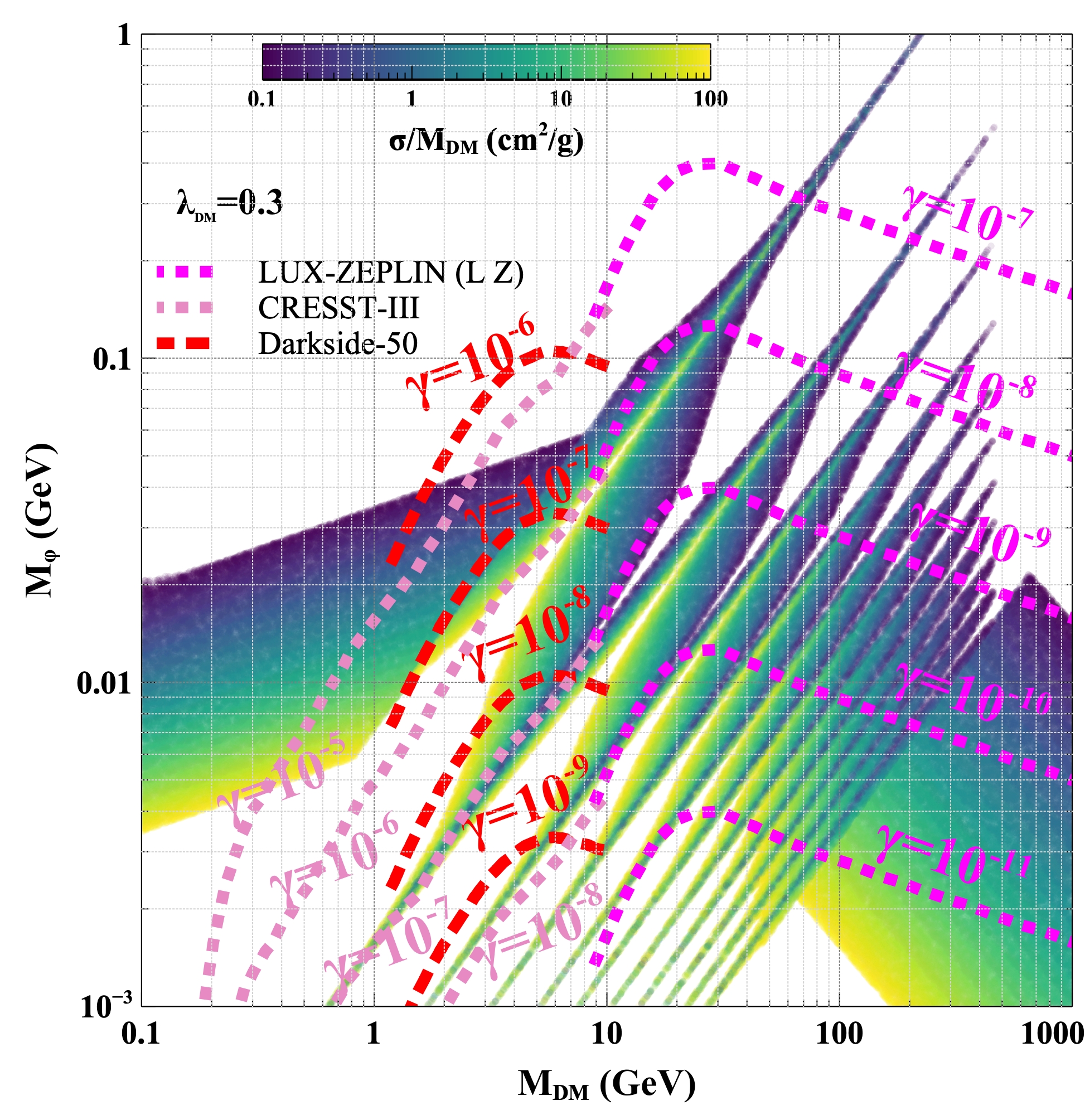}
		\caption{DM self-interaction preferred region : $\sigma/M_{\rm DM} \sim (0.1-100)~ {\rm cm}^2/{\rm g}$ in the plane of $M_{{\rm DM}}$-$M_\phi$ for a fixed value of $\lambda_{_{\rm DM}}=0.3$. Direct detection constraints from the LZ and CRESST-III experiments, along with the future sensitivity of DarkSide-50, are shown as different contours.}
		\label{fig:DDplot}
	\end{figure}
	
	Direct search experiments like the CRESST-III~\cite{CRESST:2019jnq} and
	LUX-ZEPLIN (LZ)~\cite{LUX-ZEPLIN:2022qhg}, put severe constraints on the model parameters. The LZ experiment provides the most stringent constraint on DM mass above 10 GeV, while the CRESST-III data
	constrain the mass regime below 10 GeV. In Fig.~\ref{fig:DDplot}, 
	the most stringent constraints from the CRESST-III~\cite{CRESST:2019jnq},
	LZ~\cite{LUX-ZEPLIN:2022qhg} and the anticipated sensitivity of DarkSide-50~\cite{DarkSide:2018bpj} experiments are shown against the self-interaction favored parameter space in the plane of  $M_{\rm DM}$ and $M_\phi$ assuming a fixed value of $\lambda_{\rm DM}=0.3$. The region below each of these contours is excluded for that particular value of the mixing angle $\gamma$. This can be understood as follows. Let us consider the DM-nucleon scattering cross section derived from Eqs. \eqref{eq:dd1}, \eqref{eq:dd2}, \eqref{eq:dd3} for a fixed DM mass. This cross section  $\sigma^{\rm SI}_{\rm DM-N} \propto \sin^2\gamma/ M^4_\phi$.
	As illustrated in Fig.~\ref{fig:DDplot}, for a constant $\sin\gamma$, moving from top to bottom corresponds to decreasing values of $M_\phi$. Consequently, the direct detection (DD) cross section increases as we move downward on the plot. Therefore, any point below each contour of DD constraint for a given $\sin\gamma$ represents a larger cross section and is excluded by direct detection experiments. The exclusion region depends on the magnitude of $\sin\gamma$. For small $\sin\gamma$, the DD cross section is smaller, resulting in a smaller excluded parameter space.
	For large $\sin\gamma$, the DD cross section becomes larger, leading to a more extensive excluded region in the $M_{\phi}$-$M_{\rm DM}$ plane.
	
	The mixing angle $\gamma$ is subject to constraints from both upper and lower bounds. The upper bound arises from the consideration of invisible Higgs decay, as the singlet scalar is typically lighter than the Higgs mass. On the other hand, a lower bound on $\gamma$ can be obtained by requiring the scalar particle $\phi$ to decay before the epoch of BBN, i.e., $\tau_\phi < \tau_{\rm BBN}$, where $\tau_\phi$ and $\tau_{\rm BBN}$ represent the lifetime of $\phi$ and the BBN timescale, respectively. This condition ensures that the decay of $\phi$ does not disrupt the successful predictions of primordial nucleosynthesis.
	
	\begin{figure}[h]
		\centering
		\includegraphics[scale=0.489]{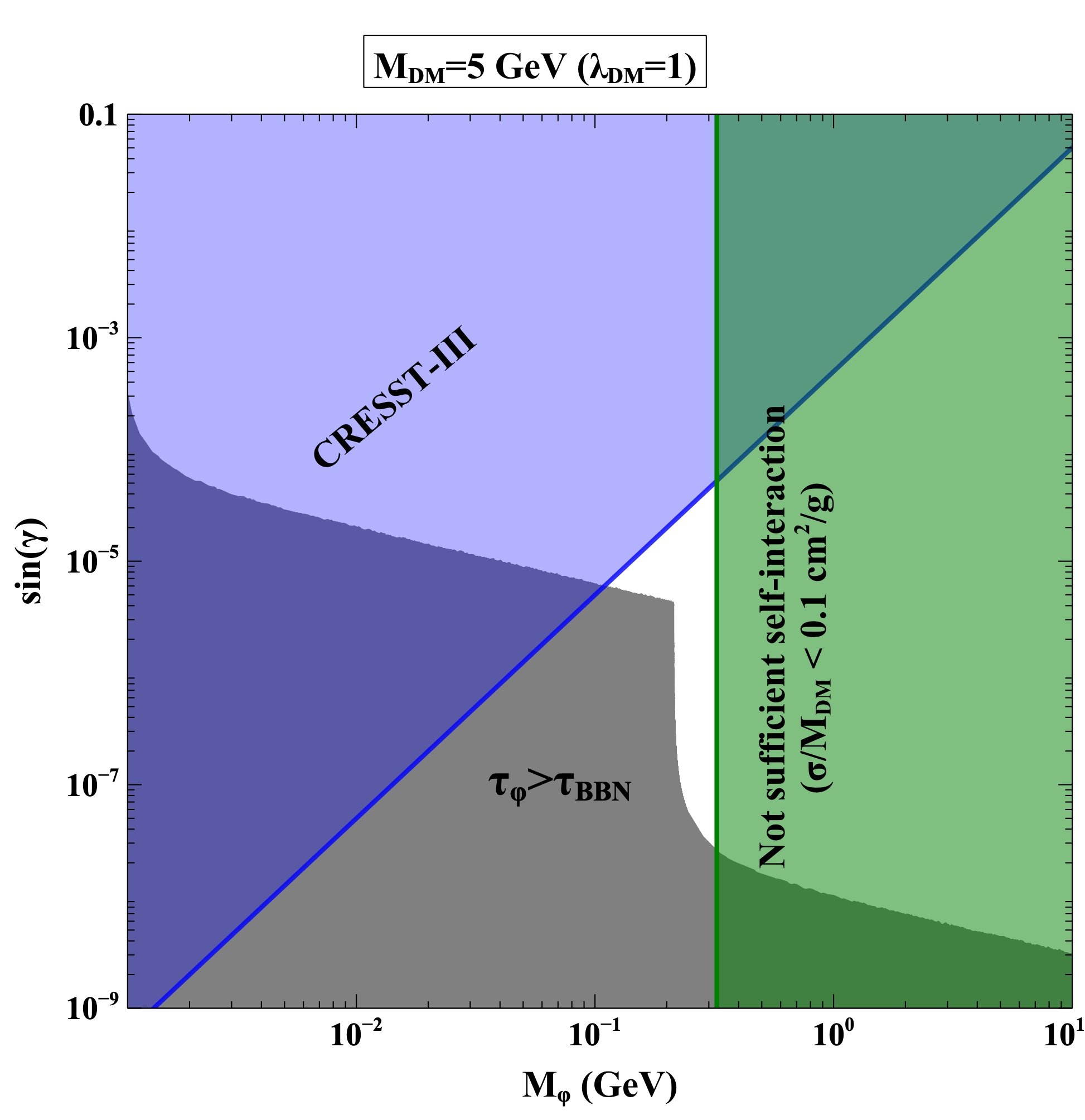}
		\caption{Constraints from DM direct detection, DM self-interaction, and BBN in the plane of $\sin(\gamma)$-$M_\phi$ for DM mass of 5 GeV.}\label{fig:DDplot2}
	\end{figure}

	In Fig. \ref{fig:DDplot2}, we have shown the allowed parameter space from the direct detection, DM self-interaction, and BBN constraints in the plane of $\sin\gamma$ and $M_{\phi}$ for a fixed DM mass, $M_{{\rm DM}}=5$ GeV, and coupling, $\lambda_{{\rm DM}}=1$. The blue-shaded region is excluded by the CRESST-III~\cite{CRESST:2019jnq} experiment due to the constraints on the DM-nucleon spin-independent scattering cross section. The gray-shaded region is disallowed by the BBN constraint on the $\phi$ decay lifetime, {\it i.e.}, in this region, $\tau_{\phi}>\tau_{\rm BBN}$. The green region is excluded due to the requirement of sufficient self-interaction among the DM.
	
\begin{figure}[t]
		\centering		\includegraphics[scale=0.45]{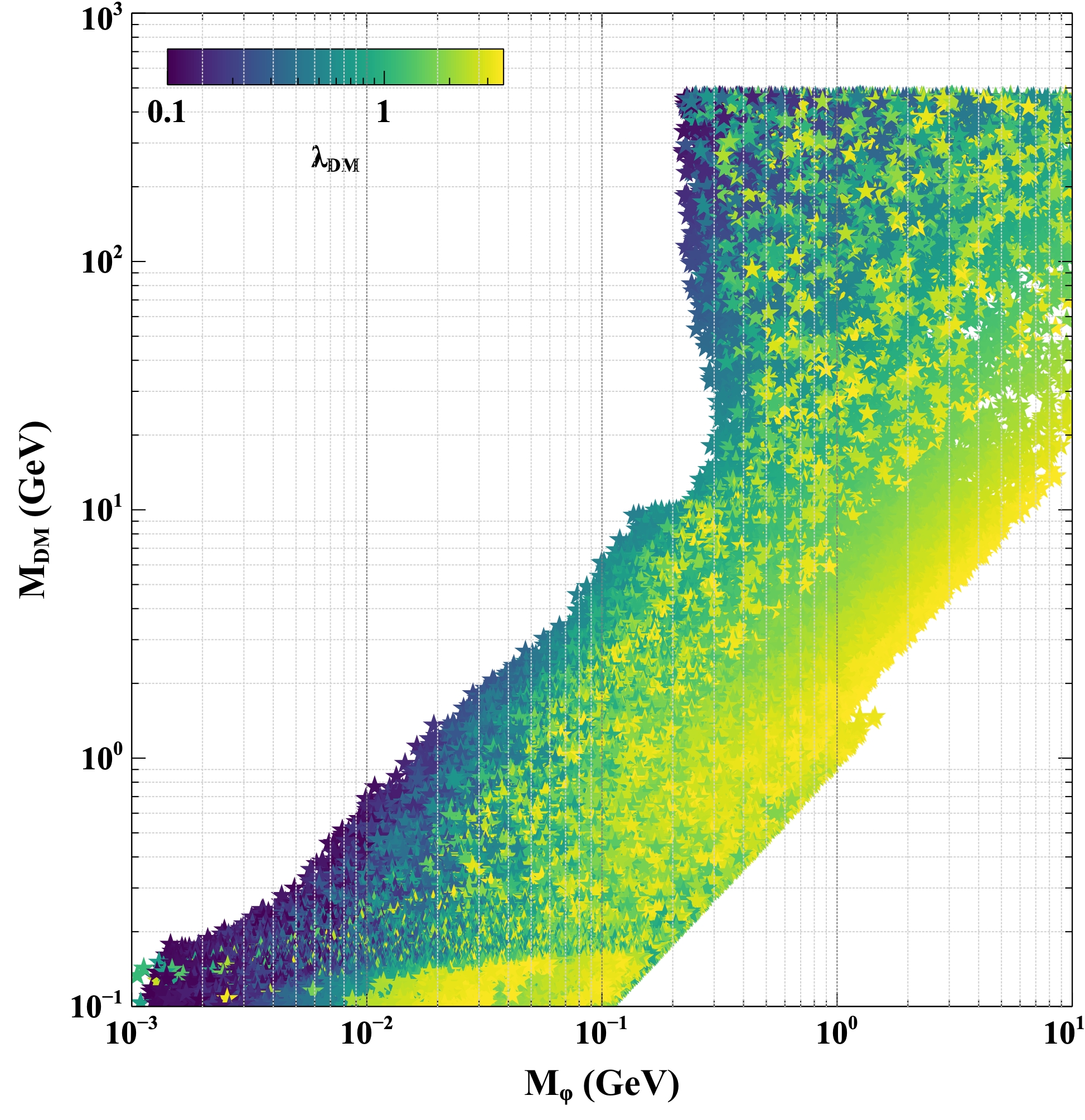}
		\caption{Parameter space in the plane of $M_{\rm DM}$ and $M_\phi$ that can give rise to the required self-interaction among the DM while being consistent with the direct detection and BBN constraints. }
		\label{fig:DDplot4}
	\end{figure}

\begin{figure}[t]
		\centering
		\includegraphics[scale=0.45]{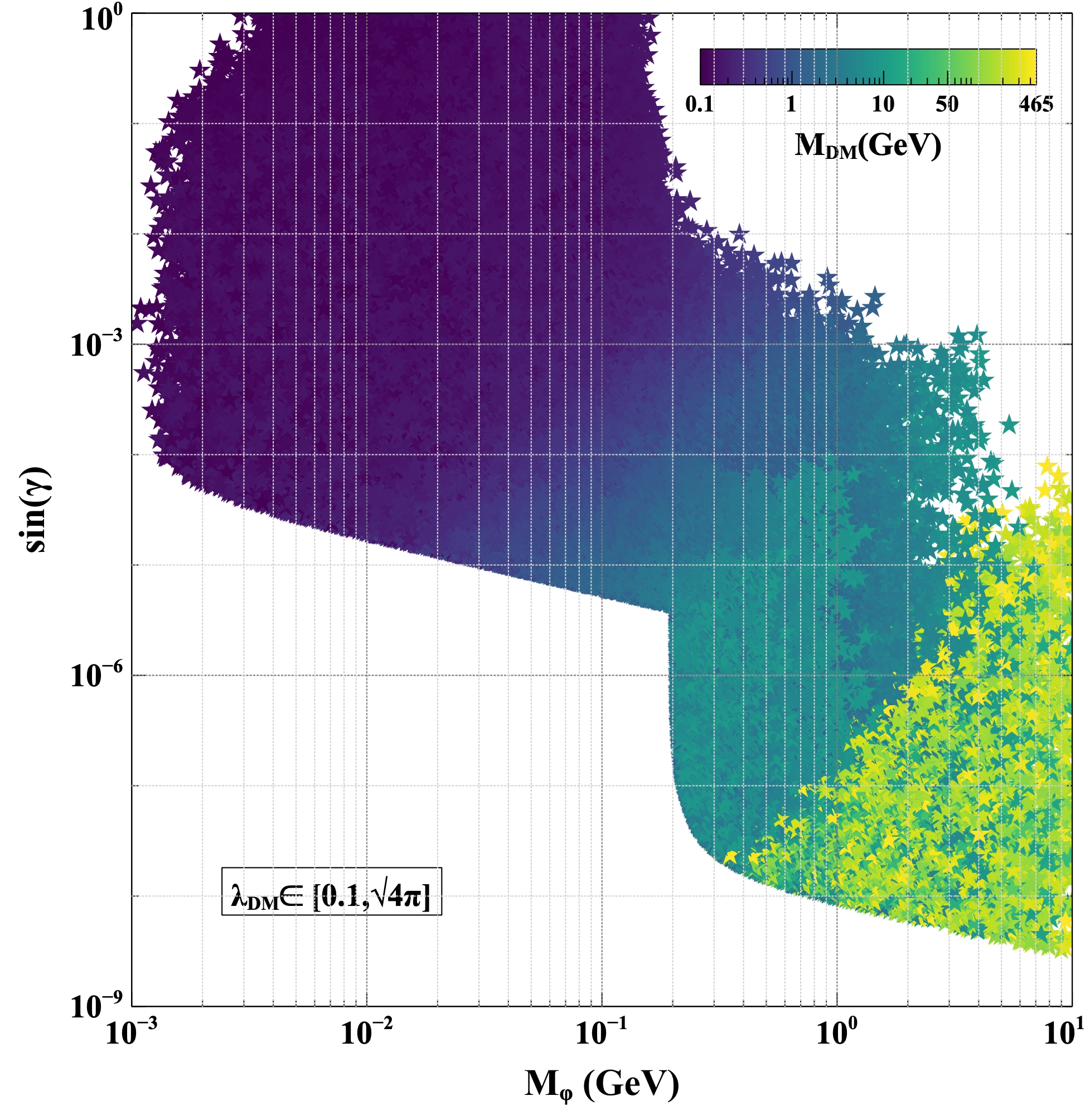}
		\caption{Parameter space in the plane of $M_\phi$ and $\sin\gamma$, which is consistent with the direct detection and BBN constraints while still giving rise to the required self-interaction.}
		\label{fig:DDplot3}
	\end{figure}

	To capture the final allowed parameter space, we have performed a scan to simultaneously satisfy the constraints from DM self-interaction [$\sigma/M_{\rm DM} \sim (0.1-100)~ {\rm cm}^2/{\rm g}$], direct detection, and BBN constraints on light mediator mass and mixing angle $\gamma$. The result of this scan is shown in Figs.\ref{fig:DDplot4} and \ref{fig:DDplot3}. 
	We have varied the free parameters within the following ranges: $M_{{\rm DM}}\in [0.1,10^3]$ GeV, $M_\phi\in[10^{-3},10]$ GeV, $\sin\gamma\in[10^{-12},0.7]$, and $\lambda_{{\rm DM}}\in[0.1,\sqrt{4\pi}]$.\footnote{We also check that changing the upper bound on $\lambda_{{\rm DM}}$ or taking other upper limits from say, unitarity considerations \cite{Walker:2013hka}  does not change the allowed mass range of DM.}
	We show the parameter space in the plane of $M_{{\rm DM}}$ and $M_{\phi}$ in Fig. \ref{fig:DDplot4}, with the color code representing $\lambda_{{\rm DM}}$. The results depict that the maximum allowed DM mass that can satisfy all the constraints is around $M_{\rm DM}\sim460$ GeV.
	In Fig. \ref{fig:DDplot3}, we represent the parameter space in the plane of $\sin\gamma$ and $M_\phi$, with the DM mass shown in the color code. 
	Clearly, a smaller mixing angle, $\gamma$, for small $M_\phi$ is disfavored by the BBN constraint, as it would lead to a lifetime of $\phi$ greater than $\tau_{\rm BBN}$. We also observe that if DM mass is light, then large $\sin\gamma$ is allowed because of weaker constraints from DD constraints, which can be probed by future experiments with enhanced sensitivities. The upper limit on $M_{\rm DM}$ can be understood from the following reasoning. As evident from Fig. \ref{fig:DDplot}, for DM masses greater than 460 GeV, the correct self-interaction cross section can only be obtained if the mediator mass ($M_\phi$) is very light, below approximately 20 MeV, falling within the classical regime as elaborated in Appendix~\ref{app:selfint}. However, such a small mediator mass results in a sizable DM-nucleon scattering cross section, which is ruled out by the direct detection constraints. Consequently, the mixing angle $\gamma$ must be decreased to adhere to the constraints from direct detection. On the other hand, these small values of $M_\phi$ and $\gamma$ are excluded by the BBN constraints on the lifetime of $\phi$ (see Appendix \ref{ap:philifetime} for details of $\phi$ lifetime). Therefore, the simultaneous imposition of constraints from DM self-interaction, direct detection experiments, and the BBN constraint on the lifetime of the light mediator limits the DM mass to be less than $460$ GeV. 
	%%%%%%%%%%%%%%%%%%%%%%%%%%%%%%%%%%%%%%%%%%%%%%%%%%%%%%%%%%%%%%%%%%%
	\subsection{Indirect detection}
	
\begin{figure}[h]
		\centering		\includegraphics[scale=0.5]{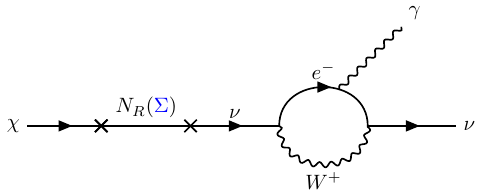}
		\includegraphics[scale=0.5]{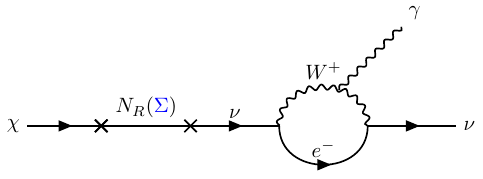}
		\caption{One-loop decay of DM into $\nu\gamma$.}\label{fig:xtonugamma}
	\end{figure}

\begin{figure}[h]
		\centering
		\includegraphics[scale=0.45]{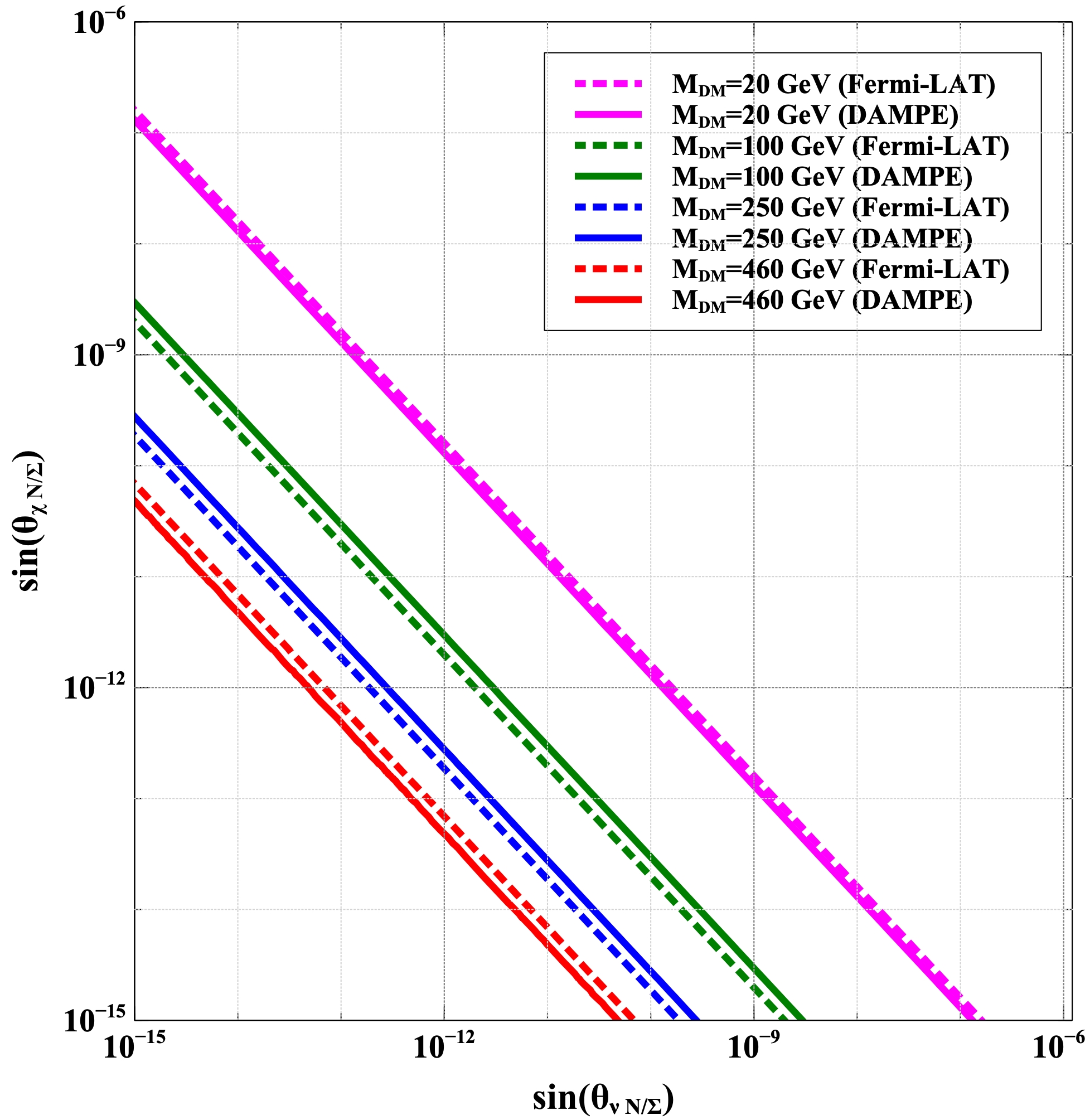}
		\caption{Constraints from the Fermi-LAT and DAMPE experiments on monochromatic gamma-ray emissions from DM in the plane of $\sin(\theta_{{\chi x}})$-$\sin(\theta_{{\nu x}})$ for different DM masses. The parameter space towards the bottom left part of each contour remains allowed for respective DM mass.}
		\label{fig:IDplot1}
	\end{figure}

	As previously discussed, due to the effective mixing between the DM and active neutrinos, a distinct decay mode for the DM can occur, resulting in the emission of a monochromatic gamma-ray line at energy $E = M_{{\rm DM}}/2$. This emission arises from a one-loop decay process, $\chi\rightarrow\nu\gamma$, as depicted in Fig. \ref{fig:xtonugamma}.
	The corresponding decay width can be estimated as \cite{Pal:1981rm,Shrock:1982sc}
	\begin{eqnarray}
		\tau_{\chi\rightarrow \nu\gamma}\simeq \bigg(\frac{9G^2_{F}\alpha_{_{\rm EM}}}{1024\pi^4}\sin(\theta_{{\chi x}})^2\sin(\theta_{{\nu x}})^2M_\chi^5\bigg)^{-1},
	\end{eqnarray}
	where $\alpha_{_{\rm EM}}=1/137$ is the fine structure constant, $G_F=1.166\times10^{-5}~\rm GeV^{-2}$ is the Fermi constant, $\theta_{{\chi x}}$ is the mixing angle between $\chi$ and $N_R(\Sigma)$, and $\theta_{{\nu x}}$ is the mixing angle between $\nu$ and $N_R(\Sigma)$. In Fig. \ref{fig:IDplot1}, we show the constraints from the gamma-ray search by the Fermi-LAT \cite{Fermi-LAT:2015kyq} and DAMPE \cite{DAMPE:2021hsz} experiments in the $\sin(\theta_{{\chi x}})$-$\sin(\theta_{{\nu x}})$ plane for different DM masses. The parameter space below these contours remains safe from such gamma-ray constraints.
	\begin{figure}[h]
		\centering		\includegraphics[scale=0.45]{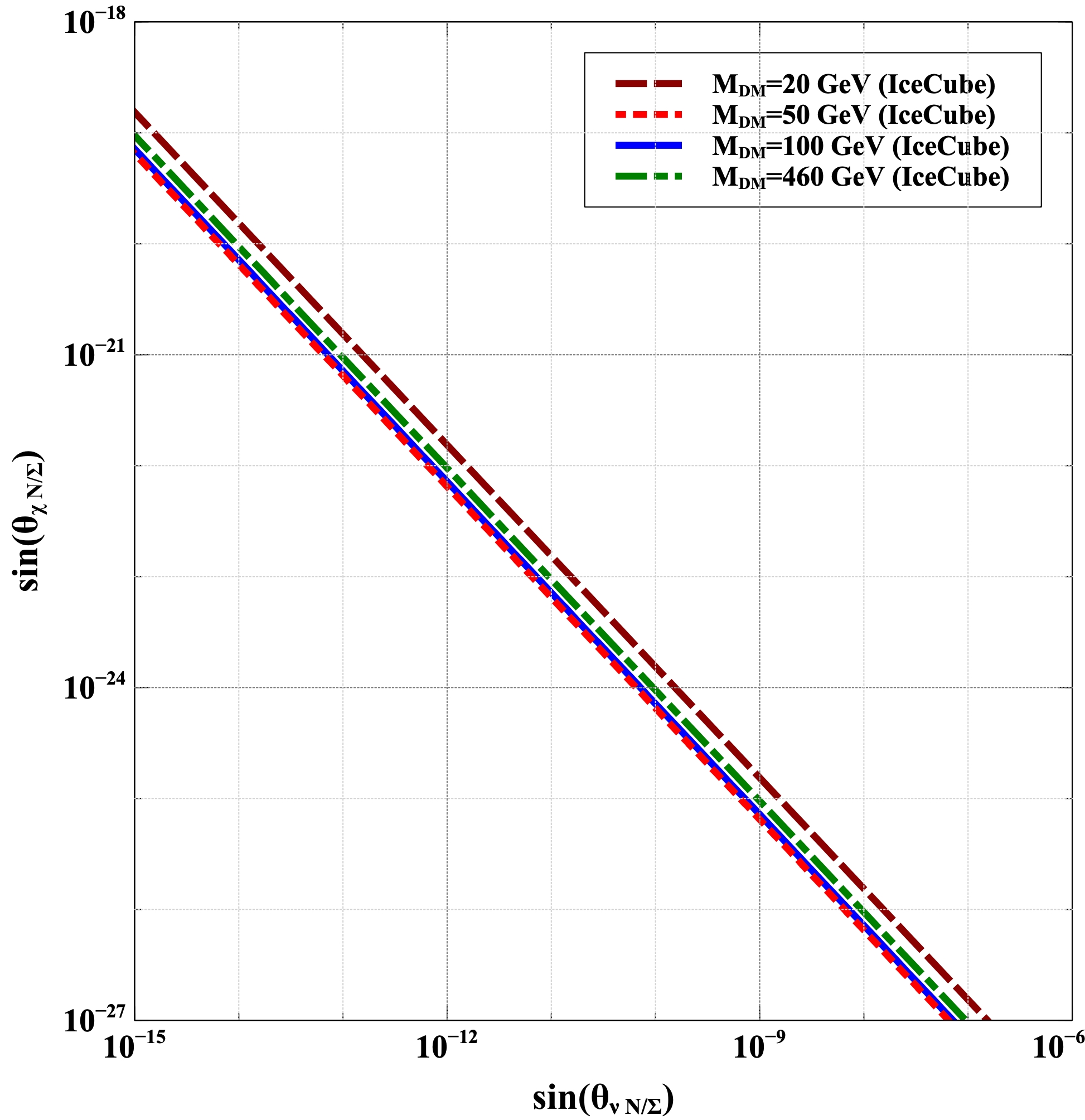}
		\caption{Constraints from the IceCube experiment on neutrino lines from DM in the plane of $\sin(\theta_{{\chi x}})$-$\sin(\theta_{{\nu x}})$ for different DM masses. The parameter space towards the bottom left part of each contour remains allowed for respective DM mass.}
		%considering NFW profile.}
	\label{fig:IDplot2}
\end{figure}

Moreover, two-body decay of the DM into neutrinos like $\chi \to \nu \phi$ in our model can give rise to monochromatic neutrino lines. This decay width is given by

\begin{eqnarray}
	\tau_{\chi\rightarrow\nu\phi}=\bigg(\frac{1}{16\pi}\lambda_{\rm DM}^2\sin(\theta_{{\chi x}})^2\sin(\theta_{{\nu x}})^2M_{N(\Sigma)}^2\frac{1}{M_{\chi}}\bigg)^{-1}\nonumber\\
\end{eqnarray}

In Fig. \ref{fig:IDplot2}, we show the constraints from the neutrino line searches by the IceCube \cite{IceCube:2023ies} experiment in the $\sin(\theta_{{\chi x}})$-$\sin(\theta_{{\nu x}})$ plane. We show contours for four different DM masses \{20 GeV, 50 GeV, 100 GeV, 460 GeV\}. We see that when the mixing angle $\sin(\theta_{{\nu x}})\sim 10^{-7}$, the $\chi$-$x$ mixing angle has to be $\sin(\theta_{{\chi x}})<10^{-27}$. In Fig. \ref{fig:IDplot1}, we see that 
when the mixing angle $\sin(\theta_{{\nu x}})\sim 10^{-7}$, the $\chi$-$x$ mixing angle has to be $\sin(\theta_{{\chi x}})<10^{-15}$ for $M_{\rm DM}=20$ GeV. Thus, the neutrino lines from the IceCube give more stringent constraints on the mixing angle than the ones from gamma-ray searches by the Fermi-LAT or DAMPE experiment.
%%%%%%%%%%%%%%%%%%%%%%%%%%%%%%%%%%%%%%%%%%%%%%%%%%%%%%%%%%%%%%%%%%%
\section{Conclusion}
\label{conclusion}

In this study, we investigate DM self-interaction and the baryon-DM coincidence problem within the context of two popular canonical seesaw frameworks, namely type-I and type-III seesaw, which explain nonzero neutrino masses. While DM self-interactions facilitated by a light mediator offer a potential solution to small-scale astrophysical structure issues of CDM, they often lead to thermally under-abundant DM relic due to strong annihilation rates of DM into light mediators. To address this challenge as well as to explain the ratio of DM to baryon density, we pursue the idea of asymmetric DM and cogenesis within type-I and type-III seesaw models. The $CP$-violating decays of RHNs (or fermion triplets) into SM leptons and the Higgs boson can produce a lepton asymmetry, while their decay into DM and a singlet scalar $\rho$ (or triplet scalar $\Delta$) simultaneously generates a DM asymmetry. This lepton asymmetry is then transformed into a baryon asymmetry, with the DM asymmetry persisting as the DM relic. DM self-interactions also ensure that the symmetric DM component annihilates away leaving only the asymmetric part, similar to baryons. While the models have promising detection prospects at direct search experiments as well as astrophysical surveys, the conventional indirect detection prospects remain low due to the absence of efficient annihilation rates of asymmetric DM in the present Universe. This can, however, change if the $\mathcal{Z}_2$ symmetry protecting DM stability is broken softly, leading to induced VEVs of $\mathcal{Z}_2$-odd scalars, thereby opening up DM decay channels into SM particles. By adjusting the induced VEVs of these scalars, it is possible to ensure the longevity of DM, making it compatible with existing DM lifetime constraints. The decay rate of DM can already saturate bounds on monochromatic gamma rays and neutrinos keeping the indirect detection prospects promising. Through an extensive analysis of the model parameter space, we identify the essential conditions for achieving successful cogenesis and adequate self-interaction while adhering to stringent phenomenological and experimental constraints. Our results demonstrate that the model is highly predictive and amenable to testing through a variety of direct and indirect DM detection. 

%%%%%%%%%%%%%%%%%%%%%%%%%%%%%%%%%%%%%%%%
\acknowledgements
S.M. acknowledges the financial support from National Research Foundation(NRF)
grant funded by the Korea government (MEST) Grant No. NRF-2022R1A2C1005050. The work of D.B. is supported by the Science and Engineering Research Board (SERB), Government of India Grants No. MTR/2022/000575 and No. CRG/2022/000603. P.K.P. would like to acknowledge the Ministry of Education, Government of India, for providing financial support for his research via the Prime Minister’s Research Fellowship (PMRF) scheme. The works of N.S. and P.S. are supported by the Department of Atomic
Energy-Board of Research in Nuclear Sciences, Government of India (Grant No. 58/14/15/2021- BRNS/37220).	

%%%%%%%%%%%%%%%%%%%%%%%%%%%%%%%%%%%%%%%%%%%%%%%%%%%%%%%%%%%%%%%%%%%

\appendix
\section{DM SELF-INTERACTION CROSS SECTION}
\label{app:selfint}
The self-interaction of non-relativistic DM can be described by a Yukawa potential given as,
\begin{eqnarray}
	V(r) = - \frac{\alpha_D}{r}e^{-M_{\phi}r}
\end{eqnarray}
where $\alpha_D=\lambda_{_{{\rm DM}}}^2/4\pi$, and $M_\phi$ is the mass of the singlet scalar, $\phi$. 
To capture the relevant physics of forward scattering divergence for the self-interaction, we define the transfer cross section $\sigma_T$ as~\cite{Feng:2009hw,Tulin:2013teo,Tulin:2017ara}
\begin{equation}
	\sigma_T = \int d\Omega (1-\cos\theta) \frac{d\sigma}{d\Omega}.
\end{equation}
In the Born limit ($\alpha_D M_\chi/M_{\phi}<< 1$), for both attractive as well as repulsive potentials, the transfer cross section is
\begin{equation}
	\sigma^{\rm Born}_T = \frac{8 \pi \alpha^2_D}{M^2_\chi v^4} \Bigg(\ln(1+ M^2_\chi v^2/M^2_{\phi} ) - \frac{M^2_\chi v^2}{M^2_{\phi}+ M^2_\chi v^2}\Bigg).
\end{equation} 
Outside the Born regime ($\alpha_D M_\chi /M_{\phi} \geq 1 $), we have two distinct regions {\it viz.} the classical region and the resonance region. In the classical limit ($M_\chi v/M_{\phi}\geq 1$), the solutions for an attractive potential is given by \cite{Tulin:2013teo,Tulin:2012wi,Khrapak:2003kjw}
{
	\begin{equation}
		\sigma^{\rm classical}_T =\left\{
		\begin{array}{l}
			\frac{4\pi}{M^2_{\phi}}\beta^2 \ln(1+\beta^{-1}) ~~~~,\beta \leqslant 10^{-1}\\
			\frac{8\pi}{M^2_{\phi}}\beta^2/(1+1.5\beta^{1.65}) ,10^{-1}\leq \beta \leqslant 10^{3}\\
			\frac{\pi}{M^2_{\phi}}( \ln\beta + 1 -\frac{1}{2}ln^{-1}\beta)^2 ,\beta \geq 10^{3}\\
		\end{array}
		\right.
\end{equation} }   
where $\beta = 2\alpha_D M_{\phi}/(M_\chi v^2)$.

In the resonance region ($ M_\chi v/M_{\phi} \leq 1$), the Yukawa potential is described by a Hulthen potential $\Big( V(r) = - \frac{\alpha_D \delta e^{-\delta r}}{1-e^{-\delta r}}\Big)$ and the solution is given as ~\cite{Tulin:2013teo}
\begin{equation}
	\sigma^{\rm Hulthen}_T = \frac{16 \pi \sin^2\delta_0}{M^2_\chi v^2}
\end{equation}
where $\delta_0$ is given by
{\scriptsize\begin{equation}
		\delta_0 ={\rm arg} \Bigg(\frac{i\Gamma \Big(\frac{i M_\chi v}{k M_{\phi}}\Big)}{\Gamma (\lambda_+)\Gamma (\lambda_-)}\Bigg), \lambda_{\pm} = 
		1+ \frac{i M_\chi v}{2 k M_{\phi}}  \pm \sqrt{\frac{\alpha_D M_\chi}{k M_{\phi}} - \frac{ M^2_\chi v^2}{4 k^2 M^2_{\phi}}}
\end{equation}}

with $k \approx 1.6$ being a dimensionless number.

\begin{figure}[h]
	\centering
	\includegraphics[scale=0.5]{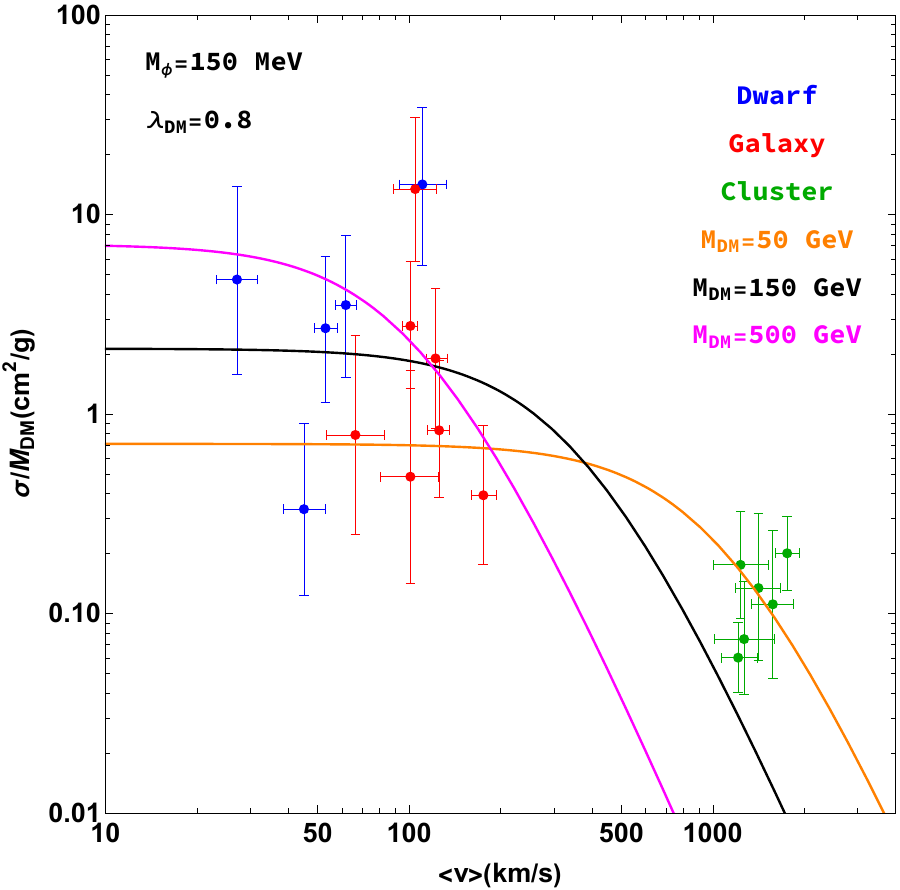}
	\caption{Self-interacting cross section per unit DM mass as a function of averaged cross section velocity. }\label{fig:crossvdv}
\end{figure}
We have shown the self-interacting cross section per unit DM mass as a function of averaged collision velocity for three different DM masses with a fixed coupling of $\lambda_{_{{\rm DM}}}$=0.8 and mediator mass, $M_\phi$=150 MeV in Fig. \ref{fig:crossvdv}, which is consistent with the observational data \cite{Kaplinghat:2015aga}. 
\label{appendix1}
%%%%%%%%%%%%%%%%%%%%%%%%%%%%%%%%%%%%%%%%%%%%%%%%%%%%%%%%%%%%%%%%%%%
\section{NEUTRINO MASS GENERATION}
\label{app:numass}
%%%%%%%%%%%%%%%%%%%%%%%%%%%%%%%%%%%%%%%%%%
\subsection{Model I}
The Lagrangian responsible for neutrino mass generation is
\begin{eqnarray}
	-\mathcal{L}_{\nu- mass} & \supset   \frac{1}{2}M_{N_R} \overline{N_R^c}N_R + y_{N} \overline{L}  \Tilde{H} N_R  +{\rm H.c.}
\end{eqnarray}
In the effective theory, the masses of light neutrinos can be obtained as:
\begin{equation}
	-\mathcal{L}_{\nu - mass}=\frac{1}{2}\begin{pmatrix}
		\overline{\nu_{L}} & \overline{N_R^c}
	\end{pmatrix} \begin{pmatrix}
		0 & \frac{y_{N} v_0}{\sqrt{2}}\\
		\frac{y_{N} v_0}{\sqrt{2}} & M_{N_R}
	\end{pmatrix} \begin{pmatrix}
		\nu_{L}^c\\
		N_R
	\end{pmatrix} + {\rm H.c.}\
\end{equation}
Diagonalizing the above mass matrix we get the Majorana mass of light neutrino  as:
\begin{equation}
	m_\nu \simeq - \frac{y_{N}^2 v_0^2}{2 M_{N_R}}.\
\end{equation}
% and 
% \begin{equation}
	% M_N \simeq m_{N_R} + \frac{y_{N}^2 v_0^2}{2 m_{N_R}}.\,
	% \end{equation}

\noindent	Considering three generations of the heavy RHN, the light neutrino mass matrix is given by
\begin{equation}
	(m_\nu)_{\alpha \beta} = (m_D)_{\alpha i} \left( M^{-1}\right)_{ij} (m_D^T)_{j\beta},\
	\label{Eq:massmatrix}
\end{equation}		
where $m_D=\frac{y_{N}v_0}{\sqrt{2}}$ is the $3\times 3$ Dirac mass matrix and $M$ is the $3\times 3$ RHN mass matrix.
Using the Casas-Ibarra parametrization \cite{Casas:2001sr}, we can calculate the Yukawa coupling matrix as
\begin{equation}
	y_{N}=-i\frac{\sqrt{2}}{v_0}(U^*_{\rm PMNS}.\sqrt{\hat{m}_\nu}.\mathcal{R}^T.\sqrt{\hat{M}_N}),\
	\label{Eq:casasibarra1}
\end{equation}
where $U_{\rm PMNS}$ is the lepton mixing matrix, $\hat{m}_\nu$ is $3\times3$ diagonal light neutrino mass matrix with eigenvalues $m_1$, $m_2$, and $m_3$; $\hat{M}_N$ is $3\times3$ diagonal RHN mass matrix with eigenvalues $M_{N_{R_1}}$, $M_{N_{R_2}}$, and $M_{N_{R_3}}$, and $\mathcal{R}$ is an arbitrary complex orthogonal matrix. 
%%%%%%%%%%%%%%%%%%%%%%%%%%%%%%%%%%%%%%%%%%
\subsection{Model II}
The Lagrangian responsible for neutrino mass generation is
\begin{eqnarray}
	-\mathcal{L}_{\nu- mass} & \supset  \frac{1}{2}M_\Sigma {\rm Tr}[\overline{\Sigma^c}\Sigma]+ \sqrt{2} y_{_\Sigma} \overline{L}  \Tilde{H} \Sigma +{\rm H.c.}
\end{eqnarray}
In the effective theory, the masses of light neutrinos can be obtained as:
\begin{equation}
	-\mathcal{L}_{\nu - mass}=\frac{1}{2}\begin{pmatrix}
		\overline{\nu_{L}} & \overline{(\Sigma^{0})^c}
	\end{pmatrix} \begin{pmatrix}
		0 & \frac{y_{_\Sigma} v_0}{\sqrt{2}}\\
		\frac{y_{_\Sigma} v_0}{\sqrt{2}} & M_\Sigma
	\end{pmatrix} \begin{pmatrix}
		(\nu_{L})^c\\
		\Sigma^0
	\end{pmatrix} +{\rm H.c.}\
\end{equation}
Diagonalizing the above mass matrix we get the Majorana mass of light neutrino as:
\begin{equation}
	m_\nu \simeq - \frac{y_{_\Sigma}^2 v_0^2}{2 M_\Sigma}.\
\end{equation}
% and 
% \begin{equation}
	% M_\Sigma \simeq m_\Sigma + \frac{y_{_\Sigma}^2 v_0^2}{2 m_\Sigma}.\,
	% \end{equation}

\noindent	Considering three generations of the heavy fermion triplets, the light neutrino mass matrix is given by
\begin{equation}
	(m_\nu)_{\alpha \beta} = (m_D)_{\alpha i} \left( M^{-1}\right)_{ij} (m_D^T)_{j\beta},\
	\label{Eq:massmatrix}
\end{equation}		
where $m_D=\frac{y_{_\Sigma}~v_0}{\sqrt{2}}$ is the $3\times 3$ Dirac mass matrix and $M$ is the $3\times 3$ triplet fermion mass matrix.
Using the Casas-Ibarra parametrization \cite{Casas:2001sr} we can calculate the Yukawa coupling matrix as
\begin{equation}
	y_{_\Sigma}=-i\frac{\sqrt{2}}{v_0}(U^*_{\rm PMNS}.\sqrt{\hat{m}_\nu}.\mathcal{R}^T.\sqrt{\hat{M}_\Sigma}),\
	\label{Eq:casasibarra}
\end{equation}
where $\hat{M}_\Sigma$ is $3\times3$ diagonal triplet fermion mass matrix with eigenvalues $M_{\Sigma_1}$, $M_{\Sigma_2}$, and $M_{\Sigma_3}$. 	
%%%%%%%%%%%%%%%%%%%%%%%%%%%%%%%%%%%%%%%%%%%%%%%%%%%%%%%%%%%%%%%%%%%
\begin{widetext}
	\section{FEYNMAN DIAGRAMS FOR LEPTOGENESIS}\label{ap:leptodiag}
	
	\subsection{$\Delta L=0$}\label{ap:delL0}
	\begin{figure}[H]
		\centering
		\includegraphics[scale=0.8]{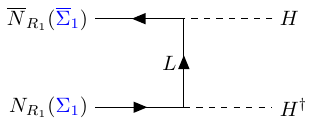}
		\includegraphics[scale=0.8]{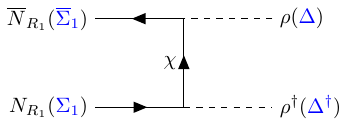}
		\includegraphics[scale=0.8]{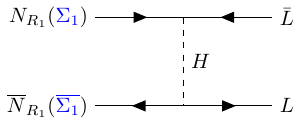}
		\includegraphics[scale=0.8]{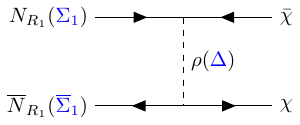}
	\end{figure}
	
	\subsection{$\Delta L=0$, transfer processes}\label{ap:delL0transfer}
	\begin{figure}[H]
		\centering
		\includegraphics[scale=0.9]{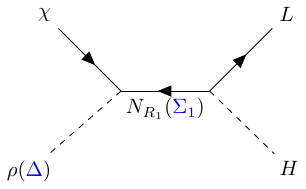}
		\includegraphics[scale=0.7]{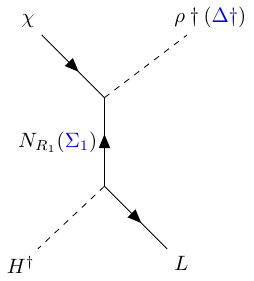}
		\includegraphics[scale=0.7]{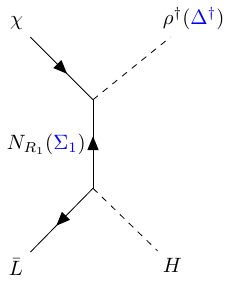}
	\end{figure}

	\subsection{$\Delta L=1$}\label{ap:delL1}
	\begin{figure}[H]
		\centering
		\includegraphics[scale=0.7]{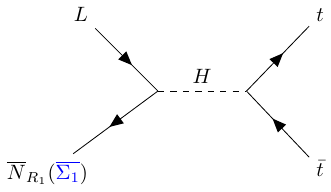}
		\includegraphics[scale=0.9]{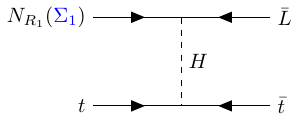}
		\includegraphics[scale=0.7]{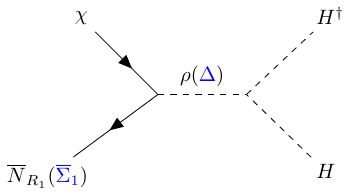}
		\includegraphics[scale=0.9]{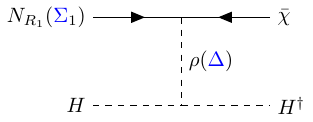}
		\includegraphics[scale=0.9]{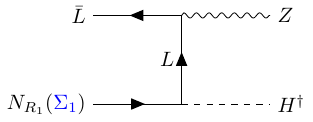}
		\includegraphics[scale=0.7]{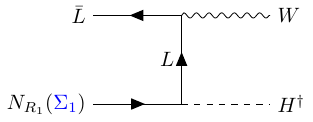}
		\includegraphics[scale=0.7]{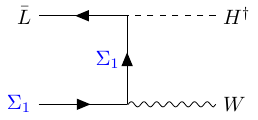}
		\includegraphics[scale=0.7]{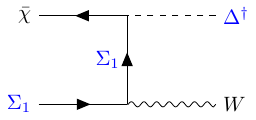}
		\includegraphics[scale=0.5]{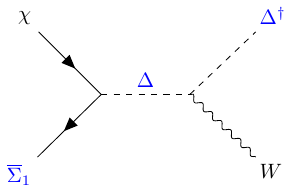}
	\end{figure}
	
	\subsection{$\Delta L=2$}\label{ap:delL2}
	\begin{figure}[H]
		\centering
		\includegraphics[scale=0.7]{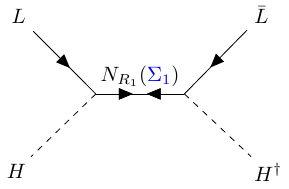}
		\includegraphics[scale=0.9]{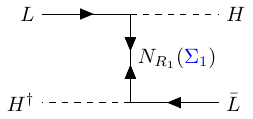}
		\includegraphics[scale=0.9]{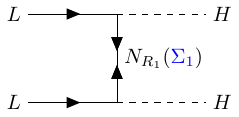}
		\includegraphics[scale=0.7]{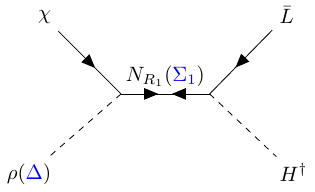}\\	
		\includegraphics[scale=0.6]{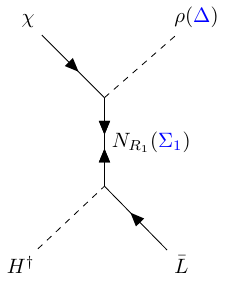}
		\includegraphics[scale=0.9]{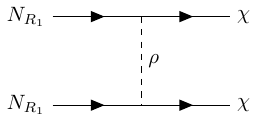}
		\includegraphics[scale=0.7]{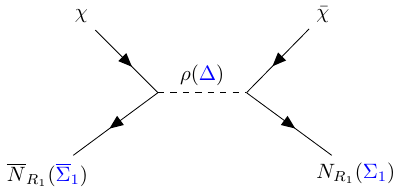}	
		\includegraphics[scale=0.9]{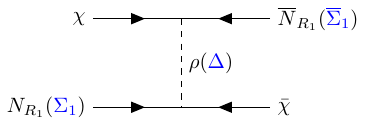}
		\includegraphics[scale=0.6]{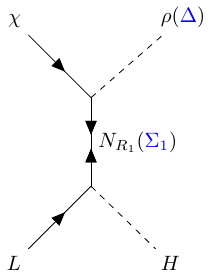}
		\includegraphics[scale=0.6]{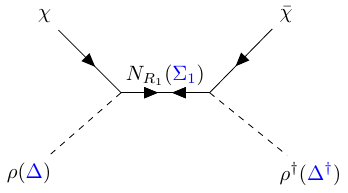}
		\includegraphics[scale=0.9]{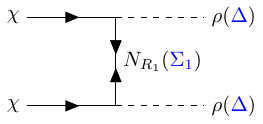}
		\includegraphics[scale=0.9]{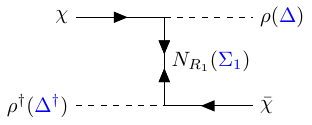}
	\end{figure}
	
	\subsection{Gauge processes}\label{ap:gaugepro}
	\begin{figure}[H]
		\centering
		\includegraphics[scale=0.6]{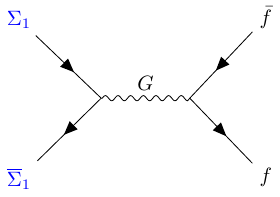}
		\includegraphics[scale=0.6]{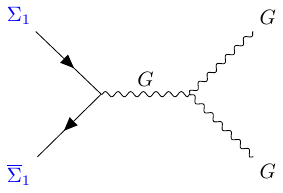}
		\includegraphics[scale=0.8]{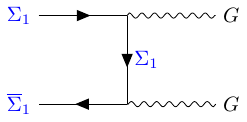}
		\includegraphics[scale=0.8]{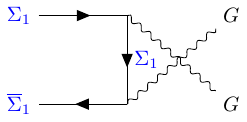}
	\end{figure}
	Here $f$ is the SM fermions and $G$ is all possible gauge bosons.
\end{widetext}
%%%%%%%%%%%%%%%%%%%%%%%%%%%%%%%%%%%%%%%%%%%%%%%%%%%%%%%%%%%%%%%%%%%
\section{LEPTON ASYMMETRY TO BARYON ASYMMETRY CONVERSION}	
\label{app:conv}
%%%%%%%%%%%%%%%%%%%%%%%%%%%%%%%%%%%%%%%%%%
\subsection{Model I}
The asymmetry in the equilibrium number densities of
particle $n_i$ over antiparticle $\bar{n}_i$ can be written in the limit $\mu_i/T\ll1$ as \cite{Harvey:1990qw}
\begin{eqnarray}
	n_i-\bar{n}_i&=&\frac{1}{6}g_i T^3 \bigg(\frac{\mu_i}{T} \bigg), ~~ \textrm{fermion}\nonumber\\&=&\frac{1}{3}g_i T^3 \bigg(\frac{\mu_i}{T} \bigg), ~~ \textrm{boson}
\end{eqnarray}
For a general derivation, we assume that the SM consists of $N$ generations of quarks and leptons, $m$ complex Higgs doublets, and we extended the SM by $N^\prime$ generations of Fermion triplets ($\Sigma$), one scalar triplet ($\Delta$), one vector-like fermion ($\chi$), and one singlet scalar ($\phi$). The chemical potentials of the SM fields are assigned as follows: $\mu_W$ for $W^-$, $\mu_0$ for $m$ $\phi^0$ Higgs fields, $\mu_{-}$ for $m$ $\phi^-$ Higgs fields, $\mu_{uL}$ for all left-handed up-quark fields, $\mu_{dL}$ for all left-handed down-quark fields,  $\mu_{u_R}$ for all the right-handed
up-quark fields, $\mu_{d_R}$ for all the right-handed
down-quark fields, $\mu_i$, for the left-handed neutrino fields,
$\mu_{iL}$ for the left-handed charged lepton fields, and $\mu_{iR}$ for
the right-handed charged lepton fields. The chemical potentials for the BSM fields are assigned as $\mu_{N}$ for all $N^\prime$ $N_R$, $\mu_\rho$ for the $\rho$, $\mu_\chi$ for $\chi$, and $\mu_\phi$ for $\phi$.  \\
Now rapid interactions in the early Universe enforces the following equilibrium relations among the chemical potentials:
{
	\begin{eqnarray}
		W^-\leftrightarrow \phi^-+\phi^0 &\Rightarrow& \mu_W=\mu_{-}+\mu_0\\
		W^-\leftrightarrow \bar{u}_L+d_L &\Rightarrow& \mu_{d_L}=\mu_W+\mu_{u_L}\\
		W^-\leftrightarrow \bar{\nu}_{iL}+e_{iL} &\Rightarrow& \mu_{iL}=\mu_W+\mu_{i}\\
		\phi^0\leftrightarrow \bar{u}_L+u_R &\Rightarrow& \mu_{u_R}=\mu_0+\mu_{u_L}\\
		\phi^0\leftrightarrow \bar{d}_R+d_L &\Rightarrow& \mu_{d_R}=-\mu_0+\mu_W+\mu_{u_L}\\
		\phi^0\leftrightarrow e_{iL}+\bar{e}_{iR} &\Rightarrow& \mu_{iR}=-\mu_0+\mu_W-\mu_{i}\\
		\phi^{0*}+\phi^{0*}\leftrightarrow \rho+\rho^{*} &\Rightarrow& \mu_{\rho}=0\\
		N_R\leftrightarrow \chi+\rho &\Rightarrow& \mu_{N}=\mu_\rho+\mu_\chi\nonumber\\&\Rightarrow& \mu_\chi=\mu_N\\
		\phi\leftrightarrow \bar{\chi}+\chi
		&\Rightarrow& \mu_{\phi}=-\mu_\chi+\mu_\chi
		\nonumber\\&\Rightarrow&\mu_\phi=0\\
		\nu_{iL}\leftrightarrow N_R+\phi^0&\Rightarrow&\mu_i=\mu_{N}+\mu_0 \label{eq:A1}
\end{eqnarray}}
From Eq. \eqref{eq:A1},
\begin{eqnarray}
	\sum_i \mu_i&=&\sum_i \mu_{N}+\sum_i\mu_0\nonumber\\&\Rightarrow&\mu=N\mu_{N}+N\mu_0\Rightarrow\mu=N\mu_{\chi}+N\mu_0\nonumber\\&\Rightarrow&\mu_\chi=\frac{\mu-N\mu_{0}}{N}
\end{eqnarray}
Now the electroweak $B+L$ anomaly implies the existence of processes that correspond to the creation of $u_Ld_Ld_L\nu_L$ from each generation out of the vacuum.	As long as these interactions are rapid, we have
\begin{eqnarray}
	N(\mu_{u_L}+2\mu_{d_L})+\sum_i\mu_i=0\Rightarrow3N\mu_{u_L}+2N\mu_W+\mu=0.\nonumber\\
\end{eqnarray}
Let us now express the baryon ($B$), lepton ($L$), charge ($Q$), and third component of weak isospin ($Q_3$) number densities as
\begin{eqnarray}
	B&=&3N(\frac{1}{3}\mu_{u_L}+\frac{1}{3}\mu_{u_R})+3N(\frac{1}{3}\mu_{d_L}+\frac{1}{3}\mu_{d_R})\nonumber\\&=&4N\mu_{u_L}+2N\mu_W
\end{eqnarray}
\begin{eqnarray}
	L&=&\sum_i(\mu_i+\mu_{iL}+\mu_{iR})+\mu_\chi\nonumber\\&=&3\mu+2N\mu_W-N{\mu_0}+\mu_\chi
\end{eqnarray}
\begin{eqnarray}
	Q&=&3N(\frac{2}{3}\mu_{u_L}-\frac{1}{3}\mu_{d_L}+\frac{2}{3}\mu_{u_R}-\frac{1}{3}\mu_{d_R})\nonumber\\&-&\big(\sum_i\mu_{iL}+\sum_i\mu_{iR}\big)-2*2\mu_W-2m\mu_-\nonumber\\&=& 2N\mu_{u_L}-2\mu-(4N+2m+4)\mu_W+(4N+2m)\mu_0\nonumber\\
\end{eqnarray}
\begin{eqnarray}
	Q_3&=&3N(\frac{1}{2}\mu_{u_L}-\frac{1}{2}\mu_{d_L})+\sum_i\big(\frac{1}{2}\mu_i-\frac{1}{2}\mu_{iL}\big)-2*2\mu_W\nonumber\\&+&2m(\frac{1}{2}\mu_+-\frac{1}{2}\mu_0)\nonumber\\&=&-(2N+m+4)\mu_W
\end{eqnarray}
Now above the critical temperature both $Q$, and $Q_3$ are zero, which will give the $B$, and $L$ in terms of $\mu_{u_L}$ as
\begin{eqnarray}
	L=\mu_{uL} N \left(\frac{4 (N+1)}{m+2 N}-3 \left(\frac{1}{N}+3\right)\right)
\end{eqnarray}
\begin{eqnarray}
	B=4 N \mu_{u_L}
\end{eqnarray}
\begin{eqnarray}
	B=-\frac{4}{3 \left(3+\frac{1}{N}\right)-\frac{4 (N+1)}{m+2 N}}L
\end{eqnarray}
%%%%%%%%%%%%%%%%%%%%%%%%%%%%%%%%%%%%%%%%%%
\subsection{Model II}
In this case, the chemical potentials for the BSM fields are assigned as: $\mu_{\Sigma^0}$ for all $N^\prime$ $\Sigma^0$, $\mu_{\Sigma^+}$ for all $\Sigma^+$, $\mu_\delta$ for the $\delta^-$, $\mu_{\delta^0}$ for $\delta^0$, $\mu_\chi$ for $\chi$, and $\mu_\phi$ for $\phi$.

Now, rapid interactions in the early Universe enforce the following equilibrium relations among the chemical potentials:
{
	\begin{eqnarray}
		\phi^0\leftrightarrow \bar{d}_R+d_L &\Rightarrow& \mu_{d_R}=-\mu_0+\mu_W+\mu_{u_L}~~~~~~\\
		\phi^0\leftrightarrow e_{iL}+\bar{e}_{iR} &\Rightarrow& \mu_{iR}=-\mu_0+\mu_W-\mu_{i}~~~~\\
		\phi^0\leftrightarrow \bar{u}_L+u_R &\Rightarrow& \mu_{u_R}=\mu_0+\mu_{u_L}\\
		W^-\leftrightarrow \phi^-+\phi^0 &\Rightarrow& \mu_W=\mu_{-}+\mu_0\\
		W^-\leftrightarrow \bar{u}_L+d_L &\Rightarrow& \mu_{d_L}=\mu_W+\mu_{u_L}\\
		W^-\leftrightarrow \bar{\nu}_{iL}+e_{iL} &\Rightarrow& \mu_{iL}=\mu_W+\mu_{i}\\
		W^-+W^-\leftrightarrow \delta^-+\delta^- &\Rightarrow& \mu_{W}+\mu_W=\mu_\delta+\mu_\delta\nonumber\\&\Rightarrow& \mu_\delta=\mu_W\\
		W^-+W^3\leftrightarrow \delta^-+\delta^0 &\Rightarrow& \mu_{W}+0=\mu_\delta+\mu_{\delta^0}\nonumber\\&\Rightarrow& \mu_{\delta^0}=0\\
		\phi\leftrightarrow \bar{\chi}+\chi&\Rightarrow& \mu_{\phi}=-\mu_\chi+\mu_\chi\nonumber\\&\Rightarrow&\mu_\phi=0\\
		\Sigma^0\leftrightarrow W^-+\Sigma^+&\Rightarrow& \mu_{\Sigma^0}=\mu_W+\mu_{\Sigma^+}\\
		\Sigma^0\leftrightarrow\delta^0+\chi&\Rightarrow&\mu_{\Sigma^0}=\mu_{\delta^0}+\mu_\chi\nonumber\\&\Rightarrow&\mu_\chi=\mu_{\Sigma^0}\\
		\nu_{iL}\leftrightarrow\Sigma^0+\phi^0&\Rightarrow&\mu_i=\mu_{\Sigma^0}+\mu_0\label{eq:A13}
\end{eqnarray}}
From Eq. \eqref{eq:A13},
\begin{eqnarray}
	\sum_i \mu_i&=&\sum_i \mu_{\Sigma^0}+\sum_i\mu_0\nonumber\\&\Rightarrow&\mu=N\mu_{\Sigma^0}+N\mu_0\Rightarrow\mu=N\mu_{\chi}+N\mu_0\nonumber\\&\Rightarrow&\mu_\chi=\frac{\mu-N\mu_{0}}{N}.
\end{eqnarray}
Now, the electroweak $B+L$ anomaly implies the existence of processes that correspond to the creation of $u_Ld_Ld_L\nu_L$ from each generation out of the vacuum.	As long as these interactions are rapid, we have
\begin{eqnarray}
	N(\mu_{u_L}+2\mu_{d_L})+\sum_i\mu_i=0\Rightarrow3N\mu_{u_L}+2N\mu_W+\mu=0.\nonumber\\
\end{eqnarray}
Let us now express the baryon ($B$), lepton ($L$), charge ($Q$), and third component of weak isospin ($Q_3$) number densities as
\begin{eqnarray}
	B&=&3N(\frac{1}{3}\mu_{u_L}+\frac{1}{3}\mu_{u_R})+3N(\frac{1}{3}\mu_{d_L}+\frac{1}{3}\mu_{d_R})\nonumber\\&=&4N\mu_{u_L}+2N\mu_W
\end{eqnarray}
\begin{eqnarray}
	L=\sum_i(\mu_i+\mu_{iL}+\mu_{iR})+\mu_\chi=3\mu+2N\mu_W-N{\mu_0}+\mu_\chi\nonumber\\
\end{eqnarray}
\begin{eqnarray}
	Q&=&3N(\frac{2}{3}\mu_{u_L}-\frac{1}{3}\mu_{d_L}+\frac{2}{3}\mu_{u_R}-\frac{1}{3}\mu_{d_R})\nonumber\\&-&\big(\sum_i\mu_{iL}+\sum_i\mu_{iR}\big)-2*2\mu_W-2m\mu_--2*2\mu_\delta\nonumber\\&+&2*2(-\mu_\delta)\nonumber\\&=& 2N\mu_{u_L}-2\mu-(4N+2m+12)\mu_W+(4N+2m)\mu_0\nonumber\\
\end{eqnarray}
\begin{eqnarray}
	Q_3&=&3N(\frac{1}{2}\mu_{u_L}-\frac{1}{2}\mu_{d_L})+\sum_i\big(\frac{1}{2}\mu_i-\frac{1}{2}\mu_{iL}\big)-2*2\mu_W\nonumber\\&+&2m(\frac{1}{2}\mu_+-\frac{1}{2}\mu_0)+2*2(-1)\mu_\delta+2*2(+1)(-\mu_\delta)\nonumber\\&=&-(2N+m+12)\mu_W
\end{eqnarray}
Now above the critical temperature both $Q$, and $Q_3$ are zero, which will give the $B$, and $L$ in terms of $\mu_{u_L}$ as
\begin{eqnarray}
	L=\mu_{uL} N \left(\frac{4 (N+1)}{m+2 N}-3 \left(\frac{1}{N}+3\right)\right)
\end{eqnarray}
\begin{eqnarray}
	B=4 N \mu_{u_L}
\end{eqnarray}
\begin{eqnarray}
	B=-\frac{4}{3 \left(3+\frac{1}{N}\right)-\frac{4 (N+1)}{m+2 N}}L
\end{eqnarray}

%%%%%%%%%%%%%%%%%%%%%%%%%%%%%%%%%%%%%%%%%%%%%%%%%%%%%%%%%%%%%%%%%%%
\begin{widetext}	
	\section{DARK MATTER DECAY WIDTH}\label{ap:decayrate}
	\small\begin{eqnarray}	
		\Gamma_{{\chi}}^{(1)} (\chi\rightarrow e W)=\frac{{G_{{\chi}}^{(1)}}^2 M^3_\chi}{16\pi M_W^2} \bigg(1+\frac{m^4_e}{M^4_\chi}-\frac{2M^4_W}{M^4_\chi}-\frac{2m^2_e}{M^2_\chi}+\frac{M^2_W}{M^2_\chi}+\frac{m^2_{e}M^2_W}{M^4_\chi}\bigg)\bigg( 1+\frac{m_e^4}{M_\chi^4} +\frac{M_W^4}{M_\chi^4}-\frac{2m_e^2}{M_\chi^2}-\frac{2M_W^2}{M_\chi^2}-\frac{2m_e^2M_W^2}{M_\chi^4}\bigg)^{\frac{1}{2}},\nonumber\\ \label{eq:2body1}
	\end{eqnarray}	
	\begin{eqnarray}
		\Gamma_{{\chi}}^{(2)}(\chi\rightarrow \nu Z)=\frac{{G_{{\chi}}^{(2)}}^2 M^3_\chi}{16\pi M_Z^2}\bigg(1+\frac{m^4_\nu}{M^4_\chi}-\frac{2M^4_Z}{M^4_\chi}-\frac{2m^2_\nu}{M^2_\chi}+\frac{M^2_Z}{M^2_\chi}+\frac{m^2_{\nu}M^2_Z}{M^4_\chi}\bigg)\bigg( 1+\frac{m_\nu^4}{M_\chi^4} +\frac{M_Z^4}{M_\chi^4}-\frac{2m_\nu^2}{M_\chi^2}-\frac{2M_Z^2}{M_\chi^2}-\frac{2m_\nu^2M_Z^2}{M_\chi^4}\bigg)^{\frac{1}{2}},\nonumber\\ \label{eq:2body2}
	\end{eqnarray}
	
	\begin{eqnarray}
		\Gamma_{{\chi}}^{(3)}(\chi\rightarrow \phi \nu)=&\frac{{G_{{\chi}}^{(3)}}^2 M_\chi}{16\pi} \bigg( 1+\frac{m_\nu^2}{M_\chi^2} -\frac{M_\phi^2}{M_\chi^2}+\frac{2m_\nu}{M_\chi}\bigg)\bigg( 1+\frac{m_\nu^4}{M_\chi^4} +\frac{M_\phi^4}{M_\chi^4}-\frac{2m_\nu^2}{M_\chi^2}-\frac{2M_\phi^2}{M_\chi^2}-\frac{2m_\nu^2M_\phi^2}{M_\chi^4}\bigg)^{\frac{1}{2}}, \label{eq:2body3}
	\end{eqnarray}
	
	\begin{eqnarray}
		\Gamma_{{\chi}}^{(4)}(\chi\rightarrow \phi\nu)&=&\frac{{G_{{\chi}}^{(4)}}^2 M_\chi}{16\pi} \bigg( 1+\frac{m_\nu^2}{M_\chi^2} -\frac{M_\phi^2}{M_\chi^2}+\frac{2m_\nu}{M_\chi}\bigg)\bigg( 1+\frac{m_\nu^4}{M_\chi^4} +\frac{M_\phi^4}{M_\chi^4}-\frac{2m_\nu^2}{M_\chi^2}-\frac{2M_\phi^2}{M_\chi^2}-\frac{2m_\nu^2M_\phi^2}{M_\chi^4}\bigg)^{\frac{1}{2}}, \label{eq:2body4}
	\end{eqnarray}	
	
	\begin{eqnarray}
		\Gamma_{{\chi}}^{(5)}(\chi\rightarrow \nu f \bar{f})&=&\frac{{G_{{\chi}}^{(5)}}^2 M_\chi^5}{192\pi^3} \bigg( \frac{1}{4}(1-4s_w^2+8s_w^4)((1-14\frac{m_f^2}{M_\chi^2}-2\frac{m_f^4}{M_\chi^4}-12\frac{m_f^6}{M_\chi^6})\sqrt{1-4\frac{m_f^2}{M_\chi^2}}+12\frac{m_f^4}{M_\chi^4}(\frac{m_f^4}{M_\chi^4}-1)\nonumber\\&&\log\bigg[\frac{1-3\frac{m_f^2}{M_\chi^2}-(1-\frac{m_f^2}{M_\chi^2})\sqrt{1-4\frac{m_f^2}{M_\chi^2}}}{\frac{m_f^2}{M_\chi^2}(1+\sqrt{1-4\frac{m_f^2}{M_\chi^2}})}\bigg])+2s_w^2(2s_w^2-1)(\frac{m_f^2}{M_\chi^2}(2+10\frac{m_f^2}{M_\chi^2}-12\frac{m_f^4}{M_\chi^4})\sqrt{1-4\frac{m_f^2}{M_\chi^2}}\nonumber\\&+&6\frac{m_f^4}{M_\chi^4}(1-2\frac{m_f^2}{M_\chi^2}+2\frac{m_f^4}{M_\chi^4})\log\bigg[\frac{1-3\frac{m_f^2}{M_\chi^2}-(1-\frac{m_f^2}{M_\chi^2})\sqrt{1-4\frac{m_f^2}{M_\chi^2}}}{\frac{m_f^2}{M_\chi^2}(1+\sqrt{1-4\frac{m_f^2}{M_\chi^2}})}\bigg])   \bigg) ,\label{eq:3body}
	\end{eqnarray}
\end{widetext}
where the
effective couplings $G_{{\chi}}^{(1)}$, $G_{{\chi}}^{(2)}$, $G_{{\chi}}^{(3)}$, $G_{{\chi}}^{(4)}$, and $G_{{\chi}}^{(5)}$ can be given as
\begin{equation}
	G_{{\chi}}^{(1)} =\big(y_\chi v_1\big)\bigg(\frac{1}{M_{\Sigma_1}}\bigg)\frac{\big(y_{_\Sigma}v_0\big)}{\sqrt{2}}\bigg(\frac{1}{M_{\chi}}\bigg)\bigg(\frac{g}{2~ \cos\theta_w}\bigg),\ \label{eq:gx1}
\end{equation}
\begin{equation}
	G_{{\chi}}^{(2)} =\big(y_\chi v_1\big)\bigg(\frac{1}{M_{\Sigma_1}}\bigg)\frac{\big(y_{_\Sigma}v_0\big)}{\sqrt{2}}\bigg(\frac{1}{M_{\chi}}\bigg)\bigg(\frac{g}{\sqrt{2}}\bigg),\ \label{eq:gx2}
\end{equation}	
\begin{equation}
	G_{{\chi}}^{(3)} =\lambda_{{\rm DM}}\bigg(\frac{1}{M_{\chi}}\bigg)\big(y_{\chi}v_1\big)\bigg(\frac{1}{M_{\Sigma_1}}\bigg)\frac{\big(y_{_\Sigma}v_0\big)}{\sqrt{2}},\ \label{eq:gx3}
\end{equation}	
\begin{equation}
	G_{{\chi}}^{(4)} =\big(y_{\chi}v_1\big)\bigg(\frac{1}{M_{\Sigma_1}}\bigg)(y_{_{\Sigma}}) (\sin(\gamma)),\ \label{eq:gx4}
\end{equation}	
\begin{equation}
	G_{{\chi}}^{(5)} =\big(y_\chi v_1\big)\bigg(\frac{1}{ M_{\Sigma_1}}\bigg)(y_{_\Sigma})\bigg(\frac{1}{M_{h_1}^2}\bigg)\bigg(\frac{m_f}{v_0}\bigg),\ \label{eq:gx5}
\end{equation}
where $v_1$ is the VEV of $\rho$ ($\Delta$).\\
%%%%%%%%%%%%%%%%%%%%%%%%%%%%%%%%%%%%%%%%%%%%%%%%%%%%%%%%%%%%%%%%%%%
\section{LIFETIME OF $\phi$}\label{ap:philifetime}
The light scalar, $\phi$, can decay to the SM fermions after mixing with the SM Higgs after the electroweak symmetry breaking. The decay width is given as
\begin{eqnarray}
	\Gamma_{\phi\rightarrow ff}=\sin^2\gamma\left(\frac{m_f}{v_0}\right)^2\frac{M_{\phi}}{8\pi}\bigg(1-\frac{4m_f^2}{M^2_{\phi}}\bigg)^{\frac{3}{2}},
\end{eqnarray}
where $m_f$ is the mass of the fermion.
The lifetime of $\phi$ is calculated as
\begin{eqnarray}
	\tau_{\phi}=(\Gamma_{\phi\rightarrow ff})^{-1}.
\end{eqnarray}

\vfill
\eject


\begin{thebibliography}{92}%
	\makeatletter
	\providecommand \@ifxundefined [1]{%
		\@ifx{#1\undefined}
	}%
	\providecommand \@ifnum [1]{%
		\ifnum #1\expandafter \@firstoftwo
		\else \expandafter \@secondoftwo
		\fi
	}%
	\providecommand \@ifx [1]{%
		\ifx #1\expandafter \@firstoftwo
		\else \expandafter \@secondoftwo
		\fi
	}%
	\providecommand \natexlab [1]{#1}%
	\providecommand \enquote  [1]{``#1''}%
	\providecommand \bibnamefont  [1]{#1}%
	\providecommand \bibfnamefont [1]{#1}%
	\providecommand \citenamefont [1]{#1}%
	\providecommand \href@noop [0]{\@secondoftwo}%
	\providecommand \href [0]{\begingroup \@sanitize@url \@href}%
	\providecommand \@href[1]{\@@startlink{#1}\@@href}%
	\providecommand \@@href[1]{\endgroup#1\@@endlink}%
	\providecommand \@sanitize@url [0]{\catcode `\\12\catcode `\$12\catcode
		`\&12\catcode `\#12\catcode `\^12\catcode `\_12\catcode `\%12\relax}%
	\providecommand \@@startlink[1]{}%
	\providecommand \@@endlink[0]{}%
	\providecommand \url  [0]{\begingroup\@sanitize@url \@url }%
	\providecommand \@url [1]{\endgroup\@href {#1}{\urlprefix }}%
	\providecommand \urlprefix  [0]{URL }%
	\providecommand \Eprint [0]{\href }%
	\providecommand \doibase [0]{https://doi.org/}%
	\providecommand \selectlanguage [0]{\@gobble}%
	\providecommand \bibinfo  [0]{\@secondoftwo}%
	\providecommand \bibfield  [0]{\@secondoftwo}%
	\providecommand \translation [1]{[#1]}%
	\providecommand \BibitemOpen [0]{}%
	\providecommand \bibitemStop [0]{}%
	\providecommand \bibitemNoStop [0]{.\EOS\space}%
	\providecommand \EOS [0]{\spacefactor3000\relax}%
	\providecommand \BibitemShut  [1]{\csname bibitem#1\endcsname}%
	\let\auto@bib@innerbib\@empty
	%</preamble>
	\bibitem [{\citenamefont {Zwicky}(1933)}]{Zwicky:1933gu}%
	\BibitemOpen
	\bibfield  {author} {\bibinfo {author} {\bibfnamefont {F.}~\bibnamefont
			{Zwicky}},\ }\href {https://doi.org/10.1007/s10714-008-0707-4} {\bibfield
		{journal} {\bibinfo  {journal} {Helv. Phys. Acta}\ }\textbf {\bibinfo
			{volume} {6}},\ \bibinfo {pages} {110} (\bibinfo {year} {1933})},\ \bibinfo
	{note} {[Gen. Rel. Grav.41,207(2009)]}\BibitemShut {NoStop}%
	%%CITATION = HPACA,6,110;%%
	\bibitem [{\citenamefont {Rubin}\ and\ \citenamefont
		{Ford}(1970)}]{Rubin:1970zza}%
	\BibitemOpen
	\bibfield  {author} {\bibinfo {author} {\bibfnamefont {V.~C.}\ \bibnamefont
			{Rubin}}\ and\ \bibinfo {author} {\bibfnamefont {W.~K.}\ \bibnamefont {Ford},
			\bibfnamefont {Jr.}},\ }\href {https://doi.org/10.1086/150317} {\bibfield
		{journal} {\bibinfo  {journal} {Astrophys. J.}\ }\textbf {\bibinfo {volume}
			{159}},\ \bibinfo {pages} {379} (\bibinfo {year} {1970})}\BibitemShut
	{NoStop}%
	%%CITATION = ASJOA,159,379;%%
	\bibitem [{\citenamefont {Clowe}\ \emph {et~al.}(2006)\citenamefont {Clowe},
		\citenamefont {Bradac}, \citenamefont {Gonzalez}, \citenamefont {Markevitch},
		\citenamefont {Randall}, \citenamefont {Jones},\ and\ \citenamefont
		{Zaritsky}}]{Clowe:2006eq}%
	\BibitemOpen
	\bibfield  {author} {\bibinfo {author} {\bibfnamefont {D.}~\bibnamefont
			{Clowe}}, \bibinfo {author} {\bibfnamefont {M.}~\bibnamefont {Bradac}},
		\bibinfo {author} {\bibfnamefont {A.~H.}\ \bibnamefont {Gonzalez}}, \bibinfo
		{author} {\bibfnamefont {M.}~\bibnamefont {Markevitch}}, \bibinfo {author}
		{\bibfnamefont {S.~W.}\ \bibnamefont {Randall}}, \bibinfo {author}
		{\bibfnamefont {C.}~\bibnamefont {Jones}},\ and\ \bibinfo {author}
		{\bibfnamefont {D.}~\bibnamefont {Zaritsky}},\ }\href
	{https://doi.org/10.1086/508162} {\bibfield  {journal} {\bibinfo  {journal}
			{Astrophys. J.}\ }\textbf {\bibinfo {volume} {648}},\ \bibinfo {pages} {L109}
		(\bibinfo {year} {2006})},\ \Eprint {https://arxiv.org/abs/astro-ph/0608407}
	{arXiv:astro-ph/0608407 [astro-ph]} \BibitemShut {NoStop}%
	%%CITATION = ASTRO-PH/0608407;%%
	\bibitem [{\citenamefont {Aghanim}\ \emph {et~al.}(2020)\citenamefont {Aghanim}
		\emph {et~al.}}]{Planck:2018vyg}%
	\BibitemOpen
	\bibfield  {author} {\bibinfo {author} {\bibfnamefont {N.}~\bibnamefont
			{Aghanim}} \emph {et~al.} (\bibinfo {collaboration} {Planck}),\ }\href
	{https://doi.org/10.1051/0004-6361/201833910} {\bibfield  {journal} {\bibinfo
			{journal} {Astron. Astrophys.}\ }\textbf {\bibinfo {volume} {641}},\
		\bibinfo {pages} {A6} (\bibinfo {year} {2020})},\ \bibinfo {note} {[Erratum:
		Astron.Astrophys. 652, C4 (2021)]},\ \Eprint
	{https://arxiv.org/abs/1807.06209} {arXiv:1807.06209 [astro-ph.CO]}
	\BibitemShut {NoStop}%
	\bibitem [{\citenamefont {Tulin}\ and\ \citenamefont
		{Yu}(2018)}]{Tulin:2017ara}%
	\BibitemOpen
	\bibfield  {author} {\bibinfo {author} {\bibfnamefont {S.}~\bibnamefont
			{Tulin}}\ and\ \bibinfo {author} {\bibfnamefont {H.-B.}\ \bibnamefont {Yu}},\
	}\href {https://doi.org/10.1016/j.physrep.2017.11.004} {\bibfield  {journal}
		{\bibinfo  {journal} {Phys. Rept.}\ }\textbf {\bibinfo {volume} {730}},\
		\bibinfo {pages} {1} (\bibinfo {year} {2018})},\ \Eprint
	{https://arxiv.org/abs/1705.02358} {arXiv:1705.02358 [hep-ph]} \BibitemShut
	{NoStop}%
	\bibitem [{\citenamefont {Bullock}\ and\ \citenamefont
		{Boylan-Kolchin}(2017)}]{Bullock:2017xww}%
	\BibitemOpen
	\bibfield  {author} {\bibinfo {author} {\bibfnamefont {J.~S.}\ \bibnamefont
			{Bullock}}\ and\ \bibinfo {author} {\bibfnamefont {M.}~\bibnamefont
			{Boylan-Kolchin}},\ }\href
	{https://doi.org/10.1146/annurev-astro-091916-055313} {\bibfield  {journal}
		{\bibinfo  {journal} {Ann. Rev. Astron. Astrophys.}\ }\textbf {\bibinfo
			{volume} {55}},\ \bibinfo {pages} {343} (\bibinfo {year} {2017})},\ \Eprint
	{https://arxiv.org/abs/1707.04256} {arXiv:1707.04256 [astro-ph.CO]}
	\BibitemShut {NoStop}%
	%%CITATION = ARXIV:1707.04256;%%
	\bibitem [{\citenamefont {Spergel}\ and\ \citenamefont
		{Steinhardt}(2000)}]{Spergel:1999mh}%
	\BibitemOpen
	\bibfield  {author} {\bibinfo {author} {\bibfnamefont {D.~N.}\ \bibnamefont
			{Spergel}}\ and\ \bibinfo {author} {\bibfnamefont {P.~J.}\ \bibnamefont
			{Steinhardt}},\ }\href {https://doi.org/10.1103/PhysRevLett.84.3760}
	{\bibfield  {journal} {\bibinfo  {journal} {Phys. Rev. Lett.}\ }\textbf
		{\bibinfo {volume} {84}},\ \bibinfo {pages} {3760} (\bibinfo {year}
		{2000})},\ \Eprint {https://arxiv.org/abs/astro-ph/9909386}
	{arXiv:astro-ph/9909386} \BibitemShut {NoStop}%
	\bibitem [{\citenamefont {Buckley}\ and\ \citenamefont
		{Fox}(2010)}]{Buckley:2009in}%
	\BibitemOpen
	\bibfield  {author} {\bibinfo {author} {\bibfnamefont {M.~R.}\ \bibnamefont
			{Buckley}}\ and\ \bibinfo {author} {\bibfnamefont {P.~J.}\ \bibnamefont
			{Fox}},\ }\href {https://doi.org/10.1103/PhysRevD.81.083522} {\bibfield
		{journal} {\bibinfo  {journal} {Phys. Rev. D}\ }\textbf {\bibinfo {volume}
			{81}},\ \bibinfo {pages} {083522} (\bibinfo {year} {2010})},\ \Eprint
	{https://arxiv.org/abs/0911.3898} {arXiv:0911.3898 [hep-ph]} \BibitemShut
	{NoStop}%
	\bibitem [{\citenamefont {Feng}\ \emph {et~al.}(2010)\citenamefont {Feng},
		\citenamefont {Kaplinghat},\ and\ \citenamefont {Yu}}]{Feng:2009hw}%
	\BibitemOpen
	\bibfield  {author} {\bibinfo {author} {\bibfnamefont {J.~L.}\ \bibnamefont
			{Feng}}, \bibinfo {author} {\bibfnamefont {M.}~\bibnamefont {Kaplinghat}},\
		and\ \bibinfo {author} {\bibfnamefont {H.-B.}\ \bibnamefont {Yu}},\ }\href
	{https://doi.org/10.1103/PhysRevLett.104.151301} {\bibfield  {journal}
		{\bibinfo  {journal} {Phys. Rev. Lett.}\ }\textbf {\bibinfo {volume} {104}},\
		\bibinfo {pages} {151301} (\bibinfo {year} {2010})},\ \Eprint
	{https://arxiv.org/abs/0911.0422} {arXiv:0911.0422 [hep-ph]} \BibitemShut
	{NoStop}%
	\bibitem [{\citenamefont {Feng}\ \emph {et~al.}(2009)\citenamefont {Feng},
		\citenamefont {Kaplinghat}, \citenamefont {Tu},\ and\ \citenamefont
		{Yu}}]{Feng:2009mn}%
	\BibitemOpen
	\bibfield  {author} {\bibinfo {author} {\bibfnamefont {J.~L.}\ \bibnamefont
			{Feng}}, \bibinfo {author} {\bibfnamefont {M.}~\bibnamefont {Kaplinghat}},
		\bibinfo {author} {\bibfnamefont {H.}~\bibnamefont {Tu}},\ and\ \bibinfo
		{author} {\bibfnamefont {H.-B.}\ \bibnamefont {Yu}},\ }\href
	{https://doi.org/10.1088/1475-7516/2009/07/004} {\bibfield  {journal}
		{\bibinfo  {journal} {JCAP}\ }\textbf {\bibinfo {volume} {07}},\ \bibinfo
		{pages} {004}},\ \Eprint {https://arxiv.org/abs/0905.3039} {arXiv:0905.3039
		[hep-ph]} \BibitemShut {NoStop}%
	\bibitem [{\citenamefont {Loeb}\ and\ \citenamefont
		{Weiner}(2011)}]{Loeb:2010gj}%
	\BibitemOpen
	\bibfield  {author} {\bibinfo {author} {\bibfnamefont {A.}~\bibnamefont
			{Loeb}}\ and\ \bibinfo {author} {\bibfnamefont {N.}~\bibnamefont {Weiner}},\
	}\href {https://doi.org/10.1103/PhysRevLett.106.171302} {\bibfield  {journal}
		{\bibinfo  {journal} {Phys. Rev. Lett.}\ }\textbf {\bibinfo {volume} {106}},\
		\bibinfo {pages} {171302} (\bibinfo {year} {2011})},\ \Eprint
	{https://arxiv.org/abs/1011.6374} {arXiv:1011.6374 [astro-ph.CO]}
	\BibitemShut {NoStop}%
	\bibitem [{\citenamefont {Zavala}\ \emph {et~al.}(2013)\citenamefont {Zavala},
		\citenamefont {Vogelsberger},\ and\ \citenamefont {Walker}}]{Zavala:2012us}%
	\BibitemOpen
	\bibfield  {author} {\bibinfo {author} {\bibfnamefont {J.}~\bibnamefont
			{Zavala}}, \bibinfo {author} {\bibfnamefont {M.}~\bibnamefont
			{Vogelsberger}},\ and\ \bibinfo {author} {\bibfnamefont {M.~G.}\ \bibnamefont
			{Walker}},\ }\href {https://doi.org/10.1093/mnrasl/sls053} {\bibfield
		{journal} {\bibinfo  {journal} {Mon. Not. Roy. Astron. Soc.}\ }\textbf
		{\bibinfo {volume} {431}},\ \bibinfo {pages} {L20} (\bibinfo {year}
		{2013})},\ \Eprint {https://arxiv.org/abs/1211.6426} {arXiv:1211.6426
		[astro-ph.CO]} \BibitemShut {NoStop}%
	\bibitem [{\citenamefont {Vogelsberger}\ \emph {et~al.}(2012)\citenamefont
		{Vogelsberger}, \citenamefont {Zavala},\ and\ \citenamefont
		{Loeb}}]{Vogelsberger:2012ku}%
	\BibitemOpen
	\bibfield  {author} {\bibinfo {author} {\bibfnamefont {M.}~\bibnamefont
			{Vogelsberger}}, \bibinfo {author} {\bibfnamefont {J.}~\bibnamefont
			{Zavala}},\ and\ \bibinfo {author} {\bibfnamefont {A.}~\bibnamefont {Loeb}},\
	}\href {https://doi.org/10.1111/j.1365-2966.2012.21182.x} {\bibfield
		{journal} {\bibinfo  {journal} {Mon. Not. Roy. Astron. Soc.}\ }\textbf
		{\bibinfo {volume} {423}},\ \bibinfo {pages} {3740} (\bibinfo {year}
		{2012})},\ \Eprint {https://arxiv.org/abs/1201.5892} {arXiv:1201.5892
		[astro-ph.CO]} \BibitemShut {NoStop}%
	\bibitem [{\citenamefont {Bringmann}\ \emph {et~al.}(2017)\citenamefont
		{Bringmann}, \citenamefont {Kahlhoefer}, \citenamefont {Schmidt-Hoberg},\
		and\ \citenamefont {Walia}}]{Bringmann:2016din}%
	\BibitemOpen
	\bibfield  {author} {\bibinfo {author} {\bibfnamefont {T.}~\bibnamefont
			{Bringmann}}, \bibinfo {author} {\bibfnamefont {F.}~\bibnamefont
			{Kahlhoefer}}, \bibinfo {author} {\bibfnamefont {K.}~\bibnamefont
			{Schmidt-Hoberg}},\ and\ \bibinfo {author} {\bibfnamefont {P.}~\bibnamefont
			{Walia}},\ }\href {https://doi.org/10.1103/PhysRevLett.118.141802} {\bibfield
		{journal} {\bibinfo  {journal} {Phys. Rev. Lett.}\ }\textbf {\bibinfo
			{volume} {118}},\ \bibinfo {pages} {141802} (\bibinfo {year} {2017})},\
	\Eprint {https://arxiv.org/abs/1612.00845} {arXiv:1612.00845 [hep-ph]}
	\BibitemShut {NoStop}%
	\bibitem [{\citenamefont {Kaplinghat}\ \emph {et~al.}(2016)\citenamefont
		{Kaplinghat}, \citenamefont {Tulin},\ and\ \citenamefont
		{Yu}}]{Kaplinghat:2015aga}%
	\BibitemOpen
	\bibfield  {author} {\bibinfo {author} {\bibfnamefont {M.}~\bibnamefont
			{Kaplinghat}}, \bibinfo {author} {\bibfnamefont {S.}~\bibnamefont {Tulin}},\
		and\ \bibinfo {author} {\bibfnamefont {H.-B.}\ \bibnamefont {Yu}},\ }\href
	{https://doi.org/10.1103/PhysRevLett.116.041302} {\bibfield  {journal}
		{\bibinfo  {journal} {Phys. Rev. Lett.}\ }\textbf {\bibinfo {volume} {116}},\
		\bibinfo {pages} {041302} (\bibinfo {year} {2016})},\ \Eprint
	{https://arxiv.org/abs/1508.03339} {arXiv:1508.03339 [astro-ph.CO]}
	\BibitemShut {NoStop}%
	\bibitem [{\citenamefont {van~den Aarssen}\ \emph {et~al.}(2012)\citenamefont
		{van~den Aarssen}, \citenamefont {Bringmann},\ and\ \citenamefont
		{Pfrommer}}]{Aarssen:2012fx}%
	\BibitemOpen
	\bibfield  {author} {\bibinfo {author} {\bibfnamefont {L.~G.}\ \bibnamefont
			{van~den Aarssen}}, \bibinfo {author} {\bibfnamefont {T.}~\bibnamefont
			{Bringmann}},\ and\ \bibinfo {author} {\bibfnamefont {C.}~\bibnamefont
			{Pfrommer}},\ }\href {https://doi.org/10.1103/PhysRevLett.109.231301}
	{\bibfield  {journal} {\bibinfo  {journal} {Phys. Rev. Lett.}\ }\textbf
		{\bibinfo {volume} {109}},\ \bibinfo {pages} {231301} (\bibinfo {year}
		{2012})},\ \Eprint {https://arxiv.org/abs/1205.5809} {arXiv:1205.5809
		[astro-ph.CO]} \BibitemShut {NoStop}%
	\bibitem [{\citenamefont {Tulin}\ \emph
		{et~al.}(2013{\natexlab{a}})\citenamefont {Tulin}, \citenamefont {Yu},\ and\
		\citenamefont {Zurek}}]{Tulin:2013teo}%
	\BibitemOpen
	\bibfield  {author} {\bibinfo {author} {\bibfnamefont {S.}~\bibnamefont
			{Tulin}}, \bibinfo {author} {\bibfnamefont {H.-B.}\ \bibnamefont {Yu}},\ and\
		\bibinfo {author} {\bibfnamefont {K.~M.}\ \bibnamefont {Zurek}},\ }\href
	{https://doi.org/10.1103/PhysRevD.87.115007} {\bibfield  {journal} {\bibinfo
			{journal} {Phys. Rev. D}\ }\textbf {\bibinfo {volume} {87}},\ \bibinfo
		{pages} {115007} (\bibinfo {year} {2013}{\natexlab{a}})},\ \Eprint
	{https://arxiv.org/abs/1302.3898} {arXiv:1302.3898 [hep-ph]} \BibitemShut
	{NoStop}%
	\bibitem [{\citenamefont {Borah}\ \emph
		{et~al.}(2022{\natexlab{a}})\citenamefont {Borah}, \citenamefont
		{Mahapatra},\ and\ \citenamefont {Sahu}}]{Borah:2022ask}%
	\BibitemOpen
	\bibfield  {author} {\bibinfo {author} {\bibfnamefont {D.}~\bibnamefont
			{Borah}}, \bibinfo {author} {\bibfnamefont {S.}~\bibnamefont {Mahapatra}},\
		and\ \bibinfo {author} {\bibfnamefont {N.}~\bibnamefont {Sahu}},\ }\href@noop
	{} {\  (\bibinfo {year} {2022}{\natexlab{a}})},\ \Eprint
	{https://arxiv.org/abs/2211.15703} {arXiv:2211.15703 [hep-ph]} \BibitemShut
	{NoStop}%
	\bibitem [{\citenamefont {Kolb}\ and\ \citenamefont
		{Turner}(1990)}]{Kolb:1990vq}%
	\BibitemOpen
	\bibfield  {author} {\bibinfo {author} {\bibfnamefont {E.~W.}\ \bibnamefont
			{Kolb}}\ and\ \bibinfo {author} {\bibfnamefont {M.~S.}\ \bibnamefont
			{Turner}},\ }\href@noop {} {\emph {\bibinfo {title} {{The Early
					Universe}}}},\ Vol.~\bibinfo {volume} {69}\ (\bibinfo {year}
	{1990})\BibitemShut {NoStop}%
	\bibitem [{\citenamefont {Jungman}\ \emph {et~al.}(1996)\citenamefont
		{Jungman}, \citenamefont {Kamionkowski},\ and\ \citenamefont
		{Griest}}]{Jungman:1995df}%
	\BibitemOpen
	\bibfield  {author} {\bibinfo {author} {\bibfnamefont {G.}~\bibnamefont
			{Jungman}}, \bibinfo {author} {\bibfnamefont {M.}~\bibnamefont
			{Kamionkowski}},\ and\ \bibinfo {author} {\bibfnamefont {K.}~\bibnamefont
			{Griest}},\ }\href {https://doi.org/10.1016/0370-1573(95)00058-5} {\bibfield
		{journal} {\bibinfo  {journal} {Phys. Rept.}\ }\textbf {\bibinfo {volume}
			{267}},\ \bibinfo {pages} {195} (\bibinfo {year} {1996})},\ \Eprint
	{https://arxiv.org/abs/hep-ph/9506380} {arXiv:hep-ph/9506380} \BibitemShut
	{NoStop}%
	\bibitem [{\citenamefont {Bertone}\ \emph {et~al.}(2005)\citenamefont
		{Bertone}, \citenamefont {Hooper},\ and\ \citenamefont
		{Silk}}]{Bertone:2004pz}%
	\BibitemOpen
	\bibfield  {author} {\bibinfo {author} {\bibfnamefont {G.}~\bibnamefont
			{Bertone}}, \bibinfo {author} {\bibfnamefont {D.}~\bibnamefont {Hooper}},\
		and\ \bibinfo {author} {\bibfnamefont {J.}~\bibnamefont {Silk}},\ }\href
	{https://doi.org/10.1016/j.physrep.2004.08.031} {\bibfield  {journal}
		{\bibinfo  {journal} {Phys. Rept.}\ }\textbf {\bibinfo {volume} {405}},\
		\bibinfo {pages} {279} (\bibinfo {year} {2005})},\ \Eprint
	{https://arxiv.org/abs/hep-ph/0404175} {arXiv:hep-ph/0404175} \BibitemShut
	{NoStop}%
	\bibitem [{\citenamefont {Feng}(2010)}]{Feng:2010gw}%
	\BibitemOpen
	\bibfield  {author} {\bibinfo {author} {\bibfnamefont {J.~L.}\ \bibnamefont
			{Feng}},\ }\href {https://doi.org/10.1146/annurev-astro-082708-101659}
	{\bibfield  {journal} {\bibinfo  {journal} {Ann. Rev. Astron. Astrophys.}\
		}\textbf {\bibinfo {volume} {48}},\ \bibinfo {pages} {495} (\bibinfo {year}
		{2010})},\ \Eprint {https://arxiv.org/abs/1003.0904} {arXiv:1003.0904
		[astro-ph.CO]} \BibitemShut {NoStop}%
	%%CITATION = ARXIV:1003.0904;%%
	\bibitem [{\citenamefont {Arcadi}\ \emph {et~al.}(2017)\citenamefont {Arcadi},
		\citenamefont {Dutra}, \citenamefont {Ghosh}, \citenamefont {Lindner},
		\citenamefont {Mambrini}, \citenamefont {Pierre}, \citenamefont {Profumo},\
		and\ \citenamefont {Queiroz}}]{Arcadi:2017kky}%
	\BibitemOpen
	\bibfield  {author} {\bibinfo {author} {\bibfnamefont {G.}~\bibnamefont
			{Arcadi}}, \bibinfo {author} {\bibfnamefont {M.}~\bibnamefont {Dutra}},
		\bibinfo {author} {\bibfnamefont {P.}~\bibnamefont {Ghosh}}, \bibinfo
		{author} {\bibfnamefont {M.}~\bibnamefont {Lindner}}, \bibinfo {author}
		{\bibfnamefont {Y.}~\bibnamefont {Mambrini}}, \bibinfo {author}
		{\bibfnamefont {M.}~\bibnamefont {Pierre}}, \bibinfo {author} {\bibfnamefont
			{S.}~\bibnamefont {Profumo}},\ and\ \bibinfo {author} {\bibfnamefont {F.~S.}\
			\bibnamefont {Queiroz}},\ }\href@noop {} {\  (\bibinfo {year} {2017})},\
	\Eprint {https://arxiv.org/abs/1703.07364} {arXiv:1703.07364 [hep-ph]}
	\BibitemShut {NoStop}%
	%%CITATION = ARXIV:1703.07364;%%
	\bibitem [{\citenamefont {Roszkowski}\ \emph {et~al.}(2018)\citenamefont
		{Roszkowski}, \citenamefont {Sessolo},\ and\ \citenamefont
		{Trojanowski}}]{Roszkowski:2017nbc}%
	\BibitemOpen
	\bibfield  {author} {\bibinfo {author} {\bibfnamefont {L.}~\bibnamefont
			{Roszkowski}}, \bibinfo {author} {\bibfnamefont {E.~M.}\ \bibnamefont
			{Sessolo}},\ and\ \bibinfo {author} {\bibfnamefont {S.}~\bibnamefont
			{Trojanowski}},\ }\href {https://doi.org/10.1088/1361-6633/aab913} {\bibfield
		{journal} {\bibinfo  {journal} {Rept. Prog. Phys.}\ }\textbf {\bibinfo
			{volume} {81}},\ \bibinfo {pages} {066201} (\bibinfo {year} {2018})},\
	\Eprint {https://arxiv.org/abs/1707.06277} {arXiv:1707.06277 [hep-ph]}
	\BibitemShut {NoStop}%
	\bibitem [{\citenamefont {Weinberg}(1979)}]{Weinberg:1979bt}%
	\BibitemOpen
	\bibfield  {author} {\bibinfo {author} {\bibfnamefont {S.}~\bibnamefont
			{Weinberg}},\ }\href {https://doi.org/10.1103/PhysRevLett.42.850} {\bibfield
		{journal} {\bibinfo  {journal} {Phys. Rev. Lett.}\ }\textbf {\bibinfo
			{volume} {42}},\ \bibinfo {pages} {850} (\bibinfo {year} {1979})}\BibitemShut
	{NoStop}%
	%%CITATION = PRLTA,42,850;%%
	\bibitem [{\citenamefont {Kolb}\ and\ \citenamefont
		{Wolfram}(1980)}]{Kolb:1979qa}%
	\BibitemOpen
	\bibfield  {author} {\bibinfo {author} {\bibfnamefont {E.~W.}\ \bibnamefont
			{Kolb}}\ and\ \bibinfo {author} {\bibfnamefont {S.}~\bibnamefont {Wolfram}},\
	}\href {https://doi.org/10.1016/0550-3213(80)90167-4,
		10.1016/0550-3213(82)90012-8} {\bibfield  {journal} {\bibinfo  {journal}
			{Nucl. Phys.}\ }\textbf {\bibinfo {volume} {B172}},\ \bibinfo {pages} {224}
		(\bibinfo {year} {1980})},\ \bibinfo {note} {[Erratum: Nucl.
		Phys.B195,542(1982)]}\BibitemShut {NoStop}%
	%%CITATION = NUPHA,B172,224;%%
	\bibitem [{\citenamefont {Fukugita}\ and\ \citenamefont
		{Yanagida}(1986)}]{Fukugita:1986hr}%
	\BibitemOpen
	\bibfield  {author} {\bibinfo {author} {\bibfnamefont {M.}~\bibnamefont
			{Fukugita}}\ and\ \bibinfo {author} {\bibfnamefont {T.}~\bibnamefont
			{Yanagida}},\ }\href {https://doi.org/10.1016/0370-2693(86)91126-3}
	{\bibfield  {journal} {\bibinfo  {journal} {Phys. Lett.}\ }\textbf {\bibinfo
			{volume} {B174}},\ \bibinfo {pages} {45} (\bibinfo {year}
		{1986})}\BibitemShut {NoStop}%
	%%CITATION = PHLTA,B174,45;%%
	\bibitem [{\citenamefont {Petraki}\ and\ \citenamefont
		{Volkas}(2013)}]{Petraki:2013wwa}%
	\BibitemOpen
	\bibfield  {author} {\bibinfo {author} {\bibfnamefont {K.}~\bibnamefont
			{Petraki}}\ and\ \bibinfo {author} {\bibfnamefont {R.~R.}\ \bibnamefont
			{Volkas}},\ }\href {https://doi.org/10.1142/S0217751X13300287} {\bibfield
		{journal} {\bibinfo  {journal} {Int. J. Mod. Phys.}\ }\textbf {\bibinfo
			{volume} {A28}},\ \bibinfo {pages} {1330028} (\bibinfo {year} {2013})},\
	\Eprint {https://arxiv.org/abs/1305.4939} {arXiv:1305.4939 [hep-ph]}
	\BibitemShut {NoStop}%
	%%CITATION = ARXIV:1305.4939;%%
	\bibitem [{\citenamefont {Zurek}(2014)}]{Zurek:2013wia}%
	\BibitemOpen
	\bibfield  {author} {\bibinfo {author} {\bibfnamefont {K.~M.}\ \bibnamefont
			{Zurek}},\ }\href {https://doi.org/10.1016/j.physrep.2013.12.001} {\bibfield
		{journal} {\bibinfo  {journal} {Phys. Rept.}\ }\textbf {\bibinfo {volume}
			{537}},\ \bibinfo {pages} {91} (\bibinfo {year} {2014})},\ \Eprint
	{https://arxiv.org/abs/1308.0338} {arXiv:1308.0338 [hep-ph]} \BibitemShut
	{NoStop}%
	%%CITATION = ARXIV:1308.0338;%%
	\bibitem [{\citenamefont {Nussinov}(1985)}]{Nussinov:1985xr}%
	\BibitemOpen
	\bibfield  {author} {\bibinfo {author} {\bibfnamefont {S.}~\bibnamefont
			{Nussinov}},\ }\href {https://doi.org/10.1016/0370-2693(85)90689-6}
	{\bibfield  {journal} {\bibinfo  {journal} {Phys. Lett.}\ }\textbf {\bibinfo
			{volume} {165B}},\ \bibinfo {pages} {55} (\bibinfo {year}
		{1985})}\BibitemShut {NoStop}%
	%%CITATION = PHLTA,165B,55;%%
	\bibitem [{\citenamefont {Kaplan}\ \emph {et~al.}(2009)\citenamefont {Kaplan},
		\citenamefont {Luty},\ and\ \citenamefont {Zurek}}]{Kaplan:2009ag}%
	\BibitemOpen
	\bibfield  {author} {\bibinfo {author} {\bibfnamefont {D.~E.}\ \bibnamefont
			{Kaplan}}, \bibinfo {author} {\bibfnamefont {M.~A.}\ \bibnamefont {Luty}},\
		and\ \bibinfo {author} {\bibfnamefont {K.~M.}\ \bibnamefont {Zurek}},\ }\href
	{https://doi.org/10.1103/PhysRevD.79.115016} {\bibfield  {journal} {\bibinfo
			{journal} {Phys. Rev. D}\ }\textbf {\bibinfo {volume} {79}},\ \bibinfo
		{pages} {115016} (\bibinfo {year} {2009})},\ \Eprint
	{https://arxiv.org/abs/0901.4117} {arXiv:0901.4117 [hep-ph]} \BibitemShut
	{NoStop}%
	\bibitem [{\citenamefont {Davoudiasl}\ and\ \citenamefont
		{Mohapatra}(2012)}]{Davoudiasl:2012uw}%
	\BibitemOpen
	\bibfield  {author} {\bibinfo {author} {\bibfnamefont {H.}~\bibnamefont
			{Davoudiasl}}\ and\ \bibinfo {author} {\bibfnamefont {R.~N.}\ \bibnamefont
			{Mohapatra}},\ }\href {https://doi.org/10.1088/1367-2630/14/9/095011}
	{\bibfield  {journal} {\bibinfo  {journal} {New J. Phys.}\ }\textbf {\bibinfo
			{volume} {14}},\ \bibinfo {pages} {095011} (\bibinfo {year} {2012})},\
	\Eprint {https://arxiv.org/abs/1203.1247} {arXiv:1203.1247 [hep-ph]}
	\BibitemShut {NoStop}%
	%%CITATION = ARXIV:1203.1247;%%
	\bibitem [{\citenamefont {Dutta~Banik}\ \emph {et~al.}(2021)\citenamefont
		{Dutta~Banik}, \citenamefont {Roshan},\ and\ \citenamefont
		{Sil}}]{DuttaBanik:2020vfr}%
	\BibitemOpen
	\bibfield  {author} {\bibinfo {author} {\bibfnamefont {A.}~\bibnamefont
			{Dutta~Banik}}, \bibinfo {author} {\bibfnamefont {R.}~\bibnamefont
			{Roshan}},\ and\ \bibinfo {author} {\bibfnamefont {A.}~\bibnamefont {Sil}},\
	}\href {https://doi.org/10.1088/1475-7516/2021/03/037} {\bibfield  {journal}
		{\bibinfo  {journal} {JCAP}\ }\textbf {\bibinfo {volume} {03}},\ \bibinfo
		{pages} {037}},\ \Eprint {https://arxiv.org/abs/2011.04371} {arXiv:2011.04371
		[hep-ph]} \BibitemShut {NoStop}%
	\bibitem [{\citenamefont {Barman}\ \emph {et~al.}(2022)\citenamefont {Barman},
		\citenamefont {Borah}, \citenamefont {Das},\ and\ \citenamefont
		{Roshan}}]{Barman:2021ost}%
	\BibitemOpen
	\bibfield  {author} {\bibinfo {author} {\bibfnamefont {B.}~\bibnamefont
			{Barman}}, \bibinfo {author} {\bibfnamefont {D.}~\bibnamefont {Borah}},
		\bibinfo {author} {\bibfnamefont {S.~J.}\ \bibnamefont {Das}},\ and\ \bibinfo
		{author} {\bibfnamefont {R.}~\bibnamefont {Roshan}},\ }\href
	{https://doi.org/10.1088/1475-7516/2022/03/031} {\bibfield  {journal}
		{\bibinfo  {journal} {JCAP}\ }\textbf {\bibinfo {volume} {03}}\bibfield
		{number} {\bibinfo  {number} { (03)},\ \bibinfo {pages} {031}},\ }\Eprint
	{https://arxiv.org/abs/2111.08034} {arXiv:2111.08034 [hep-ph]} \BibitemShut
	{NoStop}%
	\bibitem [{\citenamefont {Cui}\ and\ \citenamefont
		{Shamma}(2020)}]{Cui:2020dly}%
	\BibitemOpen
	\bibfield  {author} {\bibinfo {author} {\bibfnamefont {Y.}~\bibnamefont
			{Cui}}\ and\ \bibinfo {author} {\bibfnamefont {M.}~\bibnamefont {Shamma}},\
	}\href {https://doi.org/10.1007/JHEP12(2020)046} {\bibfield  {journal}
		{\bibinfo  {journal} {JHEP}\ }\textbf {\bibinfo {volume} {12}},\ \bibinfo
		{pages} {046}},\ \Eprint {https://arxiv.org/abs/2002.05170} {arXiv:2002.05170
		[hep-ph]} \BibitemShut {NoStop}%
	\bibitem [{\citenamefont {Falkowski}\ \emph {et~al.}(2011)\citenamefont
		{Falkowski}, \citenamefont {Ruderman},\ and\ \citenamefont
		{Volansky}}]{Falkowski:2011xh}%
	\BibitemOpen
	\bibfield  {author} {\bibinfo {author} {\bibfnamefont {A.}~\bibnamefont
			{Falkowski}}, \bibinfo {author} {\bibfnamefont {J.~T.}\ \bibnamefont
			{Ruderman}},\ and\ \bibinfo {author} {\bibfnamefont {T.}~\bibnamefont
			{Volansky}},\ }\href {https://doi.org/10.1007/JHEP05(2011)106} {\bibfield
		{journal} {\bibinfo  {journal} {JHEP}\ }\textbf {\bibinfo {volume} {05}},\
		\bibinfo {pages} {106}},\ \Eprint {https://arxiv.org/abs/1101.4936}
	{arXiv:1101.4936 [hep-ph]} \BibitemShut {NoStop}%
	\bibitem [{\citenamefont {Patel}\ \emph {et~al.}(2022)\citenamefont {Patel},
		\citenamefont {Malhotra}, \citenamefont {Patra},\ and\ \citenamefont
		{Yajnik}}]{Patel:2022xyv}%
	\BibitemOpen
	\bibfield  {author} {\bibinfo {author} {\bibfnamefont {U.}~\bibnamefont
			{Patel}}, \bibinfo {author} {\bibfnamefont {L.}~\bibnamefont {Malhotra}},
		\bibinfo {author} {\bibfnamefont {S.}~\bibnamefont {Patra}},\ and\ \bibinfo
		{author} {\bibfnamefont {U.~A.}\ \bibnamefont {Yajnik}},\ }\href@noop {} {\
		(\bibinfo {year} {2022})},\ \Eprint {https://arxiv.org/abs/2211.04722}
	{arXiv:2211.04722 [hep-ph]} \BibitemShut {NoStop}%
	\bibitem [{\citenamefont {Biswas}\ \emph {et~al.}(2019)\citenamefont {Biswas},
		\citenamefont {Choubey}, \citenamefont {Covi},\ and\ \citenamefont
		{Khan}}]{Biswas:2018sib}%
	\BibitemOpen
	\bibfield  {author} {\bibinfo {author} {\bibfnamefont {A.}~\bibnamefont
			{Biswas}}, \bibinfo {author} {\bibfnamefont {S.}~\bibnamefont {Choubey}},
		\bibinfo {author} {\bibfnamefont {L.}~\bibnamefont {Covi}},\ and\ \bibinfo
		{author} {\bibfnamefont {S.}~\bibnamefont {Khan}},\ }\href
	{https://doi.org/10.1007/JHEP05(2019)193} {\bibfield  {journal} {\bibinfo
			{journal} {JHEP}\ }\textbf {\bibinfo {volume} {05}},\ \bibinfo {pages}
		{193}},\ \Eprint {https://arxiv.org/abs/1812.06122} {arXiv:1812.06122
		[hep-ph]} \BibitemShut {NoStop}%
	\bibitem [{\citenamefont {Narendra}\ \emph {et~al.}(2018)\citenamefont
		{Narendra}, \citenamefont {Patra}, \citenamefont {Sahu},\ and\ \citenamefont
		{Shil}}]{Narendra:2018vfw}%
	\BibitemOpen
	\bibfield  {author} {\bibinfo {author} {\bibfnamefont {N.}~\bibnamefont
			{Narendra}}, \bibinfo {author} {\bibfnamefont {S.}~\bibnamefont {Patra}},
		\bibinfo {author} {\bibfnamefont {N.}~\bibnamefont {Sahu}},\ and\ \bibinfo
		{author} {\bibfnamefont {S.}~\bibnamefont {Shil}},\ }\href
	{https://doi.org/10.1103/PhysRevD.98.095016} {\bibfield  {journal} {\bibinfo
			{journal} {Phys. Rev. D}\ }\textbf {\bibinfo {volume} {98}},\ \bibinfo
		{pages} {095016} (\bibinfo {year} {2018})},\ \Eprint
	{https://arxiv.org/abs/1805.04860} {arXiv:1805.04860 [hep-ph]} \BibitemShut
	{NoStop}%
	\bibitem [{\citenamefont {Nagata}\ \emph {et~al.}(2017)\citenamefont {Nagata},
		\citenamefont {Olive},\ and\ \citenamefont {Zheng}}]{Nagata:2016knk}%
	\BibitemOpen
	\bibfield  {author} {\bibinfo {author} {\bibfnamefont {N.}~\bibnamefont
			{Nagata}}, \bibinfo {author} {\bibfnamefont {K.~A.}\ \bibnamefont {Olive}},\
		and\ \bibinfo {author} {\bibfnamefont {J.}~\bibnamefont {Zheng}},\ }\href
	{https://doi.org/10.1088/1475-7516/2017/02/016} {\bibfield  {journal}
		{\bibinfo  {journal} {JCAP}\ }\textbf {\bibinfo {volume} {02}},\ \bibinfo
		{pages} {016}},\ \Eprint {https://arxiv.org/abs/1611.04693} {arXiv:1611.04693
		[hep-ph]} \BibitemShut {NoStop}%
	\bibitem [{\citenamefont {Arina}\ and\ \citenamefont
		{Sahu}(2012)}]{Arina:2011cu}%
	\BibitemOpen
	\bibfield  {author} {\bibinfo {author} {\bibfnamefont {C.}~\bibnamefont
			{Arina}}\ and\ \bibinfo {author} {\bibfnamefont {N.}~\bibnamefont {Sahu}},\
	}\href {https://doi.org/10.1016/j.nuclphysb.2011.09.014} {\bibfield
		{journal} {\bibinfo  {journal} {Nucl. Phys. B}\ }\textbf {\bibinfo {volume}
			{854}},\ \bibinfo {pages} {666} (\bibinfo {year} {2012})},\ \Eprint
	{https://arxiv.org/abs/1108.3967} {arXiv:1108.3967 [hep-ph]} \BibitemShut
	{NoStop}%
	\bibitem [{\citenamefont {Arina}\ \emph {et~al.}(2012)\citenamefont {Arina},
		\citenamefont {Gong},\ and\ \citenamefont {Sahu}}]{Arina:2012fb}%
	\BibitemOpen
	\bibfield  {author} {\bibinfo {author} {\bibfnamefont {C.}~\bibnamefont
			{Arina}}, \bibinfo {author} {\bibfnamefont {J.-O.}\ \bibnamefont {Gong}},\
		and\ \bibinfo {author} {\bibfnamefont {N.}~\bibnamefont {Sahu}},\ }\href
	{https://doi.org/10.1016/j.nuclphysb.2012.07.029} {\bibfield  {journal}
		{\bibinfo  {journal} {Nucl. Phys. B}\ }\textbf {\bibinfo {volume} {865}},\
		\bibinfo {pages} {430} (\bibinfo {year} {2012})},\ \Eprint
	{https://arxiv.org/abs/1206.0009} {arXiv:1206.0009 [hep-ph]} \BibitemShut
	{NoStop}%
	\bibitem [{\citenamefont {Arina}\ \emph {et~al.}(2013)\citenamefont {Arina},
		\citenamefont {Mohapatra},\ and\ \citenamefont {Sahu}}]{Arina:2012aj}%
	\BibitemOpen
	\bibfield  {author} {\bibinfo {author} {\bibfnamefont {C.}~\bibnamefont
			{Arina}}, \bibinfo {author} {\bibfnamefont {R.~N.}\ \bibnamefont
			{Mohapatra}},\ and\ \bibinfo {author} {\bibfnamefont {N.}~\bibnamefont
			{Sahu}},\ }\href {https://doi.org/10.1016/j.physletb.2013.01.059} {\bibfield
		{journal} {\bibinfo  {journal} {Phys. Lett. B}\ }\textbf {\bibinfo {volume}
			{720}},\ \bibinfo {pages} {130} (\bibinfo {year} {2013})},\ \Eprint
	{https://arxiv.org/abs/1211.0435} {arXiv:1211.0435 [hep-ph]} \BibitemShut
	{NoStop}%
	\bibitem [{\citenamefont {Narendra}\ \emph {et~al.}(2021)\citenamefont
		{Narendra}, \citenamefont {Sahu},\ and\ \citenamefont
		{Shil}}]{Narendra:2019cyt}%
	\BibitemOpen
	\bibfield  {author} {\bibinfo {author} {\bibfnamefont {N.}~\bibnamefont
			{Narendra}}, \bibinfo {author} {\bibfnamefont {N.}~\bibnamefont {Sahu}},\
		and\ \bibinfo {author} {\bibfnamefont {S.}~\bibnamefont {Shil}},\ }\href
	{https://doi.org/10.1140/epjc/s10052-021-09882-3} {\bibfield  {journal}
		{\bibinfo  {journal} {Eur. Phys. J. C}\ }\textbf {\bibinfo {volume} {81}},\
		\bibinfo {pages} {1098} (\bibinfo {year} {2021})},\ \Eprint
	{https://arxiv.org/abs/1910.12762} {arXiv:1910.12762 [hep-ph]} \BibitemShut
	{NoStop}%
	\bibitem [{\citenamefont {Mahapatra}\ \emph {et~al.}(2023)\citenamefont
		{Mahapatra}, \citenamefont {Paul}, \citenamefont {Sahu},\ and\ \citenamefont
		{Shukla}}]{Mahapatra:2023dbr}%
	\BibitemOpen
	\bibfield  {author} {\bibinfo {author} {\bibfnamefont {S.}~\bibnamefont
			{Mahapatra}}, \bibinfo {author} {\bibfnamefont {P.~K.}\ \bibnamefont {Paul}},
		\bibinfo {author} {\bibfnamefont {N.}~\bibnamefont {Sahu}},\ and\ \bibinfo
		{author} {\bibfnamefont {P.}~\bibnamefont {Shukla}},\ }\href@noop {} {\
		(\bibinfo {year} {2023})},\ \Eprint {https://arxiv.org/abs/2305.11138}
	{arXiv:2305.11138 [hep-ph]} \BibitemShut {NoStop}%
	\bibitem [{\citenamefont {Borah}\ \emph
		{et~al.}(2023{\natexlab{a}})\citenamefont {Borah}, \citenamefont
		{Jyoti~Das},\ and\ \citenamefont {Okada}}]{Borah:2022qln}%
	\BibitemOpen
	\bibfield  {author} {\bibinfo {author} {\bibfnamefont {D.}~\bibnamefont
			{Borah}}, \bibinfo {author} {\bibfnamefont {S.}~\bibnamefont {Jyoti~Das}},\
		and\ \bibinfo {author} {\bibfnamefont {N.}~\bibnamefont {Okada}},\ }\href
	{https://doi.org/10.1007/JHEP05(2023)004} {\bibfield  {journal} {\bibinfo
			{journal} {JHEP}\ }\textbf {\bibinfo {volume} {05}},\ \bibinfo {pages}
		{004}},\ \Eprint {https://arxiv.org/abs/2212.04516} {arXiv:2212.04516
		[hep-ph]} \BibitemShut {NoStop}%
	\bibitem [{\citenamefont {Borah}\ \emph
		{et~al.}(2023{\natexlab{b}})\citenamefont {Borah}, \citenamefont
		{Jyoti~Das},\ and\ \citenamefont {Roshan}}]{Borah:2023qag}%
	\BibitemOpen
	\bibfield  {author} {\bibinfo {author} {\bibfnamefont {D.}~\bibnamefont
			{Borah}}, \bibinfo {author} {\bibfnamefont {S.}~\bibnamefont {Jyoti~Das}},\
		and\ \bibinfo {author} {\bibfnamefont {R.}~\bibnamefont {Roshan}},\ }\href
	{https://doi.org/10.1103/PhysRevD.108.075025} {\bibfield  {journal} {\bibinfo
			{journal} {Phys. Rev. D}\ }\textbf {\bibinfo {volume} {108}},\ \bibinfo
		{pages} {075025} (\bibinfo {year} {2023}{\natexlab{b}})},\ \Eprint
	{https://arxiv.org/abs/2305.13367} {arXiv:2305.13367 [hep-ph]} \BibitemShut
	{NoStop}%
	\bibitem [{\citenamefont {Zyla}\ \emph {et~al.}(2020)\citenamefont {Zyla} \emph
		{et~al.}}]{ParticleDataGroup:2020ssz}%
	\BibitemOpen
	\bibfield  {author} {\bibinfo {author} {\bibfnamefont {P.~A.}\ \bibnamefont
			{Zyla}} \emph {et~al.} (\bibinfo {collaboration} {Particle Data Group}),\
	}\href {https://doi.org/10.1093/ptep/ptaa104} {\bibfield  {journal} {\bibinfo
			{journal} {PTEP}\ }\textbf {\bibinfo {volume} {2020}},\ \bibinfo {pages}
		{083C01} (\bibinfo {year} {2020})}\BibitemShut {NoStop}%
	\bibitem [{\citenamefont {Dutta}\ \emph {et~al.}(2021)\citenamefont {Dutta},
		\citenamefont {Mahapatra}, \citenamefont {Borah},\ and\ \citenamefont
		{Sahu}}]{Dutta:2021wbn}%
	\BibitemOpen
	\bibfield  {author} {\bibinfo {author} {\bibfnamefont {M.}~\bibnamefont
			{Dutta}}, \bibinfo {author} {\bibfnamefont {S.}~\bibnamefont {Mahapatra}},
		\bibinfo {author} {\bibfnamefont {D.}~\bibnamefont {Borah}},\ and\ \bibinfo
		{author} {\bibfnamefont {N.}~\bibnamefont {Sahu}},\ }\href
	{https://doi.org/10.1103/PhysRevD.103.095018} {\bibfield  {journal} {\bibinfo
			{journal} {Phys. Rev. D}\ }\textbf {\bibinfo {volume} {103}},\ \bibinfo
		{pages} {095018} (\bibinfo {year} {2021})},\ \Eprint
	{https://arxiv.org/abs/2101.06472} {arXiv:2101.06472 [hep-ph]} \BibitemShut
	{NoStop}%
	\bibitem [{\citenamefont {Borah}\ \emph
		{et~al.}(2021{\natexlab{a}})\citenamefont {Borah}, \citenamefont {Dutta},
		\citenamefont {Mahapatra},\ and\ \citenamefont {Sahu}}]{Borah:2021pet}%
	\BibitemOpen
	\bibfield  {author} {\bibinfo {author} {\bibfnamefont {D.}~\bibnamefont
			{Borah}}, \bibinfo {author} {\bibfnamefont {M.}~\bibnamefont {Dutta}},
		\bibinfo {author} {\bibfnamefont {S.}~\bibnamefont {Mahapatra}},\ and\
		\bibinfo {author} {\bibfnamefont {N.}~\bibnamefont {Sahu}},\ }\href@noop {}
	{\  (\bibinfo {year} {2021}{\natexlab{a}})},\ \Eprint
	{https://arxiv.org/abs/2110.00021} {arXiv:2110.00021 [hep-ph]} \BibitemShut
	{NoStop}%
	\bibitem [{\citenamefont {Borah}\ \emph
		{et~al.}(2021{\natexlab{b}})\citenamefont {Borah}, \citenamefont {Dutta},
		\citenamefont {Mahapatra},\ and\ \citenamefont {Sahu}}]{Borah:2021rbx}%
	\BibitemOpen
	\bibfield  {author} {\bibinfo {author} {\bibfnamefont {D.}~\bibnamefont
			{Borah}}, \bibinfo {author} {\bibfnamefont {M.}~\bibnamefont {Dutta}},
		\bibinfo {author} {\bibfnamefont {S.}~\bibnamefont {Mahapatra}},\ and\
		\bibinfo {author} {\bibfnamefont {N.}~\bibnamefont {Sahu}},\ }\href@noop {}
	{\  (\bibinfo {year} {2021}{\natexlab{b}})},\ \Eprint
	{https://arxiv.org/abs/2112.06847} {arXiv:2112.06847 [hep-ph]} \BibitemShut
	{NoStop}%
	\bibitem [{\citenamefont {Borah}\ \emph
		{et~al.}(2022{\natexlab{b}})\citenamefont {Borah}, \citenamefont {Dasgupta},
		\citenamefont {Mahapatra},\ and\ \citenamefont {Sahu}}]{Borah:2021qmi}%
	\BibitemOpen
	\bibfield  {author} {\bibinfo {author} {\bibfnamefont {D.}~\bibnamefont
			{Borah}}, \bibinfo {author} {\bibfnamefont {A.}~\bibnamefont {Dasgupta}},
		\bibinfo {author} {\bibfnamefont {S.}~\bibnamefont {Mahapatra}},\ and\
		\bibinfo {author} {\bibfnamefont {N.}~\bibnamefont {Sahu}},\ }\href
	{https://doi.org/10.1103/PhysRevD.106.095028} {\bibfield  {journal} {\bibinfo
			{journal} {Phys. Rev. D}\ }\textbf {\bibinfo {volume} {106}},\ \bibinfo
		{pages} {095028} (\bibinfo {year} {2022}{\natexlab{b}})},\ \Eprint
	{https://arxiv.org/abs/2112.14786} {arXiv:2112.14786 [hep-ph]} \BibitemShut
	{NoStop}%
	\bibitem [{\citenamefont {Iminniyaz}\ \emph {et~al.}(2011)\citenamefont
		{Iminniyaz}, \citenamefont {Drees},\ and\ \citenamefont
		{Chen}}]{Iminniyaz:2011yp}%
	\BibitemOpen
	\bibfield  {author} {\bibinfo {author} {\bibfnamefont {H.}~\bibnamefont
			{Iminniyaz}}, \bibinfo {author} {\bibfnamefont {M.}~\bibnamefont {Drees}},\
		and\ \bibinfo {author} {\bibfnamefont {X.}~\bibnamefont {Chen}},\ }\href
	{https://doi.org/10.1088/1475-7516/2011/07/003} {\bibfield  {journal}
		{\bibinfo  {journal} {JCAP}\ }\textbf {\bibinfo {volume} {07}},\ \bibinfo
		{pages} {003}},\ \Eprint {https://arxiv.org/abs/1104.5548} {arXiv:1104.5548
		[hep-ph]} \BibitemShut {NoStop}%
	\bibitem [{\citenamefont {Slatyer}(2016)}]{Slatyer:2015jla}%
	\BibitemOpen
	\bibfield  {author} {\bibinfo {author} {\bibfnamefont {T.~R.}\ \bibnamefont
			{Slatyer}},\ }\href {https://doi.org/10.1103/PhysRevD.93.023527} {\bibfield
		{journal} {\bibinfo  {journal} {Phys. Rev. D}\ }\textbf {\bibinfo {volume}
			{93}},\ \bibinfo {pages} {023527} (\bibinfo {year} {2016})},\ \Eprint
	{https://arxiv.org/abs/1506.03811} {arXiv:1506.03811 [hep-ph]} \BibitemShut
	{NoStop}%
	\bibitem [{\citenamefont {Lin}\ \emph {et~al.}(2012)\citenamefont {Lin},
		\citenamefont {Yu},\ and\ \citenamefont {Zurek}}]{Lin:2011gj}%
	\BibitemOpen
	\bibfield  {author} {\bibinfo {author} {\bibfnamefont {T.}~\bibnamefont
			{Lin}}, \bibinfo {author} {\bibfnamefont {H.-B.}\ \bibnamefont {Yu}},\ and\
		\bibinfo {author} {\bibfnamefont {K.~M.}\ \bibnamefont {Zurek}},\ }\href
	{https://doi.org/10.1103/PhysRevD.85.063503} {\bibfield  {journal} {\bibinfo
			{journal} {Phys. Rev. D}\ }\textbf {\bibinfo {volume} {85}},\ \bibinfo
		{pages} {063503} (\bibinfo {year} {2012})},\ \Eprint
	{https://arxiv.org/abs/1111.0293} {arXiv:1111.0293 [hep-ph]} \BibitemShut
	{NoStop}%
	\bibitem [{\citenamefont {Baldes}\ \emph {et~al.}(2018)\citenamefont {Baldes},
		\citenamefont {Cirelli}, \citenamefont {Panci}, \citenamefont {Petraki},
		\citenamefont {Sala},\ and\ \citenamefont {Taoso}}]{Baldes:2017gzu}%
	\BibitemOpen
	\bibfield  {author} {\bibinfo {author} {\bibfnamefont {I.}~\bibnamefont
			{Baldes}}, \bibinfo {author} {\bibfnamefont {M.}~\bibnamefont {Cirelli}},
		\bibinfo {author} {\bibfnamefont {P.}~\bibnamefont {Panci}}, \bibinfo
		{author} {\bibfnamefont {K.}~\bibnamefont {Petraki}}, \bibinfo {author}
		{\bibfnamefont {F.}~\bibnamefont {Sala}},\ and\ \bibinfo {author}
		{\bibfnamefont {M.}~\bibnamefont {Taoso}},\ }\href
	{https://doi.org/10.21468/SciPostPhys.4.6.041} {\bibfield  {journal}
		{\bibinfo  {journal} {SciPost Phys.}\ }\textbf {\bibinfo {volume} {4}},\
		\bibinfo {pages} {041} (\bibinfo {year} {2018})},\ \Eprint
	{https://arxiv.org/abs/1712.07489} {arXiv:1712.07489 [hep-ph]} \BibitemShut
	{NoStop}%
	\bibitem [{\citenamefont {Petraki}\ \emph {et~al.}(2014)\citenamefont
		{Petraki}, \citenamefont {Pearce},\ and\ \citenamefont
		{Kusenko}}]{Petraki:2014uza}%
	\BibitemOpen
	\bibfield  {author} {\bibinfo {author} {\bibfnamefont {K.}~\bibnamefont
			{Petraki}}, \bibinfo {author} {\bibfnamefont {L.}~\bibnamefont {Pearce}},\
		and\ \bibinfo {author} {\bibfnamefont {A.}~\bibnamefont {Kusenko}},\ }\href
	{https://doi.org/10.1088/1475-7516/2014/07/039} {\bibfield  {journal}
		{\bibinfo  {journal} {JCAP}\ }\textbf {\bibinfo {volume} {07}},\ \bibinfo
		{pages} {039}},\ \Eprint {https://arxiv.org/abs/1403.1077} {arXiv:1403.1077
		[hep-ph]} \BibitemShut {NoStop}%
	\bibitem [{\citenamefont {Chen}\ \emph {et~al.}(2023)\citenamefont {Chen},
		\citenamefont {Ye},\ and\ \citenamefont {Zhang}}]{Chen:2023rrl}%
	\BibitemOpen
	\bibfield  {author} {\bibinfo {author} {\bibfnamefont {Z.}~\bibnamefont
			{Chen}}, \bibinfo {author} {\bibfnamefont {K.}~\bibnamefont {Ye}},\ and\
		\bibinfo {author} {\bibfnamefont {M.}~\bibnamefont {Zhang}},\ }\href
	{https://doi.org/10.1103/PhysRevD.107.095027} {\bibfield  {journal} {\bibinfo
			{journal} {Phys. Rev. D}\ }\textbf {\bibinfo {volume} {107}},\ \bibinfo
		{pages} {095027} (\bibinfo {year} {2023})},\ \Eprint
	{https://arxiv.org/abs/2303.11820} {arXiv:2303.11820 [hep-ph]} \BibitemShut
	{NoStop}%
	\bibitem [{\citenamefont {Dutta}\ \emph {et~al.}(2022)\citenamefont {Dutta},
		\citenamefont {Narendra}, \citenamefont {Sahu},\ and\ \citenamefont
		{Shil}}]{Dutta:2022knf}%
	\BibitemOpen
	\bibfield  {author} {\bibinfo {author} {\bibfnamefont {M.}~\bibnamefont
			{Dutta}}, \bibinfo {author} {\bibfnamefont {N.}~\bibnamefont {Narendra}},
		\bibinfo {author} {\bibfnamefont {N.}~\bibnamefont {Sahu}},\ and\ \bibinfo
		{author} {\bibfnamefont {S.}~\bibnamefont {Shil}},\ }\href
	{https://doi.org/10.1103/PhysRevD.106.095017} {\bibfield  {journal} {\bibinfo
			{journal} {Phys. Rev. D}\ }\textbf {\bibinfo {volume} {106}},\ \bibinfo
		{pages} {095017} (\bibinfo {year} {2022})},\ \Eprint
	{https://arxiv.org/abs/2202.04704} {arXiv:2202.04704 [hep-ph]} \BibitemShut
	{NoStop}%
	\bibitem [{\citenamefont {Ghosh}\ \emph {et~al.}(2021)\citenamefont {Ghosh},
		\citenamefont {Ghosh},\ and\ \citenamefont {Mukhopadhyay}}]{Ghosh:2021qbo}%
	\BibitemOpen
	\bibfield  {author} {\bibinfo {author} {\bibfnamefont {A.}~\bibnamefont
			{Ghosh}}, \bibinfo {author} {\bibfnamefont {D.}~\bibnamefont {Ghosh}},\ and\
		\bibinfo {author} {\bibfnamefont {S.}~\bibnamefont {Mukhopadhyay}},\ }\href
	{https://doi.org/10.1103/PhysRevD.104.123543} {\bibfield  {journal} {\bibinfo
			{journal} {Phys. Rev. D}\ }\textbf {\bibinfo {volume} {104}},\ \bibinfo
		{pages} {123543} (\bibinfo {year} {2021})},\ \Eprint
	{https://arxiv.org/abs/2103.14009} {arXiv:2103.14009 [hep-ph]} \BibitemShut
	{NoStop}%
	\bibitem [{\citenamefont {Minkowski}(1977)}]{Minkowski:1977sc}%
	\BibitemOpen
	\bibfield  {author} {\bibinfo {author} {\bibfnamefont {P.}~\bibnamefont
			{Minkowski}},\ }\href {https://doi.org/10.1016/0370-2693(77)90435-X}
	{\bibfield  {journal} {\bibinfo  {journal} {Phys. Lett. B}\ }\textbf
		{\bibinfo {volume} {67}},\ \bibinfo {pages} {421} (\bibinfo {year}
		{1977})}\BibitemShut {NoStop}%
	\bibitem [{\citenamefont {Gell-Mann}\ \emph {et~al.}(1979)\citenamefont
		{Gell-Mann}, \citenamefont {Ramond},\ and\ \citenamefont
		{Slansky}}]{GellMann:1980vs}%
	\BibitemOpen
	\bibfield  {author} {\bibinfo {author} {\bibfnamefont {M.}~\bibnamefont
			{Gell-Mann}}, \bibinfo {author} {\bibfnamefont {P.}~\bibnamefont {Ramond}},\
		and\ \bibinfo {author} {\bibfnamefont {R.}~\bibnamefont {Slansky}},\
	}\href@noop {} {\bibfield  {journal} {\bibinfo  {journal} {Conf. Proc. C}\
		}\textbf {\bibinfo {volume} {790927}},\ \bibinfo {pages} {315} (\bibinfo
		{year} {1979})},\ \Eprint {https://arxiv.org/abs/1306.4669} {arXiv:1306.4669
		[hep-th]} \BibitemShut {NoStop}%
	\bibitem [{\citenamefont {Mohapatra}\ and\ \citenamefont
		{Senjanovic}(1980)}]{Mohapatra:1979ia}%
	\BibitemOpen
	\bibfield  {author} {\bibinfo {author} {\bibfnamefont {R.~N.}\ \bibnamefont
			{Mohapatra}}\ and\ \bibinfo {author} {\bibfnamefont {G.}~\bibnamefont
			{Senjanovic}},\ }\href {https://doi.org/10.1103/PhysRevLett.44.912}
	{\bibfield  {journal} {\bibinfo  {journal} {Phys. Rev. Lett.}\ }\textbf
		{\bibinfo {volume} {44}},\ \bibinfo {pages} {912} (\bibinfo {year}
		{1980})}\BibitemShut {NoStop}%
	\bibitem [{\citenamefont {Schechter}\ and\ \citenamefont
		{Valle}(1980)}]{Schechter:1980gr}%
	\BibitemOpen
	\bibfield  {author} {\bibinfo {author} {\bibfnamefont {J.}~\bibnamefont
			{Schechter}}\ and\ \bibinfo {author} {\bibfnamefont {J.}~\bibnamefont
			{Valle}},\ }\href {https://doi.org/10.1103/PhysRevD.22.2227} {\bibfield
		{journal} {\bibinfo  {journal} {Phys. Rev. D}\ }\textbf {\bibinfo {volume}
			{22}},\ \bibinfo {pages} {2227} (\bibinfo {year} {1980})}\BibitemShut
	{NoStop}%
	\bibitem [{\citenamefont {Schechter}\ and\ \citenamefont
		{Valle}(1982)}]{Schechter:1981cv}%
	\BibitemOpen
	\bibfield  {author} {\bibinfo {author} {\bibfnamefont {J.}~\bibnamefont
			{Schechter}}\ and\ \bibinfo {author} {\bibfnamefont {J.~W.~F.}\ \bibnamefont
			{Valle}},\ }\href {https://doi.org/10.1103/PhysRevD.25.774} {\bibfield
		{journal} {\bibinfo  {journal} {Phys. Rev.}\ }\textbf {\bibinfo {volume}
			{D25}},\ \bibinfo {pages} {774} (\bibinfo {year} {1982})}\BibitemShut
	{NoStop}%
	%%CITATION = PHRVA,D25,774;%%
	\bibitem [{\citenamefont {Foot}\ \emph {et~al.}(1989)\citenamefont {Foot},
		\citenamefont {Lew}, \citenamefont {He},\ and\ \citenamefont
		{Joshi}}]{Foot:1988aq}%
	\BibitemOpen
	\bibfield  {author} {\bibinfo {author} {\bibfnamefont {R.}~\bibnamefont
			{Foot}}, \bibinfo {author} {\bibfnamefont {H.}~\bibnamefont {Lew}}, \bibinfo
		{author} {\bibfnamefont {X.}~\bibnamefont {He}},\ and\ \bibinfo {author}
		{\bibfnamefont {G.~C.}\ \bibnamefont {Joshi}},\ }\href
	{https://doi.org/10.1007/BF01415558} {\bibfield  {journal} {\bibinfo
			{journal} {Z. Phys. C}\ }\textbf {\bibinfo {volume} {44}},\ \bibinfo {pages}
		{441} (\bibinfo {year} {1989})}\BibitemShut {NoStop}%
	\bibitem [{\citenamefont {Mohapatra}\ and\ \citenamefont
		{Senjanovic}(1981)}]{Mohapatra:1980yp}%
	\BibitemOpen
	\bibfield  {author} {\bibinfo {author} {\bibfnamefont {R.~N.}\ \bibnamefont
			{Mohapatra}}\ and\ \bibinfo {author} {\bibfnamefont {G.}~\bibnamefont
			{Senjanovic}},\ }\href {https://doi.org/10.1103/PhysRevD.23.165} {\bibfield
		{journal} {\bibinfo  {journal} {Phys. Rev. D}\ }\textbf {\bibinfo {volume}
			{23}},\ \bibinfo {pages} {165} (\bibinfo {year} {1981})}\BibitemShut
	{NoStop}%
	\bibitem [{\citenamefont {Wetterich}(1981)}]{Wetterich:1981bx}%
	\BibitemOpen
	\bibfield  {author} {\bibinfo {author} {\bibfnamefont {C.}~\bibnamefont
			{Wetterich}},\ }\href {https://doi.org/10.1016/0550-3213(81)90279-0}
	{\bibfield  {journal} {\bibinfo  {journal} {Nucl. Phys.}\ }\textbf {\bibinfo
			{volume} {B187}},\ \bibinfo {pages} {343} (\bibinfo {year}
		{1981})}\BibitemShut {NoStop}%
	%%CITATION = NUPHA,B187,343;%%
	\bibitem [{\citenamefont {Lazarides}\ \emph {et~al.}(1981)\citenamefont
		{Lazarides}, \citenamefont {Shafi},\ and\ \citenamefont
		{Wetterich}}]{Lazarides:1980nt}%
	\BibitemOpen
	\bibfield  {author} {\bibinfo {author} {\bibfnamefont {G.}~\bibnamefont
			{Lazarides}}, \bibinfo {author} {\bibfnamefont {Q.}~\bibnamefont {Shafi}},\
		and\ \bibinfo {author} {\bibfnamefont {C.}~\bibnamefont {Wetterich}},\ }\href
	{https://doi.org/10.1016/0550-3213(81)90354-0} {\bibfield  {journal}
		{\bibinfo  {journal} {Nucl. Phys. B}\ }\textbf {\bibinfo {volume} {181}},\
		\bibinfo {pages} {287} (\bibinfo {year} {1981})}\BibitemShut {NoStop}%
	\bibitem [{\citenamefont {Brahmachari}\ and\ \citenamefont
		{Mohapatra}(1998)}]{Brahmachari:1997cq}%
	\BibitemOpen
	\bibfield  {author} {\bibinfo {author} {\bibfnamefont {B.}~\bibnamefont
			{Brahmachari}}\ and\ \bibinfo {author} {\bibfnamefont {R.~N.}\ \bibnamefont
			{Mohapatra}},\ }\href {https://doi.org/10.1103/PhysRevD.58.015001} {\bibfield
		{journal} {\bibinfo  {journal} {Phys. Rev.}\ }\textbf {\bibinfo {volume}
			{D58}},\ \bibinfo {pages} {015001} (\bibinfo {year} {1998})},\ \Eprint
	{https://arxiv.org/abs/hep-ph/9710371} {arXiv:hep-ph/9710371 [hep-ph]}
	\BibitemShut {NoStop}%
	%%CITATION = HEP-PH/9710371;%%
	\bibitem [{\citenamefont {Sakharov}(1967)}]{Sakharov:1967dj}%
	\BibitemOpen
	\bibfield  {author} {\bibinfo {author} {\bibfnamefont {A.~D.}\ \bibnamefont
			{Sakharov}},\ }\href {https://doi.org/10.1070/PU1991v034n05ABEH002497}
	{\bibfield  {journal} {\bibinfo  {journal} {Pisma Zh. Eksp. Teor. Fiz.}\
		}\textbf {\bibinfo {volume} {5}},\ \bibinfo {pages} {32} (\bibinfo {year}
		{1967})},\ \bibinfo {note} {[Usp. Fiz. Nauk161,no.5,61(1991)]}\BibitemShut
	{NoStop}%
	%%CITATION = ZFPRA,5,32;%%
	\bibitem [{\citenamefont {Hambye}\ \emph {et~al.}(2004)\citenamefont {Hambye},
		\citenamefont {Lin}, \citenamefont {Notari}, \citenamefont {Papucci},\ and\
		\citenamefont {Strumia}}]{Hambye:2003rt}%
	\BibitemOpen
	\bibfield  {author} {\bibinfo {author} {\bibfnamefont {T.}~\bibnamefont
			{Hambye}}, \bibinfo {author} {\bibfnamefont {Y.}~\bibnamefont {Lin}},
		\bibinfo {author} {\bibfnamefont {A.}~\bibnamefont {Notari}}, \bibinfo
		{author} {\bibfnamefont {M.}~\bibnamefont {Papucci}},\ and\ \bibinfo {author}
		{\bibfnamefont {A.}~\bibnamefont {Strumia}},\ }\href
	{https://doi.org/10.1016/j.nuclphysb.2004.06.027} {\bibfield  {journal}
		{\bibinfo  {journal} {Nucl. Phys. B}\ }\textbf {\bibinfo {volume} {695}},\
		\bibinfo {pages} {169} (\bibinfo {year} {2004})},\ \Eprint
	{https://arxiv.org/abs/hep-ph/0312203} {arXiv:hep-ph/0312203} \BibitemShut
	{NoStop}%
	\bibitem [{\citenamefont {Hambye}(2012)}]{Hambye:2012fh}%
	\BibitemOpen
	\bibfield  {author} {\bibinfo {author} {\bibfnamefont {T.}~\bibnamefont
			{Hambye}},\ }\href {https://doi.org/10.1088/1367-2630/14/12/125014}
	{\bibfield  {journal} {\bibinfo  {journal} {New J. Phys.}\ }\textbf {\bibinfo
			{volume} {14}},\ \bibinfo {pages} {125014} (\bibinfo {year} {2012})},\
	\Eprint {https://arxiv.org/abs/1212.2888} {arXiv:1212.2888 [hep-ph]}
	\BibitemShut {NoStop}%
	\bibitem [{\citenamefont {Belyaev}\ \emph {et~al.}(2013)\citenamefont
		{Belyaev}, \citenamefont {Christensen},\ and\ \citenamefont
		{Pukhov}}]{Belyaev:2012qa}%
	\BibitemOpen
	\bibfield  {author} {\bibinfo {author} {\bibfnamefont {A.}~\bibnamefont
			{Belyaev}}, \bibinfo {author} {\bibfnamefont {N.~D.}\ \bibnamefont
			{Christensen}},\ and\ \bibinfo {author} {\bibfnamefont {A.}~\bibnamefont
			{Pukhov}},\ }\href {https://doi.org/10.1016/j.cpc.2013.01.014} {\bibfield
		{journal} {\bibinfo  {journal} {Comput. Phys. Commun.}\ }\textbf {\bibinfo
			{volume} {184}},\ \bibinfo {pages} {1729} (\bibinfo {year} {2013})},\ \Eprint
	{https://arxiv.org/abs/1207.6082} {arXiv:1207.6082 [hep-ph]} \BibitemShut
	{NoStop}%
	%%CITATION = ARXIV:1207.6082;%%
	\bibitem [{\citenamefont {Kuzmin}\ \emph {et~al.}(1985)\citenamefont {Kuzmin},
		\citenamefont {Rubakov},\ and\ \citenamefont {Shaposhnikov}}]{Kuzmin:1985mm}%
	\BibitemOpen
	\bibfield  {author} {\bibinfo {author} {\bibfnamefont {V.~A.}\ \bibnamefont
			{Kuzmin}}, \bibinfo {author} {\bibfnamefont {V.~A.}\ \bibnamefont
			{Rubakov}},\ and\ \bibinfo {author} {\bibfnamefont {M.~E.}\ \bibnamefont
			{Shaposhnikov}},\ }\href {https://doi.org/10.1016/0370-2693(85)91028-7}
	{\bibfield  {journal} {\bibinfo  {journal} {Phys. Lett.}\ }\textbf {\bibinfo
			{volume} {155B}},\ \bibinfo {pages} {36} (\bibinfo {year}
		{1985})}\BibitemShut {NoStop}%
	%%CITATION = PHLTA,155B,36;%%
	\bibitem [{\citenamefont {de~Salas}\ \emph {et~al.}(2021)\citenamefont
		{de~Salas}, \citenamefont {Forero}, \citenamefont {Gariazzo}, \citenamefont
		{Mart\'\i{}nez-Mirav\'e}, \citenamefont {Mena}, \citenamefont {Ternes},
		\citenamefont {T\'ortola},\ and\ \citenamefont {Valle}}]{deSalas:2020pgw}%
	\BibitemOpen
	\bibfield  {author} {\bibinfo {author} {\bibfnamefont {P.~F.}\ \bibnamefont
			{de~Salas}}, \bibinfo {author} {\bibfnamefont {D.~V.}\ \bibnamefont
			{Forero}}, \bibinfo {author} {\bibfnamefont {S.}~\bibnamefont {Gariazzo}},
		\bibinfo {author} {\bibfnamefont {P.}~\bibnamefont {Mart\'\i{}nez-Mirav\'e}},
		\bibinfo {author} {\bibfnamefont {O.}~\bibnamefont {Mena}}, \bibinfo {author}
		{\bibfnamefont {C.~A.}\ \bibnamefont {Ternes}}, \bibinfo {author}
		{\bibfnamefont {M.}~\bibnamefont {T\'ortola}},\ and\ \bibinfo {author}
		{\bibfnamefont {J.~W.~F.}\ \bibnamefont {Valle}},\ }\href
	{https://doi.org/10.1007/JHEP02(2021)071} {\bibfield  {journal} {\bibinfo
			{journal} {JHEP}\ }\textbf {\bibinfo {volume} {02}},\ \bibinfo {pages}
		{071}},\ \Eprint {https://arxiv.org/abs/2006.11237} {arXiv:2006.11237
		[hep-ph]} \BibitemShut {NoStop}%
	\bibitem [{\citenamefont {Baring}\ \emph {et~al.}(2016)\citenamefont {Baring},
		\citenamefont {Ghosh}, \citenamefont {Queiroz},\ and\ \citenamefont
		{Sinha}}]{Baring:2015sza}%
	\BibitemOpen
	\bibfield  {author} {\bibinfo {author} {\bibfnamefont {M.~G.}\ \bibnamefont
			{Baring}}, \bibinfo {author} {\bibfnamefont {T.}~\bibnamefont {Ghosh}},
		\bibinfo {author} {\bibfnamefont {F.~S.}\ \bibnamefont {Queiroz}},\ and\
		\bibinfo {author} {\bibfnamefont {K.}~\bibnamefont {Sinha}},\ }\href
	{https://doi.org/10.1103/PhysRevD.93.103009} {\bibfield  {journal} {\bibinfo
			{journal} {Phys. Rev. D}\ }\textbf {\bibinfo {volume} {93}},\ \bibinfo
		{pages} {103009} (\bibinfo {year} {2016})},\ \Eprint
	{https://arxiv.org/abs/1510.00389} {arXiv:1510.00389 [hep-ph]} \BibitemShut
	{NoStop}%
	\bibitem [{\citenamefont {Ellis}\ \emph {et~al.}(2008)\citenamefont {Ellis},
		\citenamefont {Olive},\ and\ \citenamefont {Savage}}]{Ellis:2008hf}%
	\BibitemOpen
	\bibfield  {author} {\bibinfo {author} {\bibfnamefont {J.~R.}\ \bibnamefont
			{Ellis}}, \bibinfo {author} {\bibfnamefont {K.~A.}\ \bibnamefont {Olive}},\
		and\ \bibinfo {author} {\bibfnamefont {C.}~\bibnamefont {Savage}},\ }\href
	{https://doi.org/10.1103/PhysRevD.77.065026} {\bibfield  {journal} {\bibinfo
			{journal} {Phys. Rev. D}\ }\textbf {\bibinfo {volume} {77}},\ \bibinfo
		{pages} {065026} (\bibinfo {year} {2008})},\ \Eprint
	{https://arxiv.org/abs/0801.3656} {arXiv:0801.3656 [hep-ph]} \BibitemShut
	{NoStop}%
	\bibitem [{\citenamefont {Hoferichter}\ \emph {et~al.}(2017)\citenamefont
		{Hoferichter}, \citenamefont {Klos}, \citenamefont {Men\'endez},\ and\
		\citenamefont {Schwenk}}]{Hoferichter:2017olk}%
	\BibitemOpen
	\bibfield  {author} {\bibinfo {author} {\bibfnamefont {M.}~\bibnamefont
			{Hoferichter}}, \bibinfo {author} {\bibfnamefont {P.}~\bibnamefont {Klos}},
		\bibinfo {author} {\bibfnamefont {J.}~\bibnamefont {Men\'endez}},\ and\
		\bibinfo {author} {\bibfnamefont {A.}~\bibnamefont {Schwenk}},\ }\href
	{https://doi.org/10.1103/PhysRevLett.119.181803} {\bibfield  {journal}
		{\bibinfo  {journal} {Phys. Rev. Lett.}\ }\textbf {\bibinfo {volume} {119}},\
		\bibinfo {pages} {181803} (\bibinfo {year} {2017})},\ \Eprint
	{https://arxiv.org/abs/1708.02245} {arXiv:1708.02245 [hep-ph]} \BibitemShut
	{NoStop}%
	\bibitem [{\citenamefont {Abdelhameed}\ \emph {et~al.}(2019)\citenamefont
		{Abdelhameed} \emph {et~al.}}]{CRESST:2019jnq}%
	\BibitemOpen
	\bibfield  {author} {\bibinfo {author} {\bibfnamefont {A.~H.}\ \bibnamefont
			{Abdelhameed}} \emph {et~al.} (\bibinfo {collaboration} {CRESST}),\ }\href
	{https://doi.org/10.1103/PhysRevD.100.102002} {\bibfield  {journal} {\bibinfo
			{journal} {Phys. Rev. D}\ }\textbf {\bibinfo {volume} {100}},\ \bibinfo
		{pages} {102002} (\bibinfo {year} {2019})},\ \Eprint
	{https://arxiv.org/abs/1904.00498} {arXiv:1904.00498 [astro-ph.CO]}
	\BibitemShut {NoStop}%
	\bibitem [{\citenamefont {Aalbers}\ \emph {et~al.}(2022)\citenamefont {Aalbers}
		\emph {et~al.}}]{LUX-ZEPLIN:2022qhg}%
	\BibitemOpen
	\bibfield  {author} {\bibinfo {author} {\bibfnamefont {J.}~\bibnamefont
			{Aalbers}} \emph {et~al.} (\bibinfo {collaboration} {LUX-ZEPLIN}),\
	}\href@noop {} {\  (\bibinfo {year} {2022})},\ \Eprint
	{https://arxiv.org/abs/2207.03764} {arXiv:2207.03764 [hep-ex]} \BibitemShut
	{NoStop}%
	\bibitem [{\citenamefont {Agnes}\ \emph {et~al.}(2018)\citenamefont {Agnes}
		\emph {et~al.}}]{DarkSide:2018bpj}%
	\BibitemOpen
	\bibfield  {author} {\bibinfo {author} {\bibfnamefont {P.}~\bibnamefont
			{Agnes}} \emph {et~al.} (\bibinfo {collaboration} {DarkSide}),\ }\href
	{https://doi.org/10.1103/PhysRevLett.121.081307} {\bibfield  {journal}
		{\bibinfo  {journal} {Phys. Rev. Lett.}\ }\textbf {\bibinfo {volume} {121}},\
		\bibinfo {pages} {081307} (\bibinfo {year} {2018})},\ \Eprint
	{https://arxiv.org/abs/1802.06994} {arXiv:1802.06994 [astro-ph.HE]}
	\BibitemShut {NoStop}%
	\bibitem [{\citenamefont {Walker}(2013)}]{Walker:2013hka}%
	\BibitemOpen
	\bibfield  {author} {\bibinfo {author} {\bibfnamefont {D.~G.~E.}\
			\bibnamefont {Walker}},\ }\href@noop {} {\  (\bibinfo {year} {2013})},\
	\Eprint {https://arxiv.org/abs/1310.1083} {arXiv:1310.1083 [hep-ph]}
	\BibitemShut {NoStop}%
	\bibitem [{\citenamefont {Pal}\ and\ \citenamefont
		{Wolfenstein}(1982)}]{Pal:1981rm}%
	\BibitemOpen
	\bibfield  {author} {\bibinfo {author} {\bibfnamefont {P.~B.}\ \bibnamefont
			{Pal}}\ and\ \bibinfo {author} {\bibfnamefont {L.}~\bibnamefont
			{Wolfenstein}},\ }\href {https://doi.org/10.1103/PhysRevD.25.766} {\bibfield
		{journal} {\bibinfo  {journal} {Phys. Rev.}\ }\textbf {\bibinfo {volume}
			{D25}},\ \bibinfo {pages} {766} (\bibinfo {year} {1982})}\BibitemShut
	{NoStop}%
	%%CITATION = PHRVA,D25,766;%%
	\bibitem [{\citenamefont {Shrock}(1982)}]{Shrock:1982sc}%
	\BibitemOpen
	\bibfield  {author} {\bibinfo {author} {\bibfnamefont {R.~E.}\ \bibnamefont
			{Shrock}},\ }\href {https://doi.org/10.1016/0550-3213(82)90273-5} {\bibfield
		{journal} {\bibinfo  {journal} {Nucl. Phys. B}\ }\textbf {\bibinfo {volume}
			{206}},\ \bibinfo {pages} {359} (\bibinfo {year} {1982})}\BibitemShut
	{NoStop}%
	\bibitem [{\citenamefont {Ackermann}\ \emph {et~al.}(2015)\citenamefont
		{Ackermann} \emph {et~al.}}]{Fermi-LAT:2015kyq}%
	\BibitemOpen
	\bibfield  {author} {\bibinfo {author} {\bibfnamefont {M.}~\bibnamefont
			{Ackermann}} \emph {et~al.} (\bibinfo {collaboration} {Fermi-LAT}),\ }\href
	{https://doi.org/10.1103/PhysRevD.91.122002} {\bibfield  {journal} {\bibinfo
			{journal} {Phys. Rev. D}\ }\textbf {\bibinfo {volume} {91}},\ \bibinfo
		{pages} {122002} (\bibinfo {year} {2015})},\ \Eprint
	{https://arxiv.org/abs/1506.00013} {arXiv:1506.00013 [astro-ph.HE]}
	\BibitemShut {NoStop}%
	\bibitem [{\citenamefont {Alemanno}\ \emph {et~al.}(2022)\citenamefont
		{Alemanno} \emph {et~al.}}]{DAMPE:2021hsz}%
	\BibitemOpen
	\bibfield  {author} {\bibinfo {author} {\bibfnamefont {F.}~\bibnamefont
			{Alemanno}} \emph {et~al.} (\bibinfo {collaboration} {DAMPE}),\ }\href
	{https://doi.org/10.1016/j.scib.2021.12.015} {\bibfield  {journal} {\bibinfo
			{journal} {Sci. Bull.}\ }\textbf {\bibinfo {volume} {67}},\ \bibinfo {pages}
		{679} (\bibinfo {year} {2022})},\ \Eprint {https://arxiv.org/abs/2112.08860}
	{arXiv:2112.08860 [astro-ph.HE]} \BibitemShut {NoStop}%
	\bibitem [{\citenamefont {Abbasi}\ \emph {et~al.}(2023)\citenamefont {Abbasi}
		\emph {et~al.}}]{IceCube:2023ies}%
	\BibitemOpen
	\bibfield  {author} {\bibinfo {author} {\bibfnamefont {R.}~\bibnamefont
			{Abbasi}} \emph {et~al.} (\bibinfo {collaboration} {IceCube}),\ }\href
	{https://doi.org/10.1103/PhysRevD.108.102004} {\bibfield  {journal} {\bibinfo
			{journal} {Phys. Rev. D}\ }\textbf {\bibinfo {volume} {108}},\ \bibinfo
		{pages} {102004} (\bibinfo {year} {2023})},\ \Eprint
	{https://arxiv.org/abs/2303.13663} {arXiv:2303.13663 [astro-ph.HE]}
	\BibitemShut {NoStop}%
	\bibitem [{\citenamefont {Tulin}\ \emph
		{et~al.}(2013{\natexlab{b}})\citenamefont {Tulin}, \citenamefont {Yu},\ and\
		\citenamefont {Zurek}}]{Tulin:2012wi}%
	\BibitemOpen
	\bibfield  {author} {\bibinfo {author} {\bibfnamefont {S.}~\bibnamefont
			{Tulin}}, \bibinfo {author} {\bibfnamefont {H.-B.}\ \bibnamefont {Yu}},\ and\
		\bibinfo {author} {\bibfnamefont {K.~M.}\ \bibnamefont {Zurek}},\ }\href
	{https://doi.org/10.1103/PhysRevLett.110.111301} {\bibfield  {journal}
		{\bibinfo  {journal} {Phys. Rev. Lett.}\ }\textbf {\bibinfo {volume} {110}},\
		\bibinfo {pages} {111301} (\bibinfo {year} {2013}{\natexlab{b}})},\ \Eprint
	{https://arxiv.org/abs/1210.0900} {arXiv:1210.0900 [hep-ph]} \BibitemShut
	{NoStop}%
	\bibitem [{\citenamefont {Khrapak}\ \emph {et~al.}(2003)\citenamefont
		{Khrapak}, \citenamefont {Ivlev}, \citenamefont {Morfill},\ and\
		\citenamefont {Zhdanov}}]{Khrapak:2003kjw}%
	\BibitemOpen
	\bibfield  {author} {\bibinfo {author} {\bibfnamefont {S.~A.}\ \bibnamefont
			{Khrapak}}, \bibinfo {author} {\bibfnamefont {A.~V.}\ \bibnamefont {Ivlev}},
		\bibinfo {author} {\bibfnamefont {G.~E.}\ \bibnamefont {Morfill}},\ and\
		\bibinfo {author} {\bibfnamefont {S.~K.}\ \bibnamefont {Zhdanov}},\ }\href
	{https://doi.org/10.1103/PhysRevLett.90.225002} {\bibfield  {journal}
		{\bibinfo  {journal} {Phys. Rev. Lett.}\ }\textbf {\bibinfo {volume} {90}},\
		\bibinfo {pages} {225002} (\bibinfo {year} {2003})}\BibitemShut {NoStop}%
	\bibitem [{\citenamefont {Casas}\ and\ \citenamefont
		{Ibarra}(2001)}]{Casas:2001sr}%
	\BibitemOpen
	\bibfield  {author} {\bibinfo {author} {\bibfnamefont {J.}~\bibnamefont
			{Casas}}\ and\ \bibinfo {author} {\bibfnamefont {A.}~\bibnamefont {Ibarra}},\
	}\href {https://doi.org/10.1016/S0550-3213(01)00475-8} {\bibfield  {journal}
		{\bibinfo  {journal} {Nucl. Phys. B}\ }\textbf {\bibinfo {volume} {618}},\
		\bibinfo {pages} {171} (\bibinfo {year} {2001})},\ \Eprint
	{https://arxiv.org/abs/hep-ph/0103065} {arXiv:hep-ph/0103065} \BibitemShut
	{NoStop}%
	\bibitem [{\citenamefont {Harvey}\ and\ \citenamefont
		{Turner}(1990)}]{Harvey:1990qw}%
	\BibitemOpen
	\bibfield  {author} {\bibinfo {author} {\bibfnamefont {J.~A.}\ \bibnamefont
			{Harvey}}\ and\ \bibinfo {author} {\bibfnamefont {M.~S.}\ \bibnamefont
			{Turner}},\ }\href {https://doi.org/10.1103/PhysRevD.42.3344} {\bibfield
		{journal} {\bibinfo  {journal} {Phys. Rev. D}\ }\textbf {\bibinfo {volume}
			{42}},\ \bibinfo {pages} {3344} (\bibinfo {year} {1990})}\BibitemShut
	{NoStop}%
\end{thebibliography}
\end{document}